\documentclass[aps,prl,twocolumn,superscriptaddress]{revtex4-1}

\usepackage{amsmath,amssymb,latexsym}
\usepackage{graphicx}
\usepackage{hyperref}
\usepackage{mathrsfs}
\usepackage{amscd}
\usepackage{color}
\DeclareSymbolFont{symbols}{OMS}{cmsy}{m}{n}
\DeclareSymbolFont{largesymbols}{OMX}{cmex}{m}{n}
\renewcommand{\bm}[1]{\boldsymbol #1}

\begin{document}

\title{
Tachyonic and Plasma Instabilities of $\eta$-Pairing States Coupled to Electromagnetic Fields
}

\author{Naoto Tsuji}
\affiliation{Department of Physics, University of Tokyo, Hongo, Tokyo 113-0033, Japan}
\affiliation{RIKEN Center for Emergent Matter Science (CEMS), Wako 351-0198, Japan}
\author{Masaya Nakagawa}
\affiliation{Department of Physics, University of Tokyo, Hongo, Tokyo 113-0033, Japan}
\author{Masahito Ueda}
\affiliation{Department of Physics, University of Tokyo, Hongo, Tokyo 113-0033, Japan}
\affiliation{RIKEN Center for Emergent Matter Science (CEMS), Wako 351-0198, Japan}
\affiliation{Institute for Physics of Intelligence, University of Tokyo, Hongo, Tokyo 113-0033, Japan}

\begin{abstract}
Cooper pairs featuring a nonzero center-of-mass crystal momentum $\bm Q=(\pi,\pi, \dots)$ and an off-diagonal long-range order 
($\eta$-pairing states) constitute exact eigenstates of a Hubbard model
[C. N. Yang, Phys. Rev. Lett. {\bf 63}, 2144 (1989)].
Here we show that the $\eta$-pairing states are rendered dynamically unstable 
via coupling to dynamical electromagnetic fields.
The instability is caused by ``tachyonic'' electromagnetic fields and
unstable plasma modes for attractive and repulsive interactions, respectively.
The typical time scale of the growth of the instability
is of the order of femtoseconds for electron systems in solids, 
which places a strict bound on the lifetime of the $\eta$-pairing states.
The decay of the $\eta$-pairing states leads to enhanced light emission with frequencies shifted from the Hubbard interaction strength
for the repulsive case and unattenuated electromagnetic penetration for the attractive case.
\end{abstract}


\date{\today}

\maketitle

{\it Introduction.}---
A search for long-lived non-thermal excited states that support
macroscopic long-range order
is a modern challenge in nonequilibrium condensed matter physics. 
If such a state exists, a rich variety of possibilities arises
that could extend the landscape of long-range order
in solid-state materials.
In fact, there have been a number of experimental reports on the possible existence 
of nonequilibrium long-range order,
including light-induced \cite{Fausti2011, Kaiser2014, Hu2014, Mitrano2016, Cantaluppi2018, Buzzi2020, Budden2020} or quench-induced \cite{Oike2018} superconductivity (see also \cite{Niwa2019, Zhang2020}), photoinduced ferromagnetism \cite{Matsubara2007, McLeod2020}
and charge density waves \cite{Stojchevska2014, Vaskivskyi2015, Sun2018}.
While the experimental progress is underway, the theoretical understanding 
is yet to be made.
A major challenge is that an analysis of excited states in quantum many-body systems often requires approximations
that render the conclusion on the existence of such a state highly nontrivial.

A very exception to this situation is the $\eta$-pairing states,
which are known to be exact eigenstates of a Hubbard model
as revealed by C. N. Yang \cite{Yang1989}.
The $\eta$-pairing states exhibit a number of remarkable features.
In particular, they have an off-diagonal long-range order (ODLRO) in arbitrary dimensions
even though their eigenenergies lie much higher than the ground state.
This is to be contrasted 
with finite-temperature thermal states, which cannot show ODLROs in one and two dimensions
due to the Mermin-Wagner theorem. The non-thermal nature of the $\eta$-pairing states has also been discussed recently
in the context of quantum many-body scars \cite{Vafek2017, Mark2020, Moudgalya2020}.

The presence of such a non-thermal state suggests that the $\eta$-pairing states
with an ODLRO
(and hence superconductivity) might be realized in nonequilibrium situations. Recent theoretical studies have
demonstrated that this is indeed possible in several different setups, including 
periodic \cite{Kitamura2016, Peronaci2020, Cook2020, Tindall2021} and
pulsed \cite{Kaneko2019, Kaneko2020, Werner2019, Li2020} electric-field drives,
dissipation engineering \cite{Diehl2008, Kraus2008},
spin-dependent dephasing \cite{Bernier2013, Tindall2019},
and spontaneous light emission \cite{Nakagawa2021}.
These mechanisms will work for the Hubbard model with or without coupling to an external bath,
which may be realized in electrically neutral ultracold atoms trapped in an optical lattice.

In view of applications to real materials, 
one cannot ignore the coupling of electrons to {\it dynamical electromagnetic fields}, since electrons have electric charges.
This point is crucial for the stability of the $\eta$-pairing states supported by the long lifetime of doublons.
If doublons decay into single particles or lose their momenta, they induce local electric currents due to charge transfer,
which then generate dynamical electromagnetic fields.
The effect of the latter feedbacks to electrons, and causes collective modes of electromagnetic fields,
which accelerate the relaxation of doublons.
Such a dynamical instability 
deserves careful scrutiny in view of growing attention in nonequilibrium superconductivity.

In this Letter, we study the dynamics of the $\eta$-pairing states in the Hubbard model coupled to dynamical electromagnetic fields.
Our approach is based on the exact solution of the electromagnetic response function (or the Meissner kernel) 
$K^{\mu\nu}(\bm q,\omega)$
with full momentum ($\bm q$) and frequency ($\omega$) dependences.
In contrast, previous studies have focused on the static and uniform limit (i.e., $\bm q=\omega=0$) \cite{Su1991, Su1992, Kaneko2020}.
As we will see, the momentum and frequency dependences play a pivotal role in dynamical instabilities of $\eta$-pairing states.
Combining the obtained results with the Maxwell equations,
we rigorously prove the existence of the ``tachyonic'' and plasma instabilities for attractively and repulsively interacting systems,
respectively. The time scale of the growth of the instability is surprisingly short,
being of the order of femtoseconds or even shorter than that for ordinary materials. 
This puts a severe constraint on the lifetime of the $\eta$-pairing states in electron systems. 
Finally, we discuss that
the decay of the $\eta$-pairing states leads to intense light emission with frequencies shifted from the interaction strength in the repulsive case,
and unattenuated penetration of electromagnetic fields in the attractive case.



{\it $\eta$ pairing in the Hubbard model.---}
We consider the Hubbard model on a $d$-dimensional cubic lattice subject to the periodic boundary condition with the Hamiltonian,
\begin{align}
H
&=
-t_h\sum_{\langle ij\rangle, \sigma} (c_{i\sigma}^\dagger c_{j\sigma}+\mbox{H.c.})
+U\sum_i n_{i\uparrow}n_{i\downarrow}
-\frac{U}{2}\sum_{i\sigma} n_{i\sigma},
\label{eq: Hubbard}
\end{align}
where $t_h$ ($>0$) is the hopping amplitude, $c_{i\sigma}^\dagger$ is a creation operator of an electron
at site $i$ with spin $\sigma=\uparrow,\downarrow$, 
$\langle ij\rangle$ represents a pair of nearest-neighbor lattice sites,
$U$ is the on-site interaction strength, 
and $n_{i\sigma}=c_{i\sigma}^\dagger c_{i\sigma}$ is the particle-number operator.
Since we fix the total number of electrons throughout this Letter, the last term in Eq.~(\ref{eq: Hubbard}) is a constant.
We set the lattice constant $a=1$ and the Planck constant $\hbar=1$ unless otherwise noted.

The Hubbard model (\ref{eq: Hubbard}) has the spin SU(2) symmetry
together with the ``hidden'' $\eta$ SU(2) symmetry, 
which altogether form the symmetry of ${\rm SU(2)}\times{\rm SU(2)}/\mathbb Z_2\simeq{\rm SO(4)}$ \cite{YangZhang1990}.
The existence of $\eta$-pairing states as the exact eigenstates of the Hubbard model essentially relies on
this fact. To see the $\eta$ symmetry, we define the $\eta$ operators,
$\eta^+:=\sum_j e^{i\bm Q\cdot\bm R_j} c_{j\uparrow}^\dagger c_{j\downarrow}^\dagger$,
$\eta^-:=(\eta^+)^\dagger$, and
$\eta^z:=\frac{1}{2}\sum_j (n_{j\uparrow}+n_{j\downarrow}-1)$,
where $\bm Q=(\pi,\pi,\dots)$ is the momentum at the Brillouin-zone corner,
and $\bm R_j$ is the position vector of lattice site $j$. The $\eta$ operators satisfy the ordinary su(2) algebra,
i.e., $[\eta^+, \eta^-]=2\eta^z$ and $[\eta^z, \eta^\pm]=\pm\eta^\pm$. 
From direct calculations, one can confirm that they all commute with the Hamiltonian (\ref{eq: Hubbard}):
$[H, \eta^\alpha]=0$ ($\alpha=\pm, z$).

Using the $\eta$ operators, one can construct Yang's $\eta$-pairing states.
The simplest one is
\begin{align}
|\psi_N\rangle
&=
\frac{1}{\sqrt{\mathcal N_N}}
(\eta^+)^{\frac{N}{2}}|0\rangle,
\label{eq: psi_N}
\end{align}
where $N$ is the number of electrons which is assumed to be an even integer,
$\mathcal N_N$ is the normalization constant (such that $\langle \psi_N|\psi_N\rangle=1$),
and $|0\rangle$ is the vacuum state.
The $\eta$-pairing state $|\psi_N\rangle$ consists of $\frac{N}{2}$ doublons having momentum $\bm Q$.
Since $\eta^+$ commutes with $H$ (\ref{eq: Hubbard}),
$|\psi_N\rangle$ (\ref{eq: psi_N}) is indeed the exact eigenstate of $H$ with the eigenenergy $E_N=0$
in arbitrary dimensions.
The state $|\psi_N\rangle$ (\ref{eq: psi_N}) has the ODLRO
$\frac{1}{2}\langle \psi_N| (c_{i\uparrow}^\dagger c_{i\downarrow}^\dagger c_{j\downarrow}c_{j\uparrow}+\mbox{H.c.})|\psi_N\rangle
=e^{i\bm Q\cdot(\bm R_i-\bm R_j)}C_{M,N}$ \cite{Yang1989} 
($C_{M,N}:=\frac{\frac{N}{2}(M-\frac{N}{2})}{M(M-1)}$ and $M$ is the number of lattice sites),
which saturates the upper bound of ODLRO \cite{Yang1962, Nakagawa2021}.
Physically, $|\psi_N\rangle$ (\ref{eq: psi_N}) corresponds to the condensate of
spin-singlet Cooper pairs with the center-of-mass momentum $\bm Q$.

{\it Electromagnetic response of $\eta$-pairing states.---}
We study the electromagnetic response of the $\eta$-pairing state $|\psi_N\rangle$ (\ref{eq: psi_N})
within the linear-response regime. We focus on the three-dimensional case ($d=3$).
However, most of the results in the present Letter can straightforwardly be extended to other dimensions.
The response of the current against an external electromagnetic field with momentum $\bm q$ and frequency $\omega$ is given by 
$j^\mu(\bm q,\omega)=-K^{\mu\nu}(\bm q,\omega)A_\nu(\bm q,\omega)$ ($\mu,\nu=x,y,z$),
where $K^{\mu\nu}(\bm q,\omega)$ is the Meissner kernel \cite{SchriefferBook}
and $A_\nu(\bm q,\omega)$ is the vector potential.

In general, the kernel $K^{\mu\nu}(\bm q,\omega)$ consists of the paramagnetic and diamagnetic components \cite{SchriefferBook}.
In the case of $\eta$-pairing states, the diamagnetic component vanishes exactly, since it is proportional to
the kinetic energy \cite{kinetic_energy},
which vanishes for the $\eta$-pairing states. This is in stark contrast to ordinary superconductors,
in which perfect diamagnetism arises from the diamagnetic component of the Meissner kernel.
In the $\eta$-pairing states, the paramagnetic component takes over the role of the diamagnetic one
in ordinary superconductors.

The paramagnetic component is given by the Kubo formula,
\begin{align}
K_{\rm para}^{\mu\nu}(\bm R_j, t)
&=
-i\theta(t)\langle \psi_N| [J^\mu(\bm R_j, t), J^\nu(0, 0)] |\psi_N\rangle,
\label{eq: K_para}
\end{align}
where $\theta(t)$ is the unit-step function ($\theta(t)=1$ for $t\ge 0$ and $\theta(t)=0$ otherwise),
and $J^\mu(\bm R_j, t)$ is the local current operator at site $j$ and time $t$ in the Heisenberg picture.
The local current $J^\mu(\bm R_j, 0)=J^\mu(\bm R_j)$ is expressed explicitly as 
$J^\mu(\bm R_j)=-iet_h\sum_\sigma (c_{j+\mu,\sigma}^\dagger c_{j\sigma}
-c_{j\sigma}^\dagger c_{j+\mu,\sigma})$, 
where $e$ is the electric charge, and $j+\mu$ represents the nearest-neighbor site of $j$ in the $\mu$ direction.

We can evaluate Eq.~(\ref{eq: K_para}) exactly for arbitrary $\bm R_j$ and $t$
using the following algebraic relations:
$[J^\mu(\bm R_j), \eta^\pm]=:\pm 2J_\eta^{\mu\pm}(\bm R_j)$,
$[J_\eta^{\mu\pm}(\bm R_j), \eta^\pm]=0$, and
$[J_\eta^{\mu\pm}(\bm R_j), \eta^\mp]=\pm J^\mu(\bm R_j)$.
They allow us to reduce the $N$-particle correlation function (\ref{eq: K_para}) to that of the vacuum state \cite{supplementary},
\begin{align}
K_{\rm para}^{\mu\nu}(\bm R_j,t)
&=
-4i\theta(t) C_{M,N}
[\langle 0| J_\eta^{\mu-}(\bm R_j,t)J_\eta^{\nu+}(0,0) |0\rangle
\notag
\\
&\quad
-\langle 0| J_\eta^{\nu-}(0,0)J_\eta^{\mu+}(\bm R_j,t) |0\rangle].
\label{eq: K reduced}
\end{align}
In this way, the $N$-particle problem reduces to the two-particle problem, which is exactly solvable \cite{EsslerBook}.

We further decompose the kernel into the transverse and longitudinal components.
Without loss of generality, we assume that the momentum $\bm q$ of the vector potential points in the $z$ direction.
The transverse component is defined as $K^\perp(\bm q,\omega):=K^{\mu\mu}(\bm q,\omega)$ ($\mu=x,y$),
while the longitudinal one is $K^\parallel(\bm q,\omega):=K^{zz}(\bm q,\omega)$.
The kernel does not have the off-diagonal components ($K^{\mu\nu}(\bm q,\omega)=0$ for $\mu\neq\nu$) \cite{supplementary}.
In the two-particle dynamics involved in Eq.~(\ref{eq: K reduced}), the center-of-mass momentum 
and the relative coordinates in the $x$ and $y$ directions of the two particles are conserved.
In addition, for the transverse components the two particles never sit at the same site, making the dynamics effectively noninteracting.
These observations lead us to an analytical solution for the transverse component \cite{supplementary},
\begin{align}
K^\perp(\bm q, \omega)
&=
-8ie^2t_h
C_{M,N}\left[f_q\big(\tfrac{\omega+U}{t_h}\big)-f_q\big(\tfrac{\omega-U}{t_h}\big)\right],
\notag
\\
f_q(x)
&:=
\begin{cases}
\frac{i{\rm sgn}(x)}{\sqrt{x^2-16\sin^2\frac{q}{2}}} & {\rm for}\; |x|>4\sin\frac{q}{2};
\\
\frac{1}{\sqrt{16\sin^2\frac{q}{2}-x^2}} & {\rm for}\; |x|<4\sin\frac{q}{2},
\end{cases}
\label{eq: K_perp}
\end{align}
where $q$ $(>0)$ is the $z$ component of $\bm q$.
The longitudinal component does not have such a compact expression, but can be evaluated exactly
in a similar manner \cite{supplementary}.

The exact solution (\ref{eq: K_perp}) for the electromagnetic response function 
reveals a number of important properties of the $\eta$-pairing states. 
By taking the limit $\lim_{\bm q\to 0}\lim_{\omega\to 0} K^\perp(\bm q,\omega)=\frac{16e^2t_h^2}{U}C_{M,N}=:\frac{1}{\pi}D_s$,
one can recover the Meissner weight or superfluid stiffness $D_s$ in Refs.~\cite{Su1991, Su1992}.
By taking another limit $\lim_{\bm q\to 0}K^\perp(\bm q,\omega)=:-i\omega\sigma(\omega)$, one obtains the optical conductivity,
$\sigma(\omega)=\frac{8ie^2t_h^2 C_{M,N}}{\omega+i\delta}(\frac{1}{\omega+U+i\delta}-\frac{1}{\omega-U+i\delta})$
($\delta$ is a positive infinitesimal constant), the real part of which shows delta-function-like peaks at $\omega=0$ and $\pm U$.
If one split the optical conductivity into the singular part at $\omega=0$ and the regular part as
${\rm Re}\,\sigma(\omega)=D\delta(\omega)+\sigma_{\rm reg}(\omega)$,
one obtains the Drude weight or charge stiffness $D=\frac{16\pi e^2t_h^2}{U}C_{M,N}$ \cite{Kaneko2020}.
For the distinction between $D$ and $D_s$, we refer to Ref.~\cite{Scalapino1993}.

\begin{figure}[t]
\includegraphics[width=8cm]{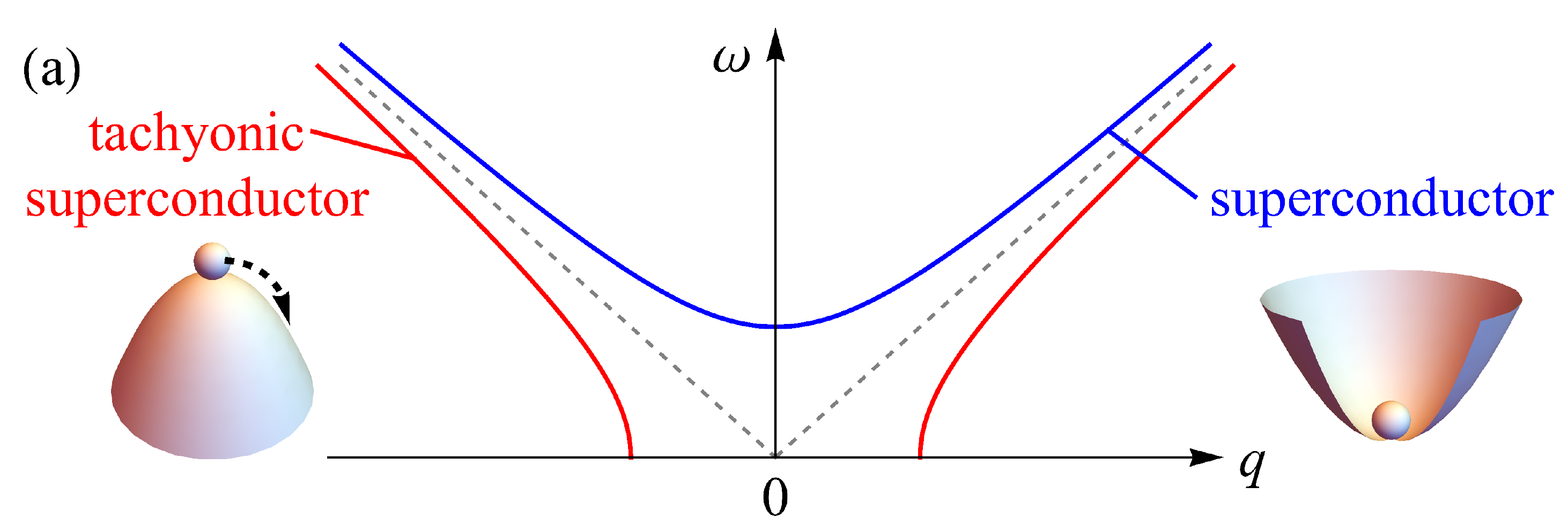}
\includegraphics[width=8cm]{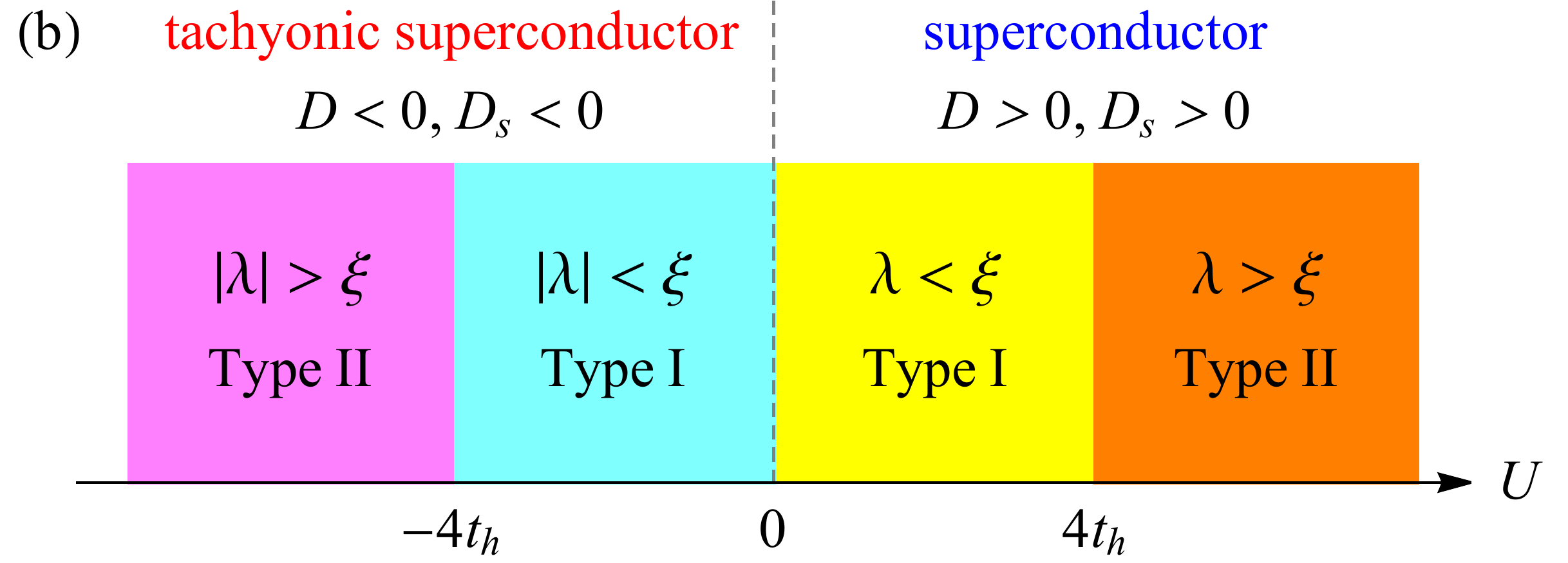}
\caption{(a) Schematic dispersion of electromagnetic fields and their effective potentials (3D images)
for ordinary superconductors and tachyonic superconductors. The dashed lines show the dispersion relation of electromagnetic fields in the vacuum.
(b) Phase diagram of the $\eta$-pairing states in the Hubbard model.}
\label{fig: phase diagram}
\end{figure}

For $U>0$, $D_s$ becomes positive, where the electromagnetic field acquires a mass due to the Anderson-Higgs mechanism
\cite{Anderson1963, ShimanoTsuji2020}
[$\omega^2\simeq c^2q^2+m^2c^4$ with $c$ the speed of light and $m^2\propto D_s>0$, see Fig.~\ref{fig: phase diagram}(a)].
On the other hand, for $U<0$, $D_s$ takes a negative value, implying that
the electromagnetic field has a negative squared mass 
[$m^2\propto D_s<0$, Fig.~\ref{fig: phase diagram}(a)].
Thus, the system is a ``tachyonic'' superconductor \cite{tachyonic_superconductor} 
[see the phase diagram in Fig.~\ref{fig: phase diagram}(b)],
in which the vacuum of the electromagnetic field lies
at the local maximum of the effective potential $V_{\rm eff}(\bm A)\propto m^2\bm A^2$
[Fig.~\ref{fig: phase diagram}(a)].
The electromagnetic field in the tachyonic superconductor becomes unstable,
and starts to grow exponentially in time.
The repulsive case ($U>0$) does not have such a tachyonic instability, 
but shows a different type of instability, as discussed below.

If we look at the electromagnetic response closer, we find that there is a {\it phase transition} at $U=\pm 4t_h$ [Fig.~\ref{fig: phase diagram}(b)].
This can be seen from the behavior of the Meissner kernel represented in real space $K^\perp(j, \omega=0)$ \cite{supplementary},
which asymptotically decays exponentially as $\sim \exp(-j/\xi)$
with $\xi=1/\cosh^{-1}(U^2/8t_h^2-1)$ for $|U|>4t_h$ \cite{supplementary}.
Here $\xi$ is Pippard's coherence length \cite{SchriefferBook}, which diverges at $U_c=\pm 4t_h$ as $\xi\sim |U-U_c|^{-1/2}$.
For $|U|<4t_h$, the kernel shows a power-law decay as $\sim j^{-1/2}$ \cite{supplementary}.

Let us compare the coherence length $\xi$ with London's penetration depth $\lambda$ defined by $\frac{1}{\lambda^2}=\frac{\mu_0}{\pi} D_s$
($\mu_0$ is the vacuum permeability). For $U>0$, we have 
$\lambda=\sqrt{\frac{U}{16\mu_0e^2t_h^2C_{M,N}}}$,
which grows smoothly as $U$ increases. In the region of $0<U<4t_h$, the penetration depth is smaller than
the coherence length ($\lambda<\xi$), and the system belongs to type-I superconductors (Fig.~\ref{fig: phase diagram}).
For $U>4t_h$, on the other hand, the relation becomes opposite ($\lambda>\xi$), and the system turns to
a type-II superconductor (Fig.~\ref{fig: phase diagram}) \cite{type_II}.
In an analogous way, we call the region $-4t_h<U<0$ ($U<-4t_h$)
a type-I (type-II) tachyonic superconductor (Fig.~\ref{fig: phase diagram}). 
They have different magnetic properties (for details, see \cite{supplementary}).

\begin{figure*}[t]
\includegraphics[width=6cm]{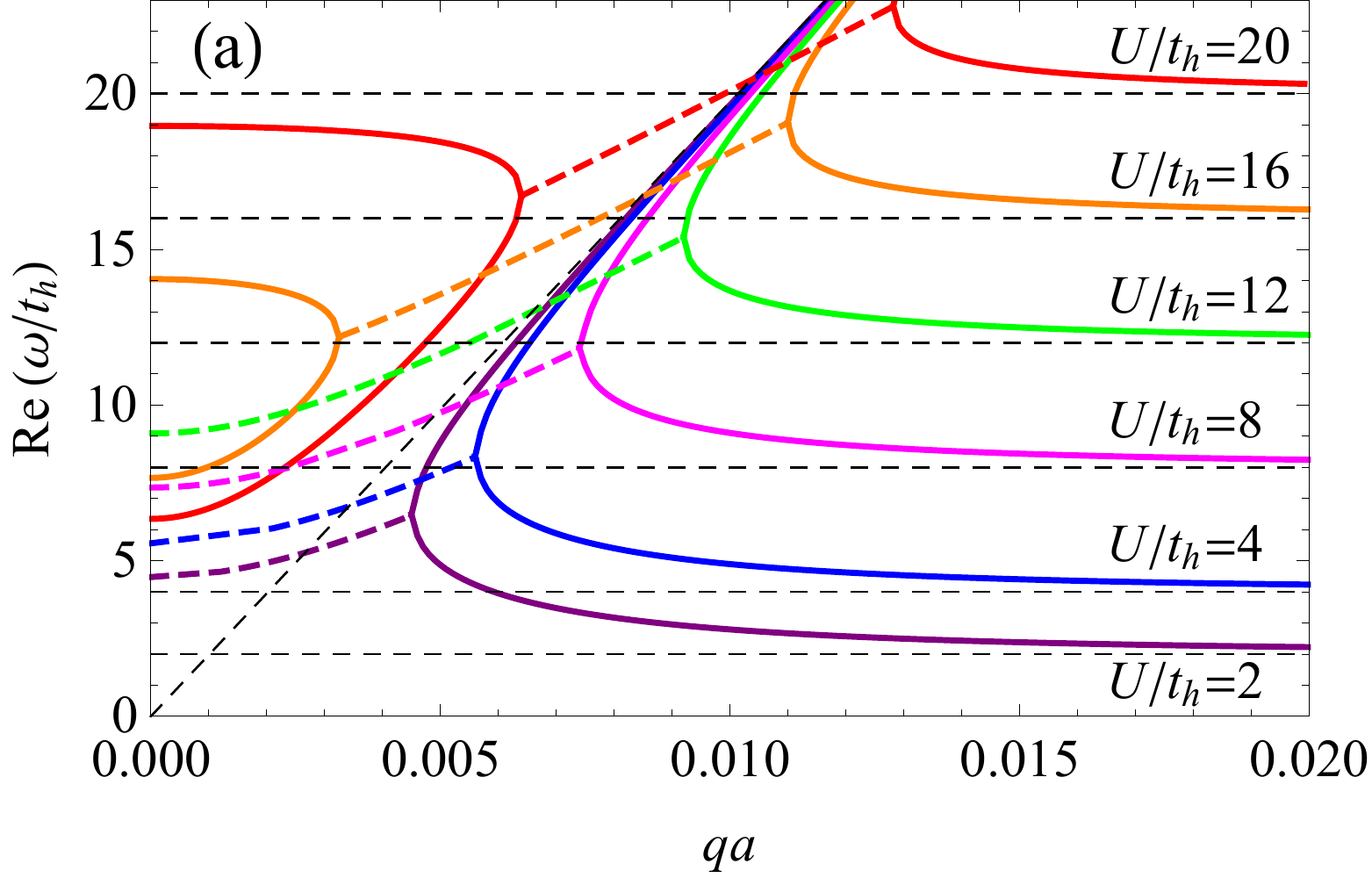}
\includegraphics[width=6cm]{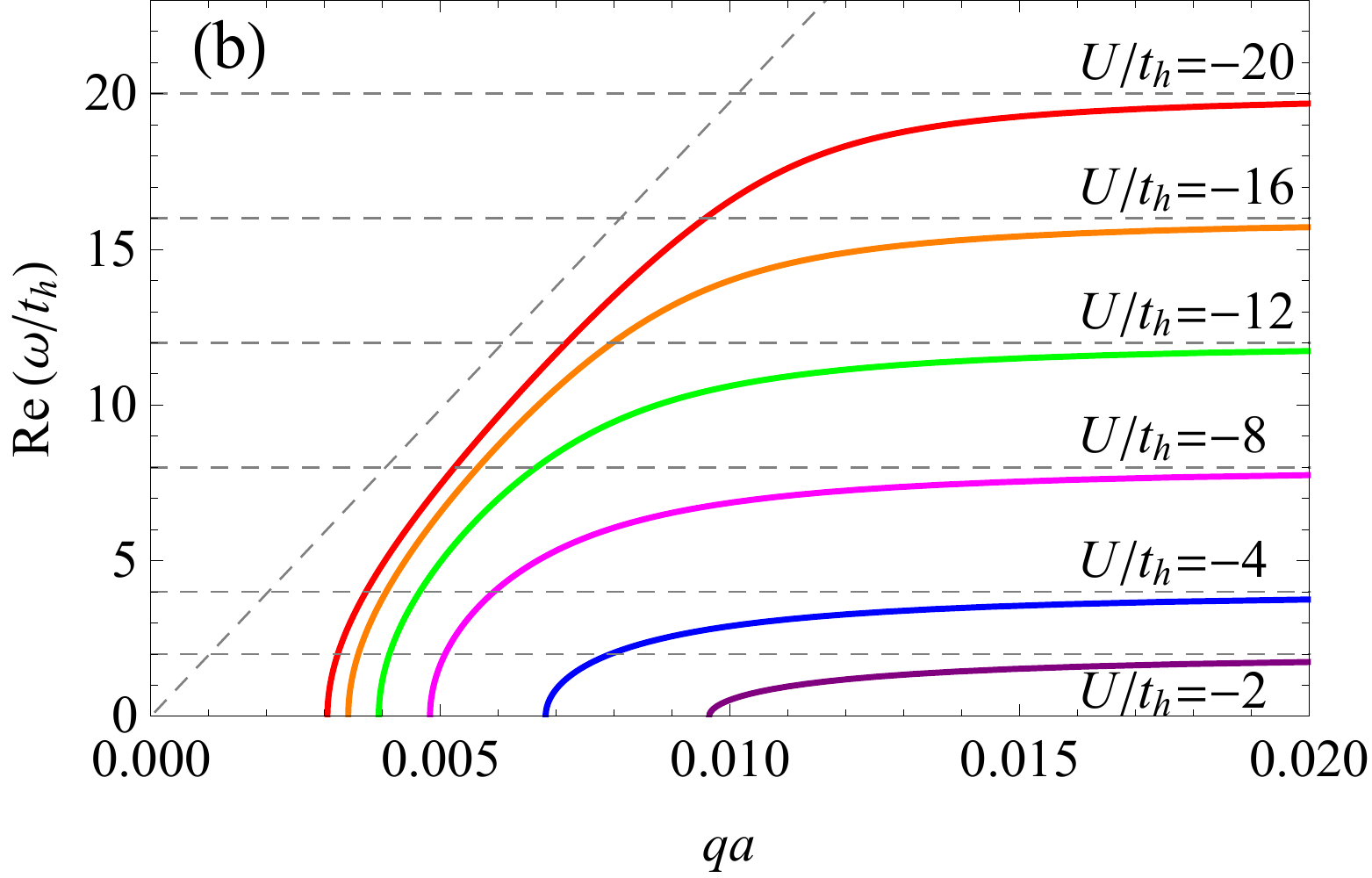}
\includegraphics[width=4.8cm]{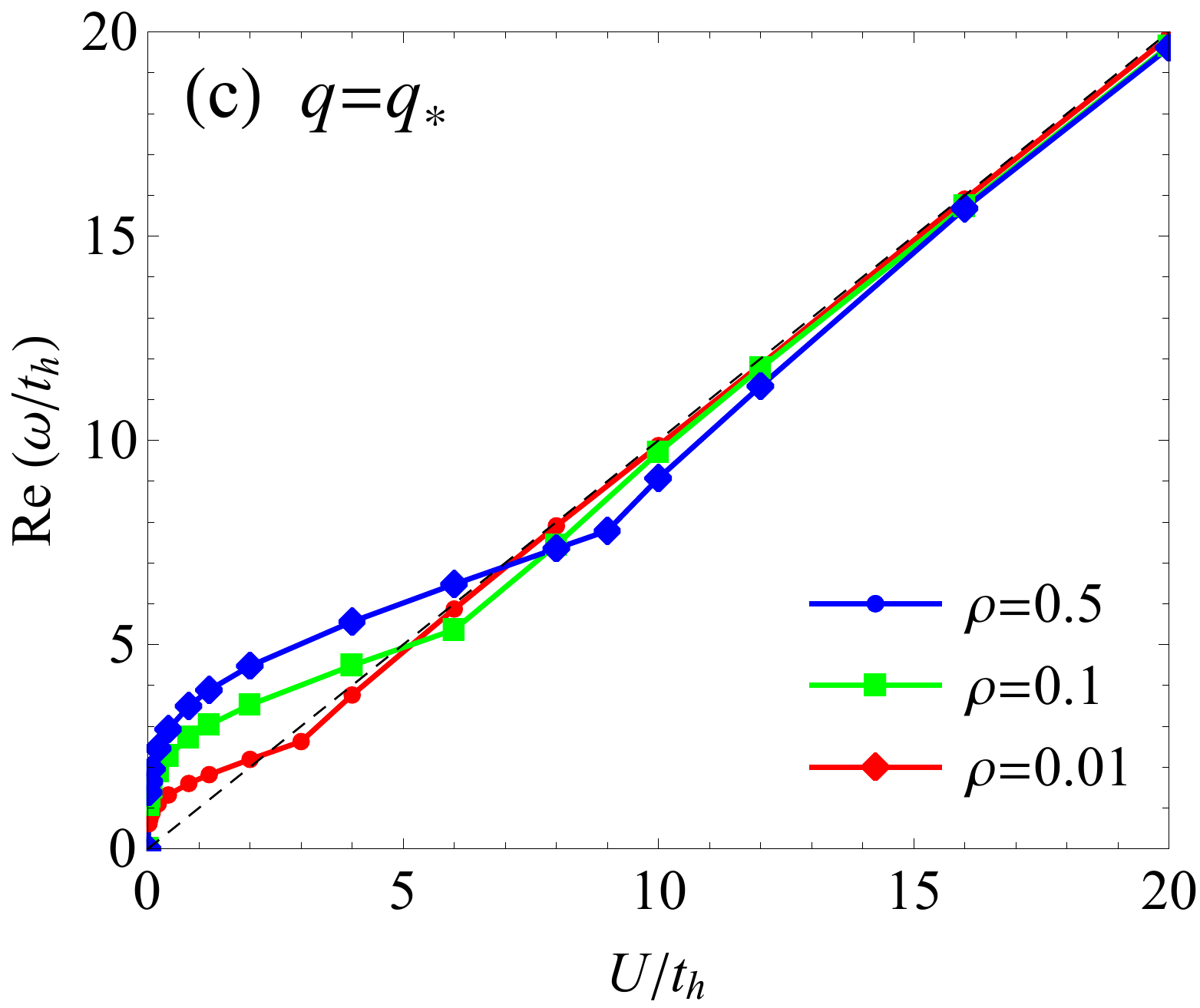}
\includegraphics[width=6cm]{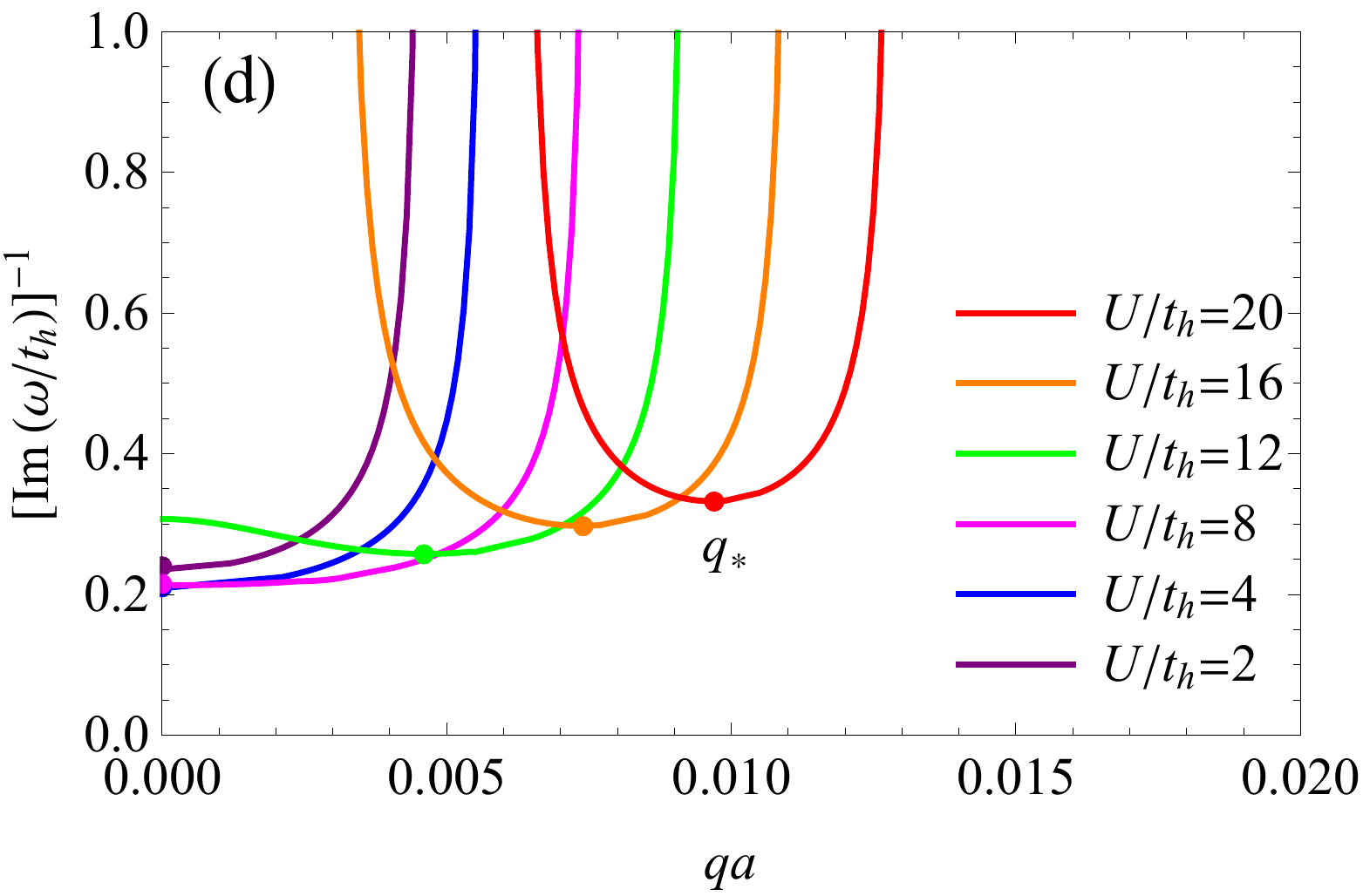}
\includegraphics[width=6cm]{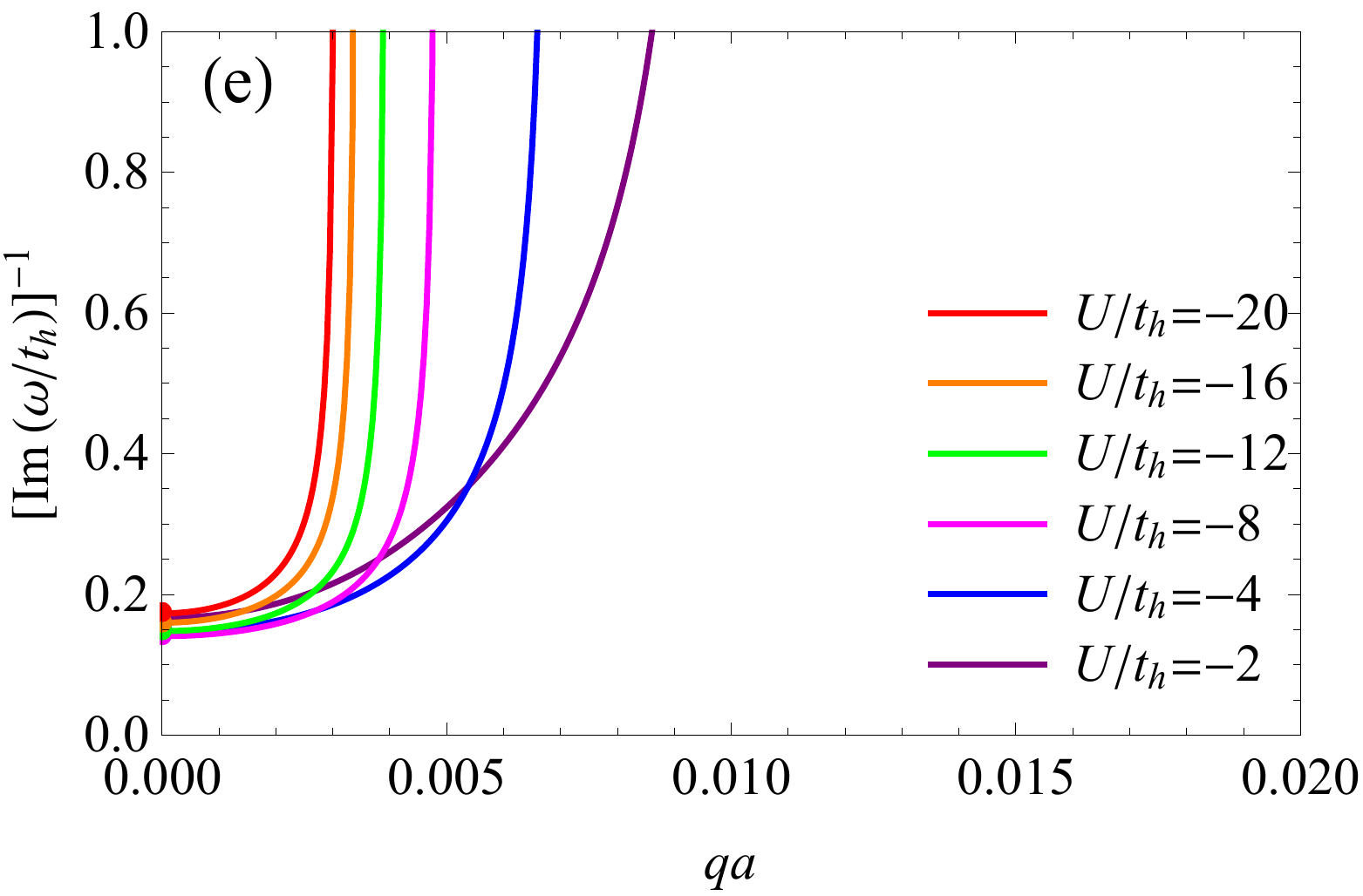}
\includegraphics[width=4.8cm]{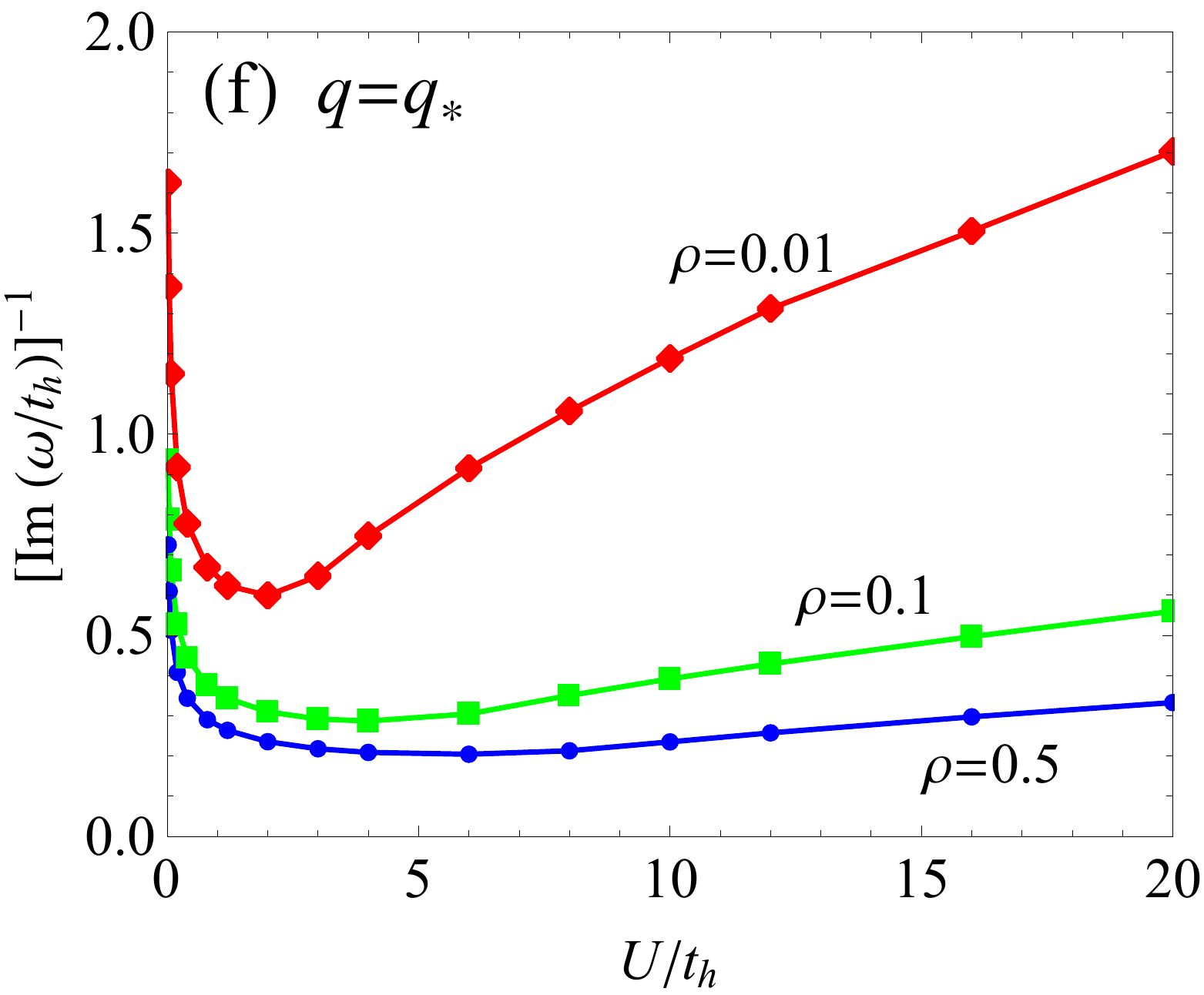}
\caption{(a), (b): Energy dispersion of the transverse electromagnetic field coupled to the $\eta$-pairing states
in the Hubbard model.
The solid curves show the real frequencies, while the dashed ones represent the real part of the complex frequencies.
The sloped dashed line shows the dispersion relation in the vacuum ($\omega=cq$), and the horizontal dashed lines correspond to $\omega=|U|$.
(c) Real part of $\omega$ at $q=q_\ast$,
where the imaginary part takes the maximal value.
The dashed line shows $\omega=U$.
(d), (e): Inverse of the imaginary part of $\omega$ corresponding to the time scale of the growth of the dynamical instability. 
(f): Inverse of the imaginary part of $\omega$ at $q=q_\ast$.
We set $\rho=0.5$ in (a), (b), (d), (e) and $t_h=1$ [eV] and $a=1$ [\AA] in (a)-(f).
}
\label{fig: positive U}
\end{figure*}

{\it Dynamical instability of $\eta$-pairing states.---}
Now, let us study the dynamics of electromagnetic fields coupled to the $\eta$-pairing states for $U>0$.
To this end, we consider the Maxwell equation in the Lorenz gauge,
$-\frac{\omega^2}{c^2}\bm A+q^2\bm A=\mu_0\bm j$, combined with the response of the $\eta$-pairing states,
$j^\mu=-K^{\mu\nu}(\bm q,\omega)A_\nu$. The equation of motion determines the energy dispersion of collective modes of
electromagnetic fields coupled with the $\eta$-pairing states.
We focus on the transverse mode, whose energy dispersion is given by
\begin{align}
\frac{\omega^2}{c^2}-q^2
&=
\mu_0 K^\perp(\bm q,\omega).
\label{eq: Maxwell}
\end{align}

At $\bm q=0$, Eq.~(\ref{eq: Maxwell}) becomes
$\frac{\omega^2}{c^2}=-\frac{16e^2t_h^2C_{M,N}U}{\omega^2-U^2}$,
which has imaginary-frequency solutions when $U^3<64\mu_0e^2c^2t_h^2C_{M,N}$.
If we input $t_h=1$ [eV] and $a=1$ [\AA] for ordinary materials, the condition reads
$U/t_h<22.6 \cdot (\rho(1-\rho))^{1/3}$, where $\rho:=(N/2)/M$ is the number of doublons per site ($0\le\rho\le 1$).
Surprisingly, the $\eta$-pairing states coupled to electromagnetic fields are dynamically unstable 
over a wide range of the parameter space against $\bm q=0$ modes.
More generally, we find that the $\eta$-pairing states are unstable for all the parameters if we take into account arbitrary $\bm q$ modes.

In Fig.~\ref{fig: positive U}, we plot the numerical solutions of Eq.~(\ref{eq: Maxwell}) for various parameters.
In Figs.~\ref{fig: positive U}(a) ($U>0$) and (b) ($U<0$), 
the solid curves show the real-frequency solutions, while the dashed curves represent the real part of the complex-frequency solutions.
When $U$ is positive and sufficiently large, there are two branches of the real solutions with the gaps near $q=0$.
As $q$ increases, the two branches merge at some point,
and turn into a conjugate pair of complex frequencies \cite{non-hermitian}.
After going across the vacuum dispersion ($\omega=cq$),
the solutions become real and split into two branches again. 
At high momentum, the two branches approach $\omega=cq$ and $\omega=U$.
For $0<U/t_h<14.3$ ($\rho=0.5$), the real branches near $q=0$ vanish as discussed above,
and only complex solutions exist at low momentum. 
For $U<0$ [Fig.~\ref{fig: positive U}(b)], the dispersion shows a tachyonic spectrum.
In general, we can prove that complex frequencies appear for all $U$ and $\rho$ \cite{supplementary},
indicating that the electromagnetic field (and hence the $\eta$-pairing state) is always dynamically unstable.


In Figs.~\ref{fig: positive U}(d) and (e), we plot $[{\rm Im}\,\omega]^{-1}$,
i.e., the time scale of the growth of the instability.
One can see that the shortest time scale among the $\bm q$ modes 
(whose momentum is denoted by $q_\ast$) is of the order of $\hbar/t_h$,
which is in the femtosecond regime.
In the decaying process, the energy of the electromagnetic field is transferred from the binding energy of doublons for $U>0$, and from
the kinetic energy of doublons for $U<0$. In the former, the doublons break up into two particles, while in the latter 
the doublons lose their momentum $\bm Q$. In both cases, the $\eta$-pairing states will eventually disappear.

For $U>0$, 
the complex frequencies have nonzero real parts [see Fig.~\ref{fig: positive U}(a)],
so that the exponential growth of the electromagnetic field is accompanied by plasma oscillations.
They induce intense light emission, where doublons' binding energies are released collectively.
In Figs.~\ref{fig: positive U}(c) and (f), we plot ${\rm Re}\,\omega$ and $[{\rm Im}\,\omega]^{-1}$ at momentum $q_\ast$,
corresponding to the characteristic frequency and the growing time scale of the dominant emitted light waves, respectively.
The characteristic frequency is shifted from $U$.
In particular, at $U\ll t_h$ it is proportional to $U^{1/4}$ with $q_\ast=0$ \cite{supplementary}. As $U$ increases, 
$q_\ast$ starts to take a nonzero value around $U/t_h\approx 9$ [Fig.~\ref{fig: positive U}(d)],
making a kink-like structure in Fig.~\ref{fig: positive U}(c). 
The time scale of the growth increases as the density decreases [Fig.~\ref{fig: positive U}(f)], but
stays within the femtosecond regime even at $\rho=0.01$.
For $U<0$, the decay of the $\eta$-pairing states is accompanied by unattenuated penetration of electromagnetic fields \cite{supplementary}
in such a way that a tachyonic field grows exponentially as in order-parameter dynamics near critical points \cite{Kibble1976, Zurek1985, PolkovnikovSenguptaSilvaVengalattore2011}.

We emphasize that the mechanism of the plasma instability at $U>0$ is different from that of spontaneous light emission,
the latter of which is caused by a quantum-mechanical effect of electromagnetic fields
and has the frequency $\omega=U$.
The decay width of spontaneous emission is evaluated by 
$\Gamma=\frac{\omega}{3\pi\varepsilon_0 \hbar c^3}\sum_n |\langle \psi_N|J^\mu|\phi_n\rangle|^2$ \cite{LoudonBook},
where $\varepsilon_0$ is the vacuum permittivity
and the sum runs over all the eigenstates of the Hubbard model.
For $\rho=0.5$, $t_h=U=1$ [eV], and $a=1$ [\AA], we have $\Gamma/M=2.3\times 10^7$ [s$^{-1}$] \cite{supplementary}.
Thus, spontaneous emission takes place for each site in the time scale of $10^2$ [ns], which is much slower than
the plasma instability.


{\it Summary and outlook.---}
We have shown that Yang's $\eta$-pairing states have the intrinsic plasma instability for $U>0$ and the tachyonic instability for $U<0$
when the system is coupled to electromagnetic fields.
The time scales of both of these instabilities are of the order of femtoseconds, 
which puts a strong constraint on the realization of the $\eta$-pairing states in real materials.
The decay of the $\eta$-pairing states leads to enhanced light emission with characteristic frequencies
shifted from the Hubbard interaction $U$ for the repulsive case, and unattenuated penetration of electromagnetic fields for the attractive case. 
While we have focused on the simplest form of the $\eta$-pairing eigenstates (\ref{eq: psi_N}),
we expect that similar instabilities might exist for more general states having unpaired particles (at least if they are dilute enough).
Stabilizing the $\eta$-pairing states coupled to electromagnetic fields is an interesting open problem, 
which merits further studies.

\acknowledgements

N.T. acknowledges support by KAKENHI Grant No. JP20K03811.
M.N. acknowledges support by KAKENHI Grant No. JP20K14383.
M.U. acknowledges support by KAKENHI Grant No. JP18H01145.

\bibliography{ref}

\begin{thebibliography}{54}
\bibitem [{kin()}]{kinetic_energy}%
  \BibitemOpen
  \href@noop {} {}\bibinfo {note} {The diamagnetic component of the
  electromagnetic response function for the $\eta$-pairing states in the
  Hubbard model is given by $K_{\rm dia}^{\mu\nu}(\bm q,\omega)$ $\propto
  \delta_{\mu\nu} \sum_{\bm k\sigma} \frac{\partial^2\varepsilon_{\bm
  k}}{\partial k_\mu^2} \langle \psi_N| c_{\bm k\sigma}^\dagger c_{\bm
  k\sigma}|\psi_N\rangle \propto \delta_{\mu\nu} \sum_{\bm k\sigma}
  \varepsilon_{\bm k} \langle \psi_N| c_{\bm k\sigma}^\dagger c_{\bm
  k\sigma}|\psi_N\rangle=0$, where $\varepsilon_{\bm k}=-2t_h\sum_{\mu=1}^3
  \cos k_\mu$ is the single-particle band dispersion and $c_{\bm
  k\sigma}=\frac{1}{\sqrt{M}}\sum_j e^{-i\bm k\cdot\bm R_j}
  c_{j\sigma}$.}\BibitemShut {Stop}%
\end{thebibliography}


\clearpage

\renewcommand{\theequation}{S\arabic{equation}}
\setcounter{equation}{0}
\renewcommand{\thefigure}{S\arabic{figure}}
\setcounter{figure}{0}

\onecolumngrid
\begin{center}
\large{\textbf{Supplemental Material for ``Tachyonic and Plasma Instabilities of $\eta$-Pairing States Coupled to Electromagnetic Fields''}}\\
\vspace{.4cm}
\normalsize{Naoto Tsuji$^{1,2}$, Masaya Nakagawa$^1$, and Masahito Ueda$^{1,2,3}$}\\
\vspace{.1cm}
\it{$^1$Department of Physics, University of Tokyo, Hongo, Tokyo 113-0033, Japan}\\
\it{$^2$RIKEN Center for Emergent Matter Science (CEMS), Wako 351-0198, Japan}\\
\it{$^3$Institute for Physics of Intelligence, University of Tokyo, Hongo, Tokyo 113-0033, Japan}\\
\rm{(Dated: \today)}
\end{center}

\vspace{.5cm}

\twocolumngrid

\section{I. Derivation of the electromagnetic response function}

In this section, we describe how to analytically evaluate the electromagnetic response function (Meissner kernel) 
[Eq.~(\ref{eq: K_para}) in the main text],
\begin{align}
K^{\mu\nu}(\bm R_j, t)
&=
-i\theta(t) \langle \psi_N| [J^\mu(\bm R_j, t), J^\nu(0,0)] |\psi_N\rangle,
\label{eq: kernel}
\end{align}
at arbitrary lattice coordinate $\bm R_j$ and time $t$ 
for Yang's $\eta$-pairing state $|\psi_N\rangle$ [Eq.~(\ref{eq: K reduced})] in the Hubbard model.

\subsection{A. Reduction to the two-particle correlation function}

The first step is to reduce the correlation function of $N$ particles (\ref{eq: kernel}) to
that of two particles by shifting all the $\eta$ operators in $|\psi_N\rangle$
to the left of the current operators using the commutation relation
\begin{align}
[J^\mu(\bm R_j), \eta^+]
&=
-2iet_h e^{i\bm Q\cdot\bm R_j}
(c_{j+\mu\uparrow}^\dagger c_{j\downarrow}^\dagger
+c_{j\uparrow}^\dagger c_{j+\mu\downarrow}^\dagger).
\label{eq: [J, eta^+]}
\end{align}
For convenience, we define the operator that appeared on the right-hand side of Eq.~(\ref{eq: [J, eta^+]}) as
\begin{align}
J_\eta^{\mu+}(\bm R_j)
&:=
-iet_h e^{i\bm Q\cdot\bm R_j}
(c_{j+\mu\uparrow}^\dagger c_{j\downarrow}^\dagger
+c_{j\uparrow}^\dagger c_{j+\mu\downarrow}^\dagger).
\end{align}
With this definition, we can write 
\begin{align}
[J^\mu(\bm R_j), \eta^+]
&=
2J_\eta^{\mu+}(\bm R_j).
\label{eq: [J^mu, eta^+]}
\end{align}
Since $J_\eta^{\mu+}(\bm R_j)$ only involves creation operators, $J_\eta^{\mu+}(\bm R_j)$ commutes with $\eta^+$.
Therefore, we can repeatedly use the relation (\ref{eq: [J^mu, eta^+]}) to obtain
\begin{align}
[J^\mu(\bm R_j), (\eta^+)^n]
&=
2n(\eta^+)^{n-1}J_\eta^{\mu+}(\bm R_j)
\label{eq: [J_eta, eta^+]}
\end{align}
for $n=1,2,\dots$. By using Eq.~(\ref{eq: [J_eta, eta^+]}), 
we can evaluate the current-current correlation function as
\begin{align}
&
\langle \psi_N| J^\mu(\bm R_j, t) J^\nu(0,0)|\psi_N\rangle
\notag
\\
&=
\frac{1}{\sqrt{\mathcal N_N}}\langle \psi_N| J^\mu(\bm R_j, t) J^\nu(0,0) (\eta^+)^{\frac{N}{2}}|0\rangle
\notag
\\
&=
\frac{1}{\sqrt{\mathcal N_N}}N
\langle \psi_N| J^\mu(\bm R_j, t) (\eta^+)^{\frac{N}{2}-1} J_\eta^{\nu+}(0) |0\rangle
\notag
\\
&=
\frac{1}{\sqrt{\mathcal N_N}}N
\langle \psi_N| (\eta^+)^{\frac{N}{2}-1} J^\mu(\bm R_j, t) J_\eta^{\nu+}(0) |0\rangle
\notag
\\
&\quad
+\frac{1}{\sqrt{\mathcal N_N}}N(N-2)
\langle \psi_N| (\eta^+)^{\frac{N}{2}-2} J_\eta^{\mu+}(\bm R_j, t)  J_\eta^{\nu+}(0) |0\rangle.
\label{eq: current-current1}
\end{align}
Then, straightforward calculations show that
\begin{align}
\langle \psi_N| (\eta^+)^{\frac{N}{2}-1}
&=
\frac{1}{\sqrt{\mathcal N_N}} \frac{(\frac{N}{2})!(M-1)!}{(M-\frac{N}{2})!}\langle 0|\eta^-,
\\
\langle \psi_N| (\eta^+)^{\frac{N}{2}-2}
&=
\frac{1}{\sqrt{\mathcal N_N}} \frac{(\frac{N}{2})!(M-2)!}{2(M-\frac{N}{2})!}\langle 0|(\eta^-)^2,
\end{align}
which can be used to rewrite the correlation function (\ref{eq: current-current1}) as
\begin{align}
&
\langle \psi_N| J^\mu(\bm R_j, t) J^\nu(0,0)|\psi_N\rangle
\notag
\\
&=
\frac{N}{\mathcal N_N}\frac{(\frac{N}{2})!(M-1)!}{(M-\frac{N}{2})!}
\langle 0| \eta^- J^\mu(\bm R_j, t) J_\eta^{\nu+}(0) |0\rangle
\notag
\\
&
+\frac{N(N-2)}{\mathcal N_N}\frac{(\frac{N}{2})!(M-2)!}{2(M-\frac{N}{2})!}
\langle 0| (\eta^-)^2 J_\eta^{\mu+}(\bm R_j, t)  J_\eta^{\nu+}(0) |0\rangle.
\end{align}
To further simplify the expression, we use the commutation relations,
\begin{align}
[\eta^-, J^\mu(\bm R_j)]
&=
2J_\eta^{\mu-}(\bm R_j),
\\
[\eta^-, J_\eta^{\mu+}(\bm R_j)]
&=
-J^\mu(\bm R_j),
\end{align}
where $J_\eta^{\mu-}(\bm R_j):=[J_\eta^{\mu+}(\bm R_j)]^\dagger$.
Thus, the correlation function becomes
\begin{align}
&
\langle \psi_N| J^\mu(\bm R_j, t) J^\nu(0,0)|\psi_N\rangle
\notag
\\
&=
\frac{2N}{\mathcal N_N}\frac{(\frac{N}{2})!(M-1)!}{(M-\frac{N}{2})!}
\langle 0| J_\eta^{\mu-}(\bm R_j, t) J_\eta^{\nu+}(0) |0\rangle
\notag
\\
&\quad
-\frac{N(N-2)}{\mathcal N_N}\frac{(\frac{N}{2})!(M-2)!}{2(M-\frac{N}{2})!}
\langle 0| \eta^- J^\mu(\bm R_j, t)  J_\eta^{\nu+}(0) |0\rangle
\notag
\\
&=
\frac{2N(\frac{N}{2})!(M-2)!}{(M-\frac{N}{2}-1)!\mathcal N_N}
\langle 0| J_\eta^{\mu-}(\bm R_j, t)  J_\eta^{\nu+}(0) |0\rangle.
\label{eq: current-current2}
\end{align}
Here, let us recall that the normalization constant for the $\eta$-pairing state is explicitly given by
\begin{align}
\mathcal N_N
&=
\frac{(\frac{N}{2})!M!}{(M-\frac{N}{2})!}.
\label{eq: normalization}
\end{align}
We use Eq.~(\ref{eq: current-current2}) to reduce the current-current correlation function of $N$ particles to that of two particles,
\begin{align}
&
\langle \psi_N| J^\mu(\bm R_j, t) J^\nu(0,0)|\psi_N\rangle
\notag
\\
&=
4C_{M,N}
\langle 0| J_\eta^{\mu-}(\bm R_j, t)  J_\eta^{\nu+}(0) |0\rangle,
\label{eq: two-particle}
\end{align}
where
\begin{align}
C_{M,N}
&:=
\frac{\frac{N}{2}(M-\frac{N}{2})}{M(M-1)}.
\end{align}
Equation (\ref{eq: two-particle}) leads to Eq.~(\ref{eq: K reduced}) in the main text.

The technique used here (i.e., reduction of $N$-particle to few-particle correlation functions) can be applied
not only to the electromagnetic response function (\ref{eq: kernel}) but also to arbitrary correlation functions
constructed from few-body operators.

\subsection{B. Evaluation of the two-particle dynamics}

In the previous subsection, we have shown that the $N$-particle correlation function (\ref{eq: kernel})
can be reduced to the two-particle correlation function (\ref{eq: two-particle}).
Since the two-particle problem in the Hubbard model is exactly solvable, we can evaluate the two-particle correlation function exactly.
Here we describe the details of the evaluation.

First, we Fourier transform Eq.~(\ref{eq: two-particle}) to obtain
\begin{align}
&
\sum_{j} e^{i\bm q\cdot \bm R_j} \langle \psi_N| J^\mu(\bm R_j, t) J^\nu(0,0)|\psi_N\rangle
\notag
\\
&=
\frac{4}{M}C_{M,N} \langle 0| J_\eta^{\mu-}(\bm q, t) J_\eta^{\nu+}(\bm q)|0\rangle,
\label{eq: Fourier}
\end{align}
where
\begin{align}
J_\eta^{\mu+}(\bm q)
&:=
\sum_j e^{-i\bm q\cdot\bm R_j} J_\eta^{\mu+}(\bm R_j),
\\
J_\eta^{\mu-}(\bm q)
&:=
[J_\eta^{\mu+}(\bm q)]^\dagger.
\end{align}
Acting $J_\eta^{\nu+}(\bm q)$ on the vacuum state, we obtain
\begin{align}
&
J_\eta^{\nu+}(\bm q)|0\rangle
=
-iet_h\sum_j e^{i(\bm Q-\bm q)\cdot\bm R_j}(c_{j+\nu\uparrow}^\dagger c_{j\downarrow}^\dagger
+c_{j\uparrow}^\dagger c_{j+\nu\downarrow}^\dagger)|0\rangle
\notag
\\
&=
-iet_h\sum_j e^{i(\bm Q-\bm q)\cdot\bm R_j}(|\bm R_j+\bm e_\nu, \bm R_j\rangle
+|\bm R_j, \bm R_j+\bm e_\nu\rangle),
\label{eq: J_eta^nu+|0>}
\end{align}
where we have introduced the notation $|\bm R_j, \bm R_k\rangle:=c_{j\uparrow}^\dagger c_{k\downarrow}^\dagger |0\rangle$ 
to represent a two-particle state with $\uparrow$ spin at site $j$ and $\downarrow$ spin at site $k$,
and $\bm e_\nu$ is the unit vector in the $\nu$ direction.

Since the system has the (discrete) translation symmetry, the center-of-mass (crystal) momentum of two particles under consideration is conserved.
Let us define the translation operator $T_\mu$ that shifts two particles by one lattice site in the $\mu$ direction.
We express the eigenstates of $T_\mu$ in terms of the center-of-mass momentum $\bm K$ and the relative coordinate $\bm r$
of the two particles as
\begin{align}
|\bm K, \bm r\rangle
&=
\frac{1}{\sqrt{M}} \sum_j e^{i\bm K\cdot\bm R_j} |\bm R_j+\bm r, \bm R_j\rangle,
\label{eq: |K,r>}
\end{align}
which satisfies $T_\mu |\bm K, \bm r\rangle=e^{-i\bm K\cdot\bm e_\mu}|\bm K, \bm r\rangle$.
Using the eigenstates (\ref{eq: |K,r>}), $J_\eta^{\nu+}(\bm q)|0\rangle$ (\ref{eq: J_eta^nu+|0>}) can be written as
\begin{align}
J_\eta^{\nu+}(\bm q)|0\rangle
&=
-iet_h\sqrt{M}(|\bm Q-\bm q, +\bm e_\nu\rangle
\notag
\\
&\quad
+e^{-i(\bm Q-\bm q)\cdot\bm e_\nu}|\bm Q-\bm q, -\bm e_\nu\rangle),
\label{eq: J_eta^nu+|0>2}
\end{align}
which has the center-of-mass momentum $\bm Q-\bm q$.
The action of the Hamiltonian in Eq.~(\ref{eq: Hubbard}) in the main text on the state $|\bm K, \bm r\rangle$ (\ref{eq: |K,r>}) is given by
\begin{align}
H|\bm K, \bm r\rangle
&=
-t_h \sum_\mu [(1+e^{i\bm K\cdot\bm e_\mu})|\bm K, \bm r+\bm e_\mu\rangle
\notag
\\
&\quad
+(1+e^{-i\bm K\cdot\bm e_\mu})|\bm K, \bm r-\bm e_\mu\rangle]
\notag
\\
&\quad
+U(\delta_{\bm r,0}-1)|\bm K, \bm r\rangle.
\end{align}
If we define operators $\Delta_\mu^\pm$ that shift the relative coordinate of two particles by $\pm\bm e_\mu$,
the Hamiltonian can be represented in the Hilbert subspace of two particles with the center-of-mass momentum $\bm K$ as
\begin{align}
H(\bm K)
&=
-t_h \sum_\mu [(1+e^{i\bm K\cdot\bm e_\mu})\Delta_\mu^+
+(1+e^{-i\bm K\cdot\bm e_\mu})\Delta_\mu^-]
\notag
\\
&\quad
+U(\delta_{\bm r,0}-1).
\label{eq: H(K)}
\end{align}

Using the representation (\ref{eq: H(K)}), the correlation function (\ref{eq: Fourier}) can be written as
\begin{align}
&
\sum_{j} e^{i\bm q\cdot \bm R_j} \langle \psi_N| J^\mu(\bm R_j, t) J^\nu(0,0)|\psi_N\rangle
\notag
\\
&=
\frac{4}{M}C_{M,N} \langle 0| J_\eta^{\mu-}(\bm q) e^{-iH(\bm Q-\bm q)t} J_\eta^{\nu+}(\bm q)|0\rangle.
\end{align}
Without loss of generality, we assume that $\bm q$ is parallel to the $z$ direction [$\bm q=(0,0,q)$].
Then,
\begin{align}
H(\bm Q-\bm q)
&=
-t_h [(1-e^{-iq})\Delta_z^+
+(1-e^{iq})\Delta_z^-]
\notag
\\
&\quad
+U(\delta_{\bm r,0}-1).
\label{eq: H(Q-q)}
\end{align}
Hence, the two-particle dynamics that we have to consider is essentially a one-dimensional problem,
in which the relative coordinates $r_x$ and $r_y$ are conserved. 
Below, we decompose the electromagnetic response function into the transverse component
$K^\perp(\bm q,\omega):=K^{\mu\mu}(\bm q,\omega)$ ($\mu=x,y$)
and the longitudinal one $K^\parallel(\bm q,\omega):=K^{zz}(\bm q,\omega)$.
The off-diagonal components are absent, i.e., $K^{\mu\nu}(\bm q,\omega)=0$ for $\mu\neq\nu$,
since the state $e^{-iH(\bm Q-\bm q)t}J_\eta^{\nu+}(\bm q)|0\rangle$ never has an overlap 
with the state $J_\eta^{\mu+}|0\rangle$ for $\mu\neq \nu$.

\subsection{C. Transverse component}

For the transverse component, the two-particle state $J_\eta^{\nu+}(\bm q)|0\rangle$ (\ref{eq: J_eta^nu+|0>2})
has the relative coordinates, $(r_x,r_y)\neq (0,0)$. Since $r_x$ and $r_y$ are conserved during the time evolution,
the two particles do not sit on the same site. Therefore, they do not interact with each other, and the dynamics
becomes effectively noninteracting. The correlation function (\ref{eq: Fourier}) now reads
\begin{align}
&
\sum_{j} e^{i\bm q\cdot \bm R_j} \langle \psi_N| J^\mu(\bm R_j, t) J^\mu(0,0)|\psi_N\rangle
\quad
(\mu=x,y)
\notag
\\
&=
4e^2t_h^2 C_{M,N}
\sum_{s=\pm} \langle \bm Q-\bm q, s\bm e_\mu| e^{-iH(\bm Q-\bm q)t} |\bm Q-\bm q, s\bm e_\mu\rangle,
\end{align}
where
\begin{align}
H(\bm Q-\bm q)
&=
-t_h [(1-e^{-iq})\Delta_z^+
+(1-e^{iq})\Delta_z^-]-U
\end{align}
is the noninteracting Hamiltonian which can be diagonalized by Fourier transformation
with respect to the relative coordinate $r_z$. The result is
\begin{align}
&
\sum_{j} e^{i\bm q\cdot \bm R_j} \langle \psi_N| J^\mu(\bm R_j, t) J^\mu(0,0)|\psi_N\rangle
\quad
(\mu=x,y)
\notag
\\
&=
8e^2t_h^2 C_{M,N}
\frac{e^{iUt}}{M}\sum_{\bm k} e^{2it_h(\cos k_z-\cos(k_z+q))t}
\notag
\\
&=
8e^2t_h^2 C_{M,N} e^{iUt}
\int_{-\pi}^\pi \frac{dk_z}{2\pi} e^{2it_h(\cos k_z-\cos(k_z+q))t}
\notag
\\
&=
8e^2t_h^2 C_{M,N} e^{iUt}
J_0\left(4t_h\sin\frac{q}{2}t\right),
\end{align}
where $J_0(x)$ is the zeroth-order Bessel function of the first kind.
We thus obtain the transverse component of the electromagnetic response function as
\begin{align}
K^\perp(\bm q,t)
&=
-8i\theta(t) e^2t_h^2 C_{M,N} (e^{iUt}-e^{-iUt})
\notag
\\
&\quad\times
J_0\left(4t_h\sin\frac{q}{2}t\right).
\end{align}
Using the integral formula for the Bessel function ($a>0$),
\begin{align}
\int_0^\infty dt e^{i\omega t} J_0(at)
&=
\begin{cases}
\frac{i {\rm sgn}(\omega)}{\sqrt{\omega^2-a^2}} & \mbox{for}\; |\omega|>a;
\\
\frac{1}{\sqrt{a^2-\omega^2}} & \mbox{for}\; |\omega|<a,
\end{cases}
\end{align}
we obtain
\begin{align}
&
K^\perp(\bm q, \omega)
=
-8i e^2t_h C_{M,N}\left[f_q\left(\frac{\omega+U}{t_h}\right)-f_q\left(\frac{\omega-U}{t_h}\right)\right],
\label{eq: K_perp suppl}
\end{align}
where
\begin{align}
f_q(x)
&:=
\begin{cases}
\frac{i{\rm sgn}(x)}{\sqrt{x^2-16\sin^2\frac{q}{2}}} & \mbox{for}\; |x|>4\sin\frac{q}{2};
\\
\frac{1}{\sqrt{16\sin^2\frac{q}{2}-x^2}} & \mbox{for}\; |x|<4\sin\frac{q}{2}.
\end{cases}
\end{align}
This is the final result for the transverse component in Eq.~(\ref{eq: K_perp}) in the main text.

\begin{figure}[t]
\includegraphics[width=4.2cm]{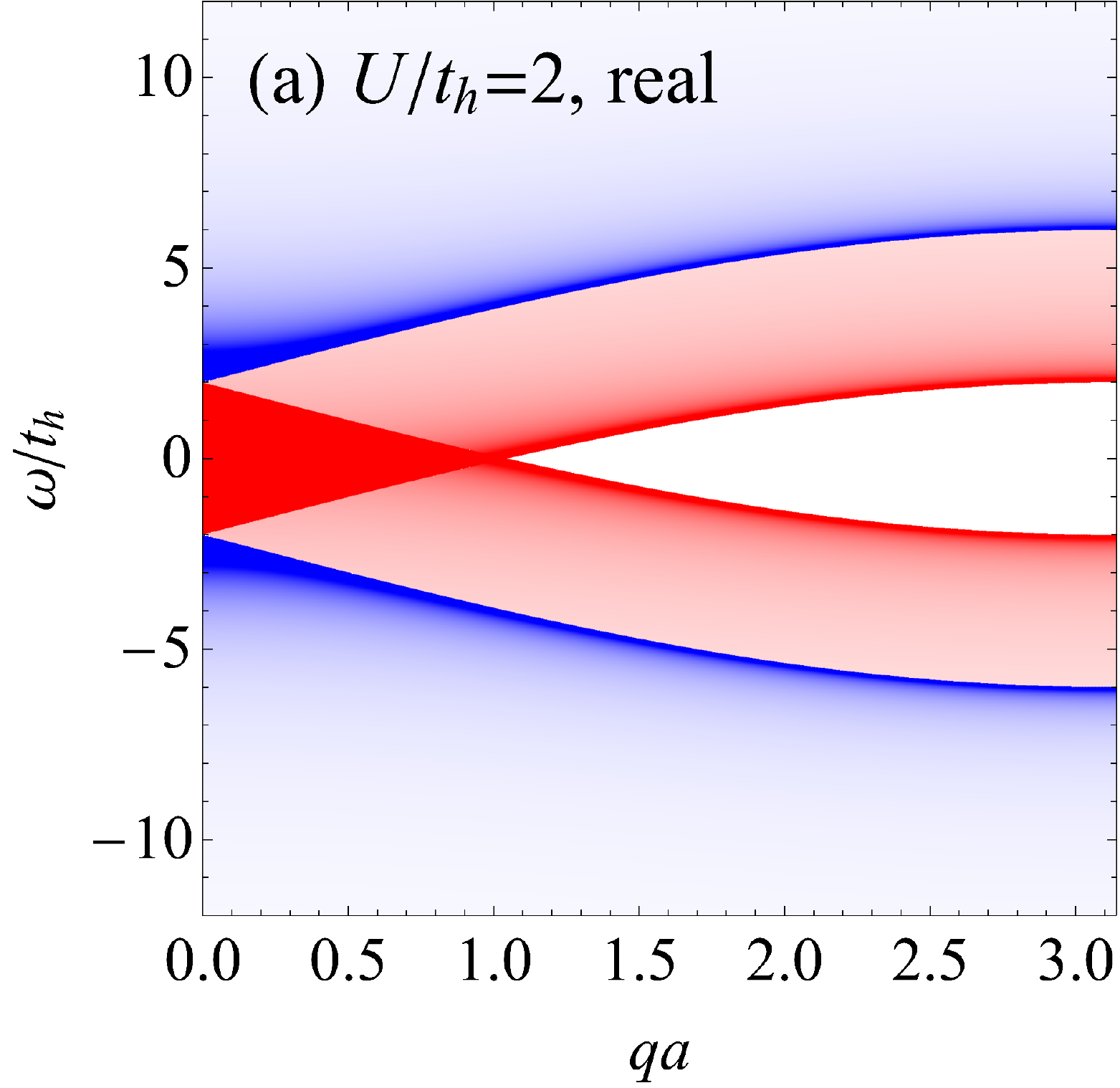}
\vspace{.2cm}
\includegraphics[width=4.2cm]{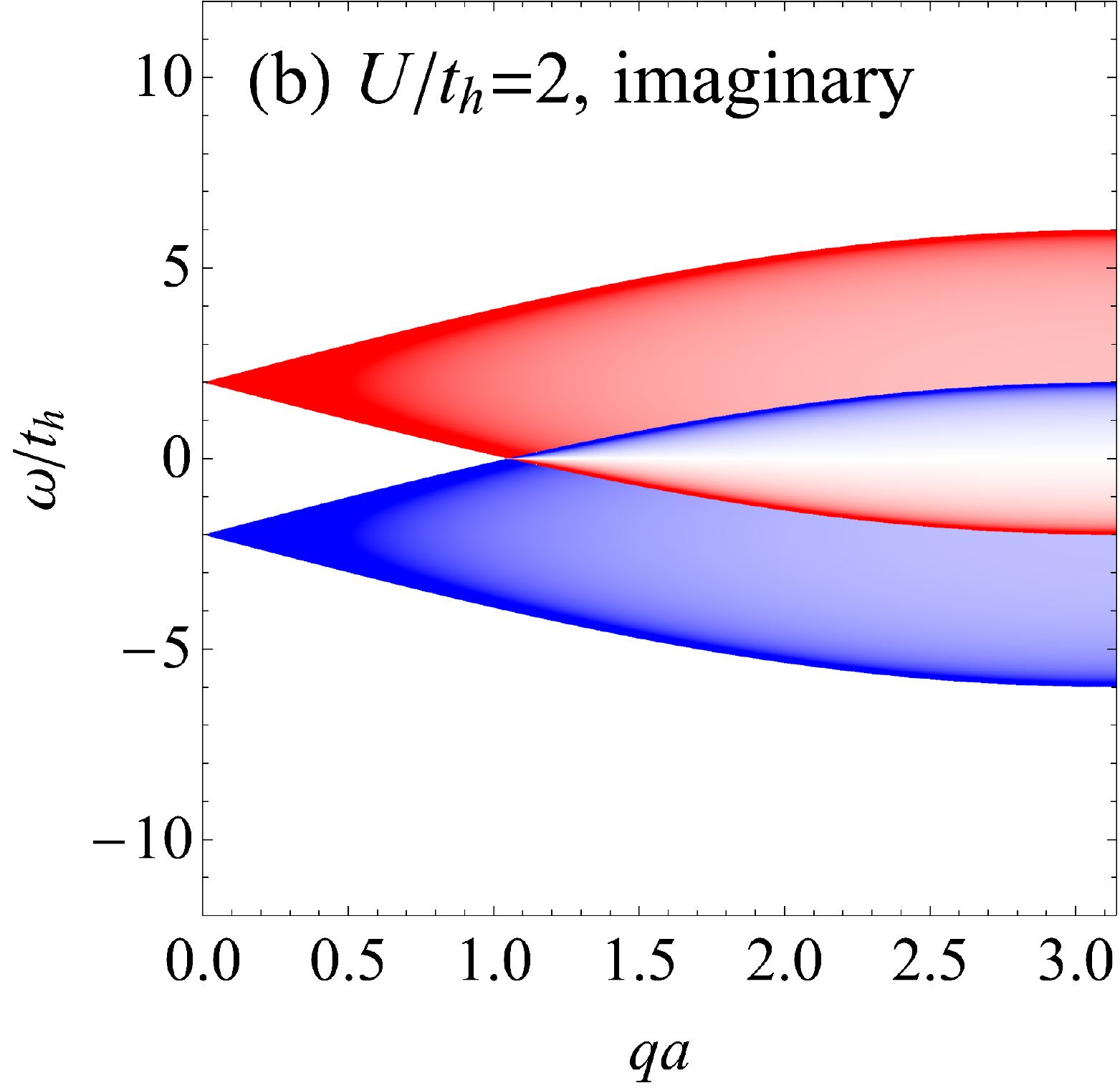}
\vspace{.2cm}
\includegraphics[width=4.2cm]{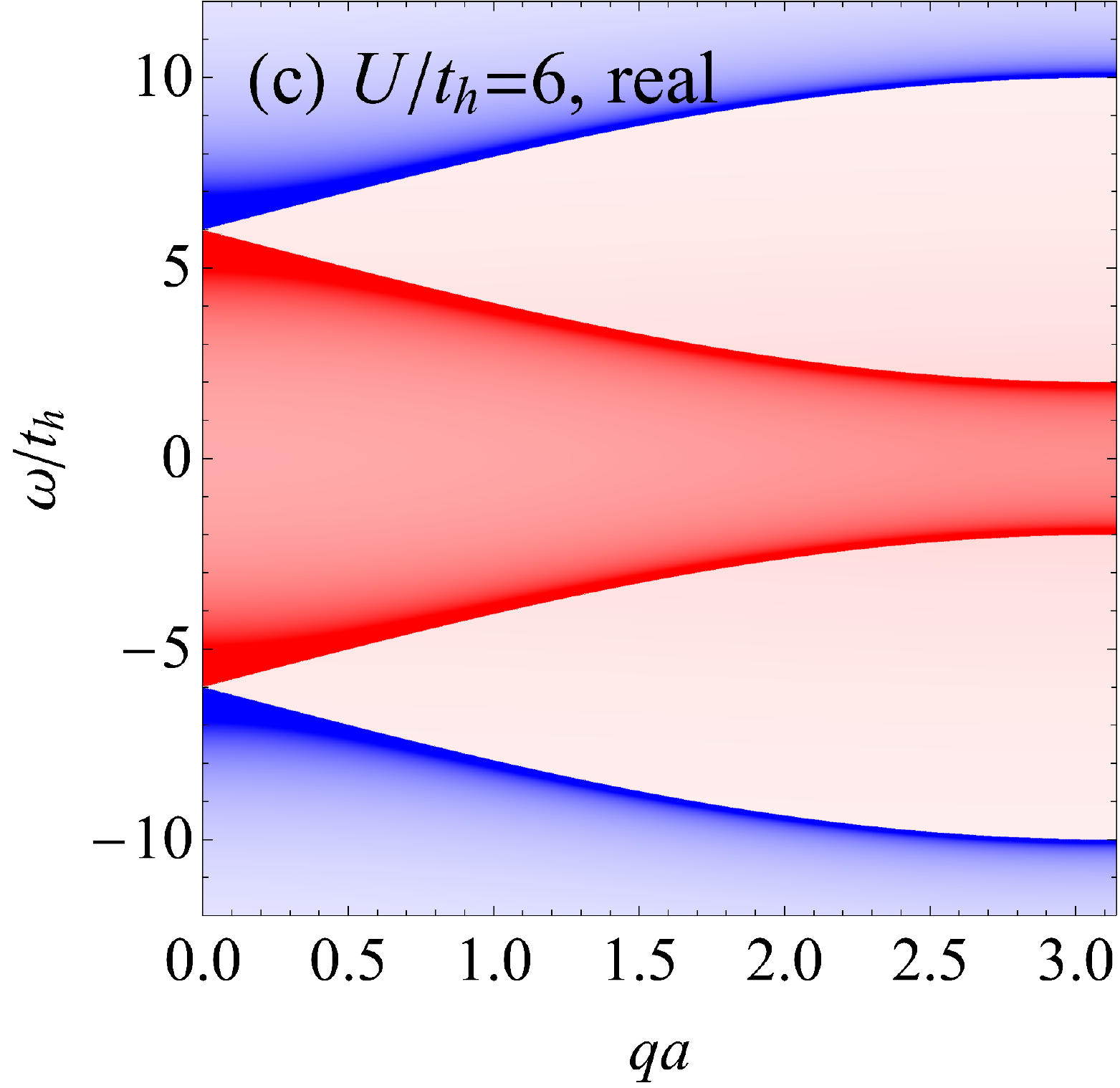}
\includegraphics[width=4.2cm]{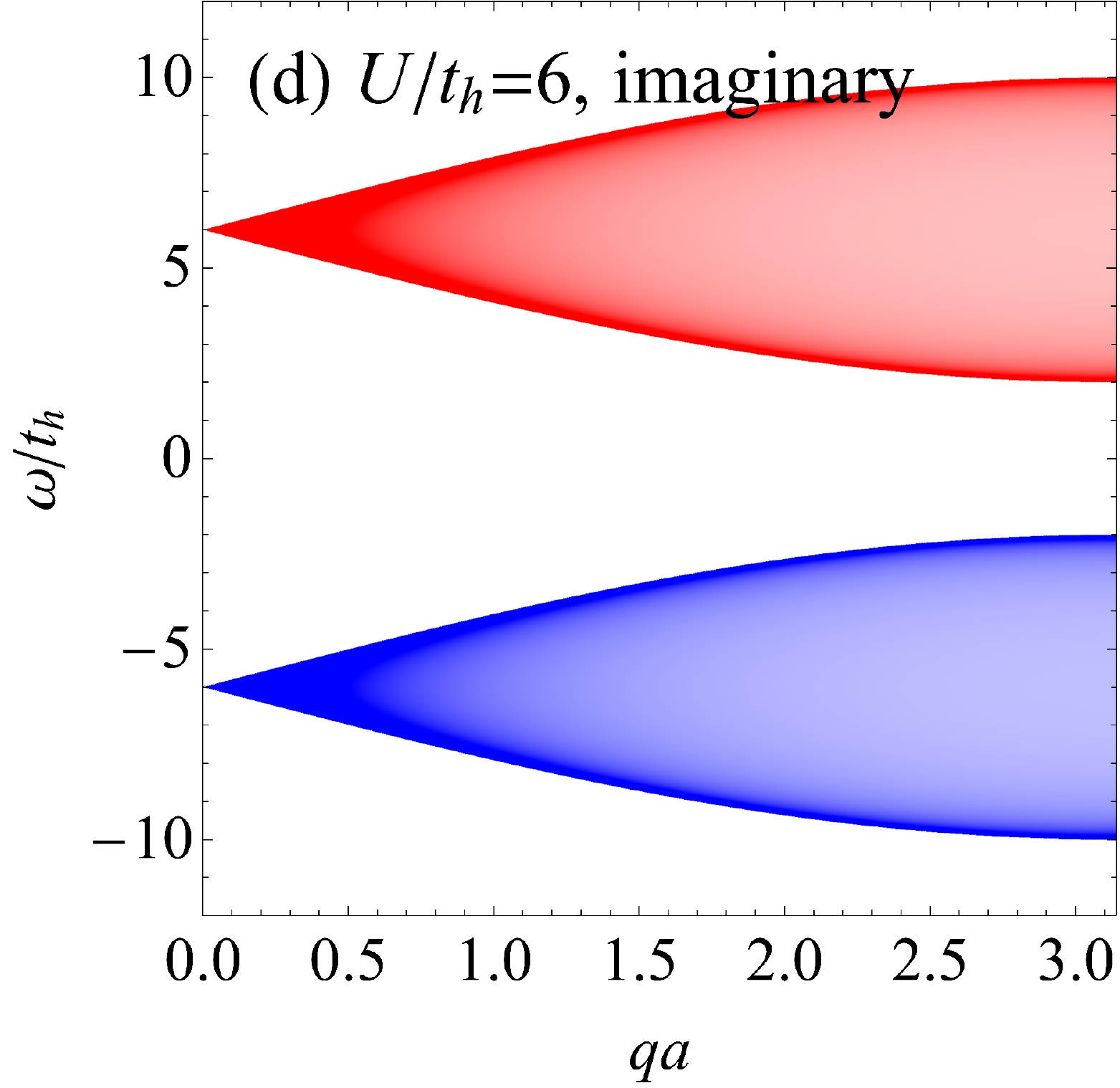}
\includegraphics[width=4.2cm]{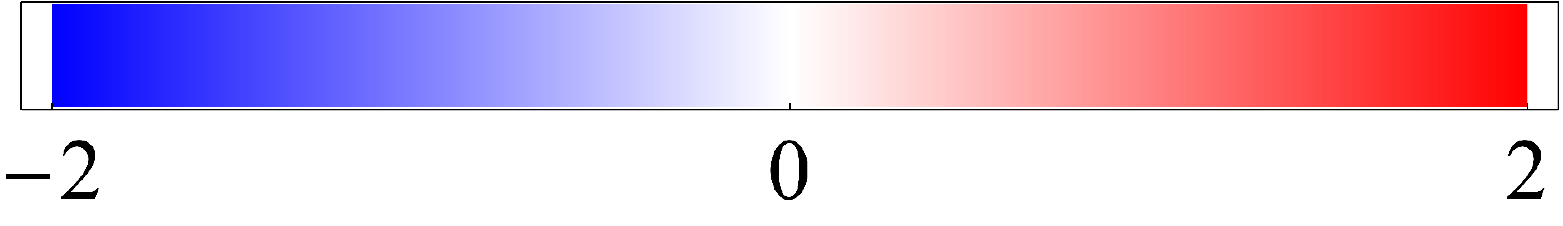}
\caption{Transverse component $K^\perp(\bm q,\omega)$ of the electromagnetic response function 
for the $\eta$-pairing states in the Hubbard model in units of $4e^2t_hC_{M,N}$.
(a), (b): Real (a) and imaginary (b) parts of $K^\perp(\bm q,\omega)$ for $U/t_h=2$.
(c), (d): Real (c) and imaginary (d) parts of $K^\perp(\bm q,\omega)$ for $U/t_h=6$.}
\label{fig: Kxx}
\end{figure}

In Fig.~\ref{fig: Kxx}, we plot $K^\perp(\bm q,\omega)$ for $U/t_h=2$ and $6$.
The imaginary part of $K^\perp(\bm q,\omega)$ represents the absorption (emission) of light for the $\eta$-pairing states.
When a doublon decays by emitting light with momentum $\bm q$, quasiparticles with momentum $\bm k$ and $-\bm k+\bm Q-\bm q$
are created. As discussed above, these particles never interact with each other, so that the total energy of the two particles is given by
$-2t_h\sum_\mu [(\cos k_\mu)+\cos(-k_\mu+Q_\mu-q_\mu)]=-4t_h\sin\frac{q}{2}\sin(k_z+\frac{q}{2})$.
The energy of quasiparticles ranges from $-4t_h\sin\frac{q}{2}$ to $4t_h\sin\frac{q}{2}$.
Since the energy of a single doublon is $U$, the condition for light emission to take place is 
$|\omega-U|<4t_h\sin\frac{q}{2}$. This is exactly the condition of ${\rm Im}\, K^\perp(\bm q,\omega)>0$ (for $U/t_h>4$).
The gap for the electromagnetic response function closes at $q=\pi$ when $U/t_h=4$.

\subsection{D. Asymptotic behavior at long distance}

Here we derive the asymptotic behavior of the transverse electromagnetic response function (\ref{eq: K_perp suppl}) at long distance
and low frequency,
which is related to Pippard's coherence length.
In the low-frequency limit, the kernel is given by
\begin{align}
&
K^\perp(\bm q, \omega=0)
\notag
\\
&=
16e^2t_h^2 C_{M,N}
\begin{cases}
\frac{{\rm sgn}(U)}{\sqrt{U^2-16t_h^2\sin^2\frac{q}{2}}} & \mbox{for}\; |U|>4t_h\sin\frac{q}{2};
\\
0 & \mbox{for}\; |U|<4t_h\sin\frac{q}{2}.
\end{cases}
\label{eq: K_perp(omega=0)}
\end{align}
We consider the problem in two distinct regimes: $|U|>4t_h$ and $|U|<4t_h$.

In the first case ($|U|>4t_h$),
the Fourier transform of $K^\perp(\bm q,\omega=0)$ (\ref{eq: K_perp(omega=0)}) to real space in the $z$ direction reads
\begin{align}
&
K^\perp(j, \omega=0)
\notag
\\
&=
16e^2t_h^2 C_{M,N}
\int_{-\pi}^\pi \frac{dq}{2\pi} e^{iqj}
\frac{{\rm sgn}(U)}{\sqrt{U^2-16t_h^2\sin^2\frac{q}{2}}}.
\label{eq: q integral}
\end{align}
Precisely speaking, Eq.~(\ref{eq: q integral}) shows the electromagnetic response function
for $q_x=q_y=0$, $R_z=j$, and $\omega=0$. The asymptotic behavior of $K^\perp(j,\omega=0)$ for $j\gg 1$
is qualitatively similar to that of $K^\perp(\bm R_j, \omega=0)$ (i.e., the full Fourier transform of $K^\perp(\bm q,\omega=0$)
in all directions) for $|\bm R_j|\gg 1$, since the dominant contribution in 
$K^\perp(j,\omega=0)=\sum_{R_x,R_y} K^\perp(R_x, R_y, R_z=j, \omega=0)$ for $j\gg 1$ arises near $R_x\sim R_y\sim 0$.
In particular, if $K^\perp(\bm R_j,\omega=0)$ decays exponentially for $|\bm R_j|\gg 1$, then $K^\perp(j,\omega=0)$ (\ref{eq: q integral}) also decays exponentially for $j\gg 1$ with the same correlation length.

The integral in Eq.~(\ref{eq: q integral}) can be evaluated analytically as
\begin{align}
K^\perp(j, \omega=0)
&=
16e^2t_h^2 C_{M,N}
\frac{{\rm sgn}(U)}{\sqrt{U^2-16t_h^2}}
\notag
\\
&\quad\times
\frac{{}_3F_2(\tfrac{1}{2},\tfrac{1}{2},1; 1-j,1+j; \tfrac{1}{1-U^2/16t_h^2})}{\Gamma(1-j)\Gamma(1+j)},
\label{eq: 3F2}
\end{align}
where ${}_3F_2(a_1, a_2, a_3; b_1, b_2; z)$ is the generalized hypergeometric function \cite{GradshteynRyzhikBook},
and $\Gamma(z)$ is the gamma function. The explicit expression (\ref{eq: 3F2}), however, does not directly
tell us about the long-distance behavior. Hence we take a different approach.

\begin{figure}[t]
\includegraphics[width=7cm]{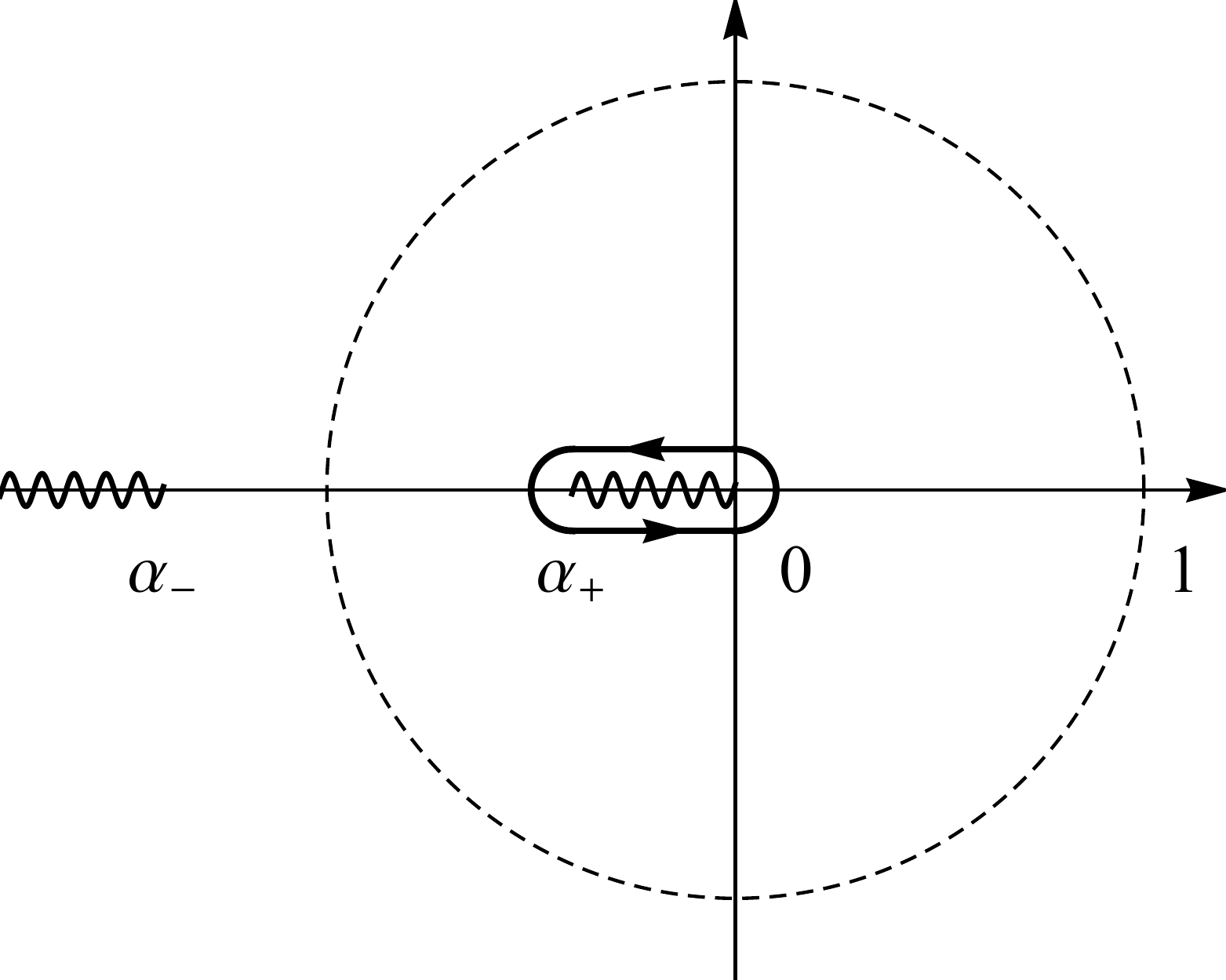}
\caption{Contour for the integral (\ref{eq: contour}) in the complex plane depicted by the dashed circle,
which can continuously be deformed to the solid curve without crossing the branch cuts
shown by the wavy lines.}
\label{fig: contour1}
\end{figure}

The kernel (\ref{eq: q integral}) can be written as
\begin{align}
K^\perp(j, \omega=0)
&=
4e^2t_h C_{M,N} {\rm sgn}(U) I_j(u),
\end{align}
where
\begin{align}
u
&=
\frac{|U|}{4t_h}
\end{align}
is the normalized interaction strength, and
\begin{align}
I_j(u)
&:=
\int_{-\pi}^\pi \frac{dq}{2\pi} e^{-iqj} \frac{1}{\sqrt{u^2-\sin^2\frac{q}{2}}}
\label{eq: I_j(u) u>1}
\end{align}
for $u>1$.
By putting $z=e^{-iq}$, we transform the integral (\ref{eq: I_j(u) u>1}) to a complex contour integral,
\begin{align}
I_j(u)
&=
\oint \frac{dz}{2\pi i} z^{j-1} \frac{1}{\sqrt{u^2-\frac{1}{2}(1-\frac{1}{2}(z+z^{-1}))}}
\notag
\\
&=
2\oint \frac{dz}{2\pi i} \frac{z^{j-\frac{1}{2}}}{\sqrt{z^2+2(2u^2-1)z+1}},
\label{eq: contour}
\end{align}
where the contour is taken to be the circle around the origin with unit radius $|z|=1$ (dashed curve in Fig.~\ref{fig: contour1}).
In the case of $|U|>4t_h$, we have $u>1$. The roots of the quadratic polynomial in the denominator of (\ref{eq: contour}) are given by
\begin{align}
\alpha_\pm
&=
1-2u^2\pm\sqrt{(2u^2-1)^2-1},
\end{align}
which are real numbers. These roots satisfy $\alpha_-<-1<\alpha_+<0$ and $\alpha_+\alpha_-=1$. Using $\alpha_\pm$,
we can write the contour integral (\ref{eq: contour}) as
\begin{align}
I_j(u)
&=
2\oint \frac{dz}{2\pi i} \frac{z^{j-\frac{1}{2}}}{\sqrt{(z-\alpha_+)(z-\alpha_-)}}.
\label{eq: denominator decomposed}
\end{align}

In Fig.~\ref{fig: contour1}, we show the branch cuts that we adopt in evaluating Eq.~(\ref{eq: denominator decomposed})
by wavy lines. With this configuration, the contour can be smoothly deformed 
to the solid curve in Fig.~\ref{fig: contour1} without crossing the branch cuts, 
where the integral is evaluated as
\begin{align}
I_j(u)
&=
2(e^{i(j-\frac{1}{2})\pi+i\pi}-e^{i(j-\frac{1}{2})\pi})
\notag
\\
&\quad\times
\int_{\alpha_+}^0 \frac{dx}{2\pi i} \frac{(-x)^{j-\frac{1}{2}}}{\sqrt{(x-\alpha_+)(x-\alpha_-)}}
\notag
\\
&=
-4e^{i(j-\frac{1}{2})\pi}
\int_{\alpha_+}^0 \frac{dx}{2\pi i} \frac{(-x)^{j-\frac{1}{2}}}{\sqrt{(x-\alpha_+)(x-\alpha_-)}}.
\label{eq: I_j(u) evaluated}
\end{align}
The asymptotic behavior of $I_j(u)$ can be read off as follows: for $j\gg 1$, 
the function $\frac{(-x)^{j-\frac{1}{2}}}{\sqrt{x-\alpha_+}}$ in the integrand has a concentrated contribution near $x=\alpha_+$,
while $\frac{1}{\sqrt{x-\alpha_-}}$ is a smooth function in $(\alpha_+, 0]$. Therefore, one can replace
$\frac{1}{\sqrt{x-\alpha_-}}$ by $\frac{1}{\sqrt{\alpha_+-\alpha_-}}$ in the integrand of Eq.~(\ref{eq: I_j(u) evaluated}), obtaining
\begin{align}
I_j(u)
&\approx
-4e^{i(j-\frac{1}{2})\pi}
\frac{1}{\sqrt{\alpha_+-\alpha_-}}
\int_{\alpha_+}^0 \frac{dx}{2\pi i} \frac{(-x)^{j-\frac{1}{2}}}{\sqrt{x-\alpha_+}}
\end{align}
for $j\gg 1$. The rest of the integral can be evaluated as
\begin{align}
I_j(u)
&\approx
-4e^{i(j-\frac{1}{2})\pi}
\frac{1}{\sqrt{\alpha_+-\alpha_-}} \frac{1}{2\pi i}
\frac{\sqrt{\pi}\Gamma(j+\frac{1}{2})}{\Gamma(j+1)} (-\alpha_+)^{j}
\notag
\\
&=
\frac{2}{\sqrt{\pi}}
\frac{1}{\sqrt{\alpha_+-\alpha_-}}
\frac{\Gamma(j+\frac{1}{2})}{\Gamma(j+1)} \alpha_+^{j}.
\end{align}
Using Stirling's formula, we obtain the asymptotic form of $I_j(u)$ as
\begin{align}
I_j(u)
&\approx
\frac{2}{\sqrt{\pi}}
\frac{1}{\sqrt{\alpha_+-\alpha_-}}
\frac{1}{\sqrt{j}} \alpha_+^{j}.
\end{align}
Thus, the transverse electromagnetic response function behaves 
in the long distance ($j\to \infty$) as
\begin{align}
K^\perp(j, \omega=0)
&\approx
\mbox{cosnt.}\times \frac{\alpha_+^j}{\sqrt{j}}.
\label{eq: K_perp asymptotic}
\end{align}

Since $|\alpha_+|<1$, $K^\perp(j, \omega=0)$ decays exponentially in space
with Pippard's coherence length $\xi$ defined by
\begin{align}
K^\perp(j, \omega=0)
&\approx
\mbox{const.}\times \exp\left(-\frac{j}{\xi}\right).
\end{align}
Physically, $\xi$ represents the length scale over which a response against a local perturbation of electromagnetic fields propagates
in space (Pippard's nonlocal electrodynamics \cite{SchriefferBook}).


From the result (\ref{eq: K_perp asymptotic}), $\xi$ is identified as
\begin{align}
\xi
&=
-\frac{a}{\ln |\alpha_+|}
=
\frac{a}{\cosh^{-1}(2u^2-1)},
\end{align}
which does not depend on the doublon density $\rho$.
At $|U|=4t_h=: U_c$, the coherence length diverges as
\begin{align}
\xi
&\propto
\frac{1}{|U-U_c|^{\frac{1}{2}}}.
\end{align}
This is exactly the point where the electromagnetic gap closes at $q=\pi$.

\begin{figure}[t]
\includegraphics[width=6cm]{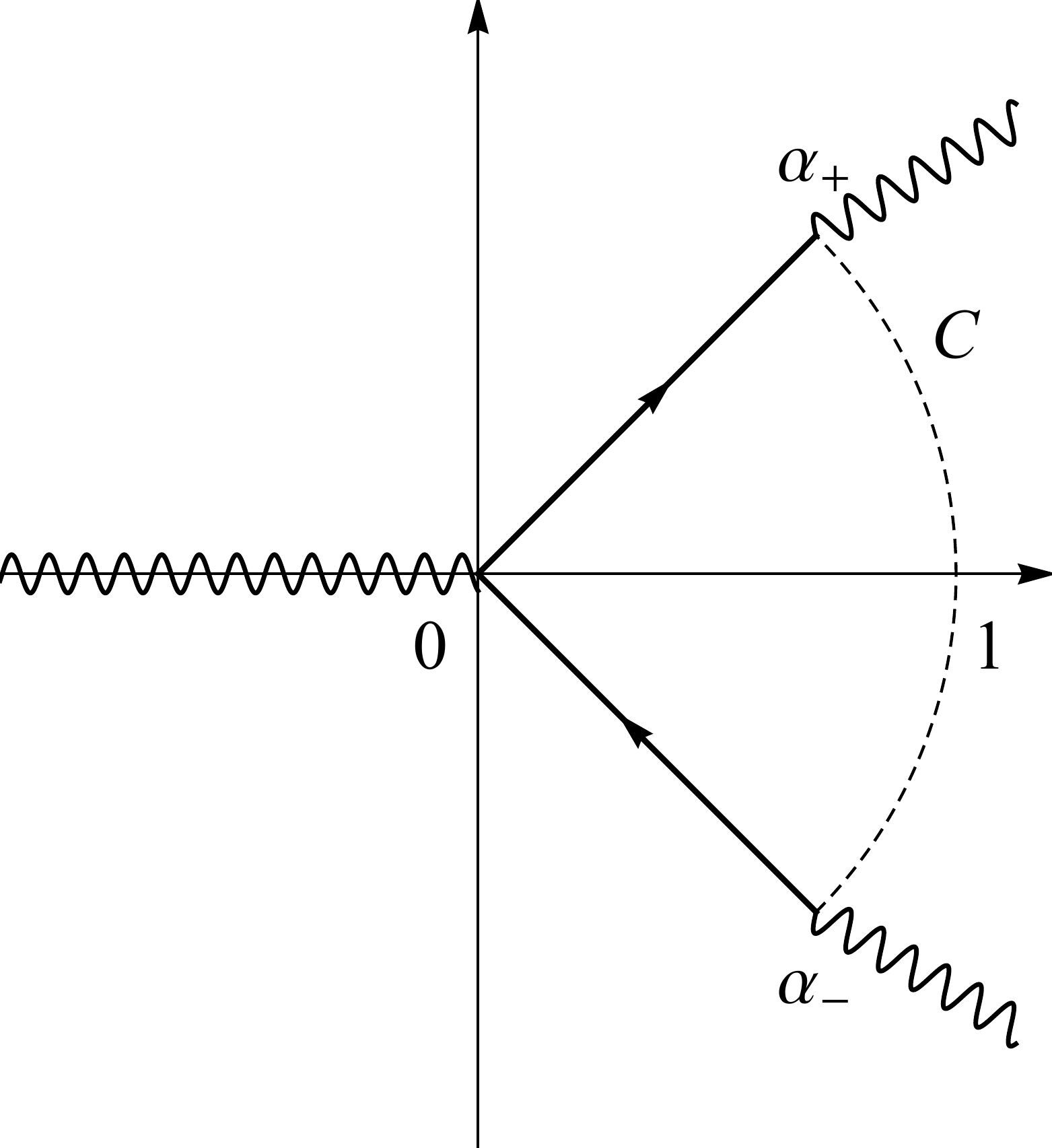}
\caption{Contour $C$ for the integral (\ref{eq: contour2}) in the complex plane depicted
by the dashed arc, which can continuously be deformed to the solid lines without crossing the branch cuts
as shown by the wavy lines.}
\label{fig: contour2}
\end{figure}

In the second case ($|U|<4t_h$), the Meissner kernel in the low-frequency limit is
expressed as
\begin{align}
K^\perp(j, \omega=0)
&=
4e^2 t_h C_{M,N}{\rm sgn}(U)I_j(u)
\end{align}
with
\begin{align}
I_j(u)
&=
\int_{u\ge |\sin\frac{q}{2}|} \frac{dq}{2\pi} e^{-iqj}
\frac{1}{\sqrt{u^2-\sin^2\frac{q}{2}}}
\label{eq: I_j(u) u<1}
\end{align}
for $u<1$. Similarly to the first case, we put $z=e^{-iq}$ to rewrite the integral (\ref{eq: I_j(u) u<1}) as
\begin{align}
I_j(u)
&=
2\int_C \frac{dz}{2\pi i} \frac{z^{j-\frac{1}{2}}}{\sqrt{(z-\alpha_+)(z-\alpha_-)}},
\label{eq: contour2}
\end{align}
where the roots in the denominator are given by
\begin{align}
\alpha_\pm
&=
1-2u^2\pm i\sqrt{1-(2u^2-1)^2},
\end{align}
and the contour $C$ is taken to be the arc of the circle with the unit radius connecting $\alpha_-$ and $\alpha_+$
(dashed curve in Fig.~\ref{fig: contour2}). We choose the branch cuts in the integrand of Eq.~(\ref{eq: contour2})
as shown by wavy lines in Fig.~\ref{fig: contour2}.

Following the steepest descent method, we deform the contour from $C$ to the solid lines in Fig.~\ref{fig: contour2}
without crossing the branch cuts, where we put $z=r\alpha_\pm$. Now the integral (\ref{eq: contour2}) can be
evaluated as
\begin{align}
I_j(u)
&=
2\int_0^1 \frac{dr}{2\pi i} \alpha_+ \frac{(r\alpha_+)^{j-\frac{1}{2}}}{\sqrt{(r\alpha_+-\alpha_+)(r\alpha_+-\alpha_-)}}
\notag
\\
&\quad
-2\int_0^1 \frac{dr}{2\pi i} \alpha_- \frac{(r\alpha_-)^{j-\frac{1}{2}}}{\sqrt{(r\alpha_--\alpha_+)(r\alpha_--\alpha_-)}}.
\label{eq: }
\end{align}

For $j\gg 1$, the function $\frac{r^{j-\frac{1}{2}}}{\sqrt{r\alpha_+-\alpha_+}}$ in the first integral
is dominantly contributed from a region near $r=1$, which allows us to replace
$\frac{1}{\sqrt{r\alpha_+-\alpha_-}}$ by $\frac{1}{\sqrt{\alpha_+-\alpha_-}}$ in the integrand.
A similar approximation can be applied to the second term. Taking care of the branch cuts, we obtain
\begin{align}
I_j(u)
&\approx
\frac{2(\alpha_+^j e^{\frac{i\pi}{4}}-\alpha_-^j e^{-\frac{i\pi}{4}})}{\sqrt{|\alpha_+-\alpha_-|}}
\int_0^1 \frac{dr}{2\pi i} \frac{r^{j-\frac{1}{2}}}{\sqrt{1-r}}.
\end{align}
If we define $\alpha_\pm=: e^{\pm i\varphi}$, the integral (\ref{eq: I_j(u) u<1}) can be approximated as
\begin{align}
I_j(u)
&\approx
\frac{2\sin(j\varphi+\frac{\pi}{4})}{\sqrt{\pi}\sqrt{|\alpha_+-\alpha_-|}}
\frac{\Gamma(j+\frac{1}{2})}{\Gamma(j+1)}.
\end{align}
Using Stirling's formula, the asymptotic form of $I_j(u)$ for $j\gg 1$ is given by
\begin{align}
I_j(u)
&\approx
\frac{2\sin(j\varphi+\frac{\pi}{4})}{\sqrt{\pi}\sqrt{|\alpha_+-\alpha_-|}}
\frac{1}{\sqrt{j}}.
\end{align}
Therefore, the kernel decays in a long distance according to a power law as
\begin{align}
K^\perp(j, \omega=0)
&\approx
\mbox{const.}\times \frac{1}{\sqrt{j}}.
\end{align}
This means that the coherence length diverges ($\xi=\infty$) for $|U|<4t_h$.

\begin{figure}[t]
\includegraphics[width=8cm]{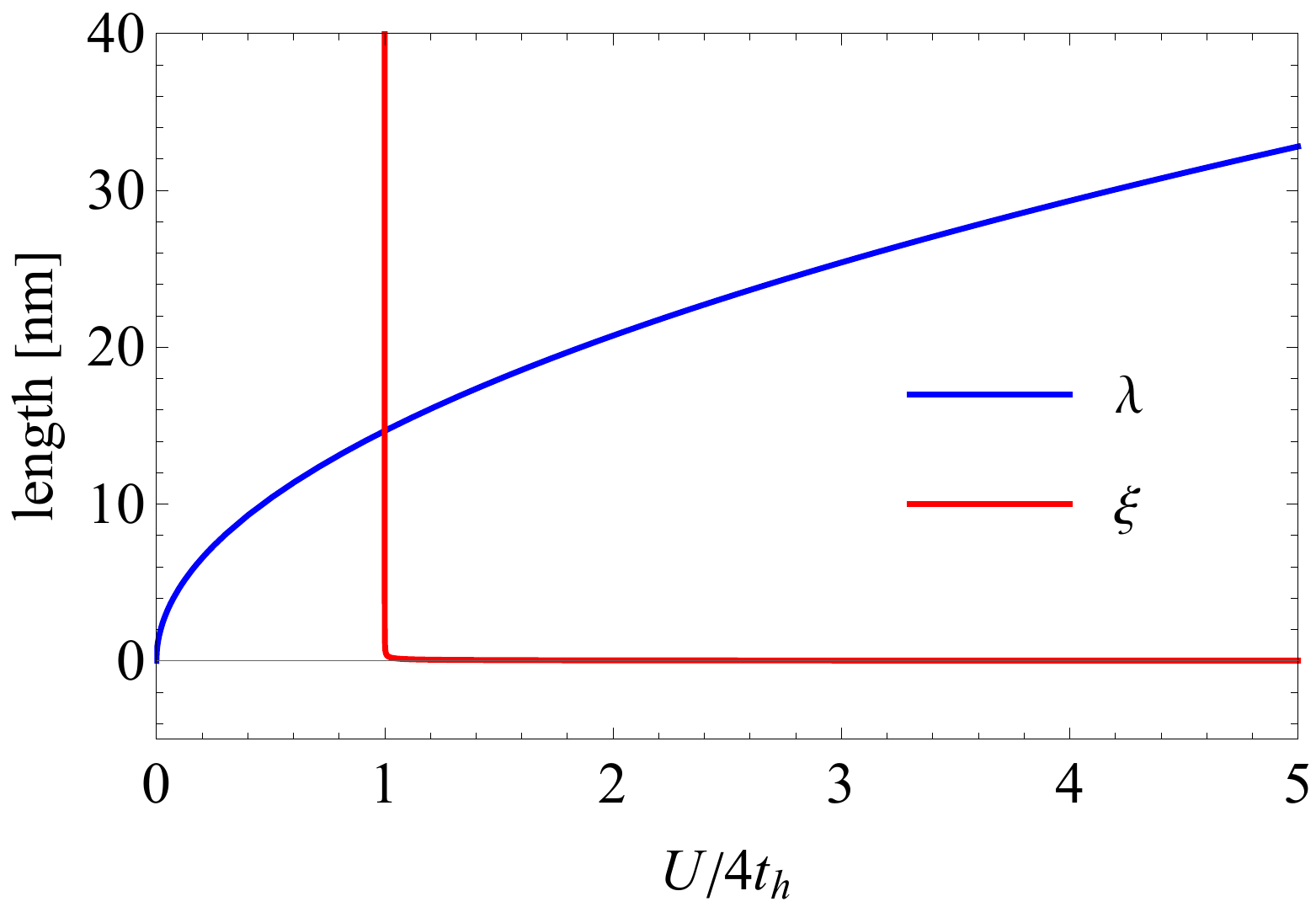}
\caption{Comparison between London's penetration depth $\lambda$ and Pippard's coherence length $\xi$
for the $\eta$-pairing states in the Hubbard model with $t_h=1$ [eV], $a=1$ [\AA], and $\rho=0.5$.}
\label{fig: lambda}
\end{figure}

In Fig.~\ref{fig: lambda}, we plot the coherence length $\xi$ for the $\eta$-pairing states with $t_h=1$ [eV], $a=1$ [\AA],
and $\rho=0.5$ in comparison with London's penetration depth
defined by $\lambda=\sqrt{\frac{\hbar^2a U}{16\mu_0e^2t_h^2C_{M,N}}}$.
At $0\le U<4t_h$, $\lambda$ grows smoothly as a function of $U$, and satisfies $\lambda<\xi=\infty$.
When $U$ exceeds $4t_h$, $\xi$ immediately decays to the order of 1 [\AA], whereas
$\lambda$ stays on the order of 10 [nm]. The point at which $\lambda$ becomes equal to $\xi$ 
is very close to $U=4t_h$, beyond which $\lambda$ becomes larger than $\xi$.
Thus, for $0\le U<4t_h$ the $\eta$-pairing state is a type-I superconductor,
whereas for $U>4t_h$ the $\eta$-pairing state is classified to a type-II superconductor.

For $U<0$, we analytically continue $\lambda$ to complex values, which has a physical meaning as discussed in Sec.~\ref{sec: magnetic properties}.
We will see that the $\eta$-pairing state has different magnetic properties depending on whether
$|\lambda|$ is larger than $\xi$ or not.
For $-4t_h<U<0$, we have the relation $|\lambda|<\xi$, where the $\eta$-pairing state is called a type-I tachyonic superconductor
(see the main text). For $U<-4t_h$, we have $|\lambda|>\xi$, where
the $\eta$-pairing state is called a type-II tachyonic superconductor (see the phase diagram in Fig.~\ref{fig: phase diagram}(b) in the main text).

\subsection{E. Longitudinal component}

The longitudinal component of the electromagnetic response function is defined by 
$K^\parallel(\bm q,\omega):=K^{zz}(\bm q,\omega)$, where we take $\bm q\parallel\bm e_z$.
The $zz$ component of the current-current correlation function (\ref{eq: two-particle}) reads
\begin{align}
&
\sum_{j} e^{i\bm q\cdot \bm R_j} \langle \psi_N| J^z(\bm R_j, t) J^z(0,0)|\psi_N\rangle
\notag
\\
&=
4e^2t_h^2 C_{M,N}
(\langle \bm Q-\bm q, +\bm e_z|-e^{-iq}\langle \bm Q-\bm q, -\bm e_z|) 
\notag
\\
&\quad\times
e^{-iH(\bm Q-\bm q)t} (|\bm Q-\bm q, +\bm e_z\rangle-e^{iq}|\bm Q-\bm q, -\bm e_z\rangle),
\end{align}
where $H(\bm Q-\bm q)$ is given in Eq.~(\ref{eq: H(Q-q)}).
During the time evolution, the relative coordinate of two particles changes only in the $z$ direction.
Therefore, what we need to solve is essentially a one-dimensional two-particle problem, 
which can be solved exactly in the spirit of the Bethe ansatz \cite{EsslerBook}. 
Here we do not go into details of analytical solutions,
since we can easily diagonalize the Hamiltonian (\ref{eq: H(Q-q)}) numerically for a large system size.

\begin{figure}[t]
\includegraphics[width=4.2cm]{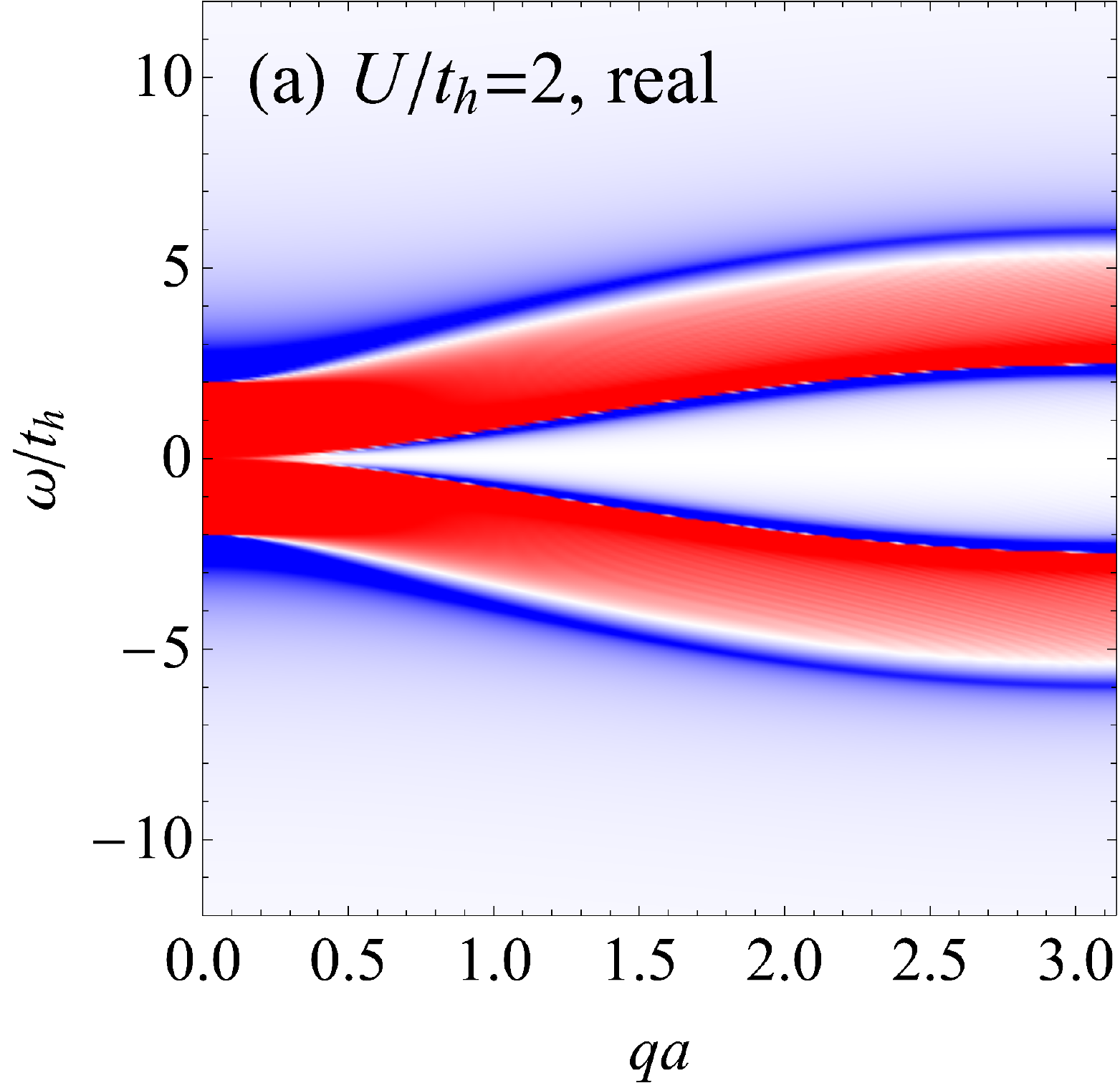}
\vspace{.2cm}
\includegraphics[width=4.2cm]{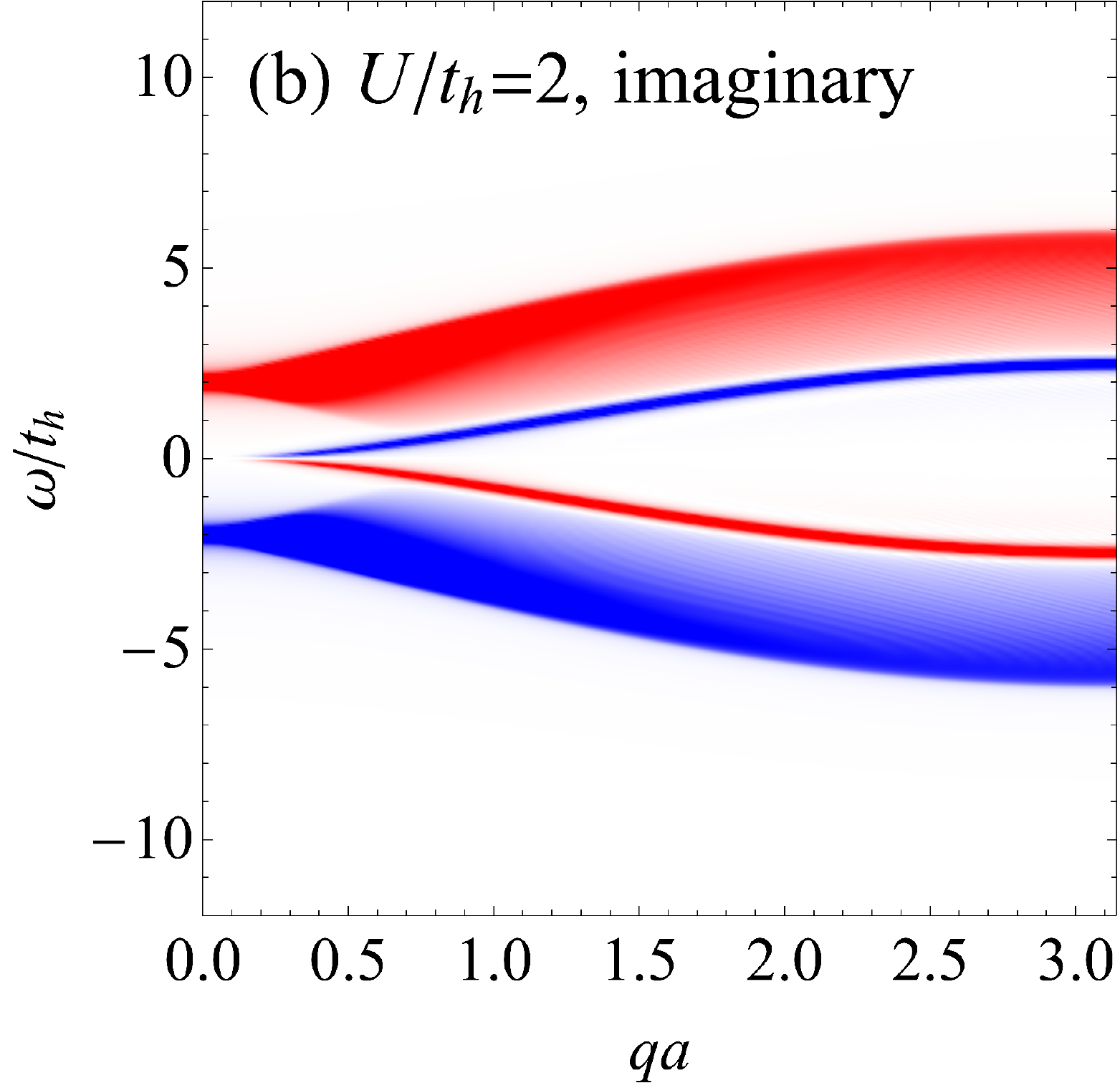}
\includegraphics[width=4.2cm]{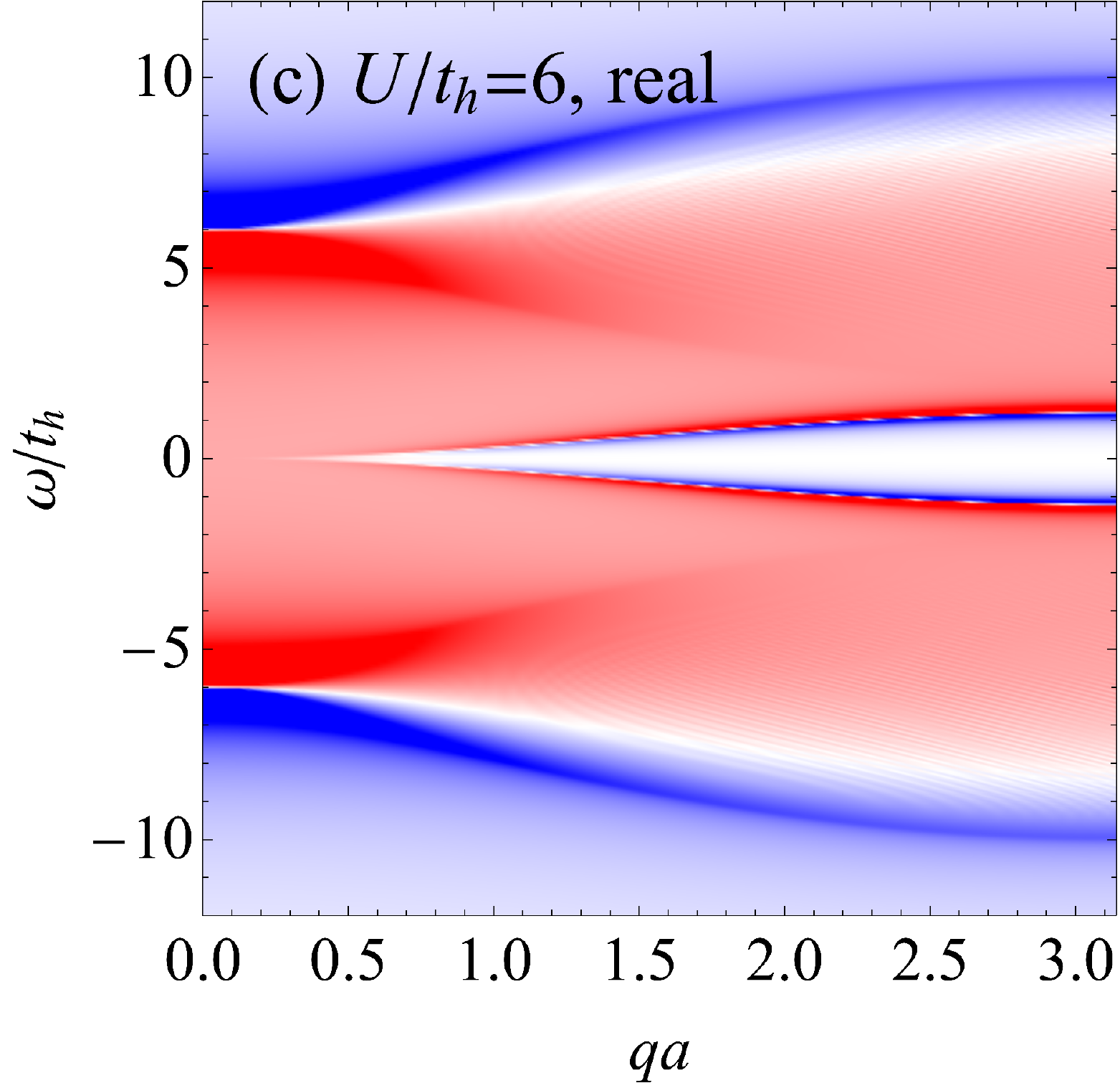}
\vspace{.2cm}
\includegraphics[width=4.2cm]{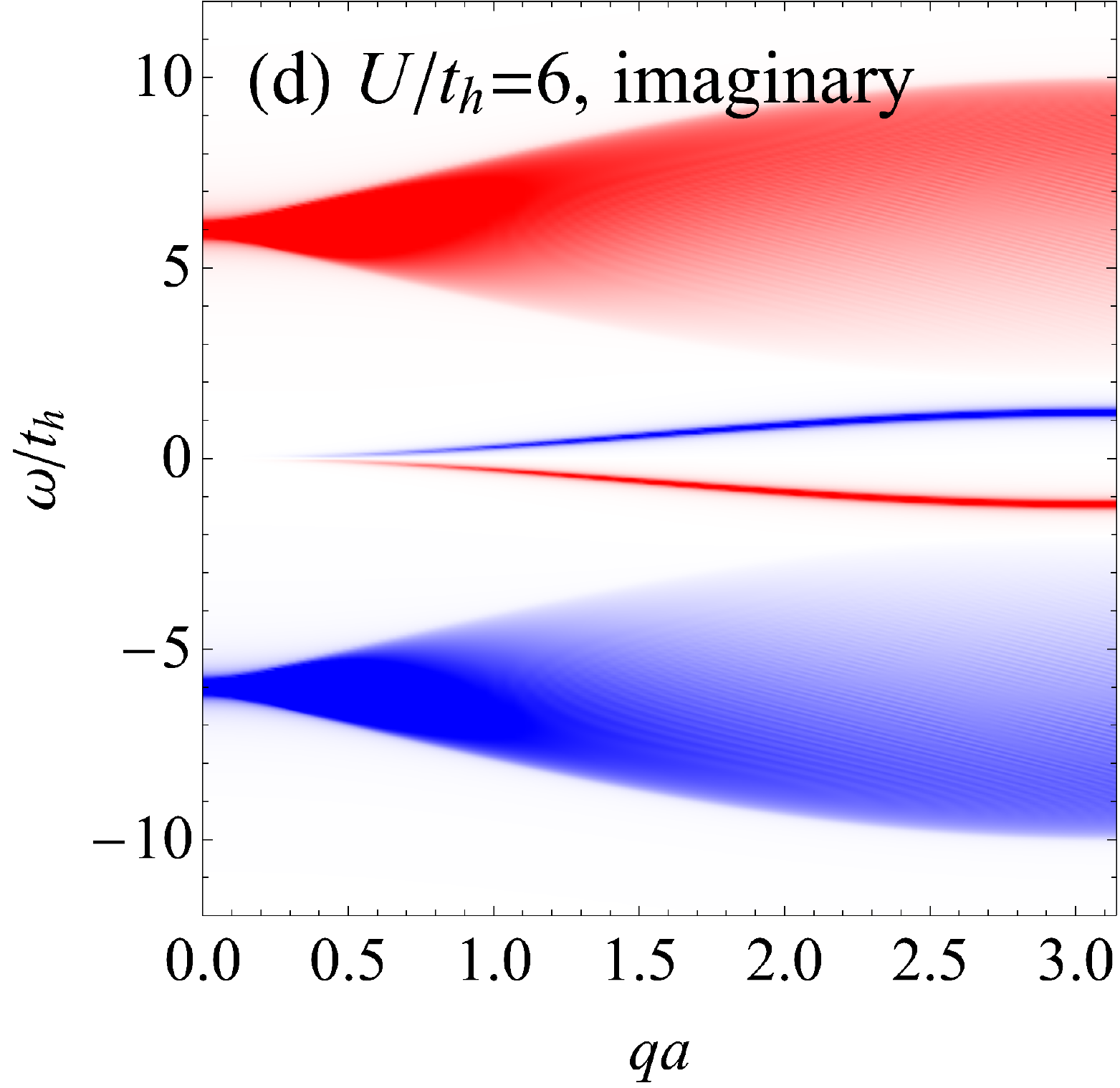}
\includegraphics[width=4.2cm]{color-bar.pdf}
\caption{Longitudinal component $K^\parallel(\bm q,\omega)$ of the electromagnetic response function
for the $\eta$-pairing states in the Hubbard model in units of $4e^2t_hC_{M,N}$.
(a), (b): Real (a) and imaginary (b) parts of $K^\parallel(\bm q,\omega)$ for $U/t_h=2$.
(c), (d): Real (c) and imaginary (d) parts of $K^\parallel(\bm q,\omega)$ for $U/t_h=6$.}
\label{fig: Kzz}
\end{figure}

In Fig.~\ref{fig: Kzz}, we plot $K^\parallel(\bm q,\omega)$ for $U/t_h=2$ and $6$.
Compared with the transverse component (Fig.~\ref{fig: Kxx}), there appear sideband structures which are shifted by $U$ from the original bands
in the longitudinal component due to the effect of the interaction. Otherwise, both of them
have similar spectral features. In the low-frequency limit, the longitudinal component vanishes,
\begin{align}
\lim_{\omega\to 0} K^\parallel(\bm q,\omega)
&=
0,
\end{align}
as required by charge conservation (see Sec.~\ref{sec: charge conservation}).
In the low-momentum limit, the longitudinal component agrees with the transverse one,
\begin{align}
\lim_{\bm q\to 0} K^\parallel(\bm q,\omega)
&=
\lim_{\bm q\to 0} K^\perp(\bm q,\omega)
\notag
\\
&=
8e^2t_h^2C_{M,N}\left(\frac{1}{\omega+U}-\frac{1}{\omega-U}\right),
\label{eq: K q->0}
\end{align}
since the hopping in the $z$ direction is suppressed in the limit of $\bm q\to 0$ as can be seen from Eq.~(\ref{eq: H(Q-q)}).

\section{II. Charge conservation}
\label{sec: charge conservation}

In the Hubbard model [Eq.~(\ref{eq: Hubbard}) in the main text], electric charge is conserved due to the charge $U(1)$ symmetry.
This imposes a nontrivial constraint on the electromagnetic response function \cite{SchriefferBook}.
To see this, we introduce the four-vector form of the electromagnetic response function defined by
\begin{align}
K^{\mu\nu}(\bm q,\omega)
&=
-i\theta(t)\langle\psi_N| [J^\mu(\bm R_j, t), J^\nu(0,0)] |\psi_N\rangle
\label{eq: K^munu}
\end{align}
$(\mu,\nu=0,x,y,z)$, where $J^\mu(\bm R_j)=(c\rho(\bm R_j), \bm J(\bm R_j))$ is the four-vector current,
and
\begin{align}
\rho(\bm R_j)
&=
e\sum_\sigma c_{j\sigma}^\dagger c_{j\sigma}
\end{align}
is the local density operator. In the following, we use the metric convention $\eta^{\mu\nu}={\rm diag}(-,+,+,+)$.
The linear response in the four-vector form reads $J^\mu(\bm q,\omega)=-K^{\mu\nu}(\bm q,\omega)A_\nu(\bm q,\omega)$,
where $A_\nu=\eta_{\nu\lambda}A^\lambda=(-\frac{\phi}{c}, \bm A)$ and $\phi$ is the scalar potential.

\subsection{A. Charge response function}

The charge response function for the $\eta$-pairing state is given by
\begin{align}
K^{00}(\bm R_j, t)
&=
-ic^2\theta(t)\langle \psi_N| [\rho(\bm R_j, t), \rho(0,0)] |\psi_N\rangle.
\end{align}
As before, the density-density correlation function for $N$ particles
can be reduced to the two-particle correlation function. To see this, we define local $\eta$ operators,
\begin{align}
\eta^+(\bm R_j)
&:=
e^{i\bm Q\cdot\bm R_j} c_{j\uparrow}^\dagger c_{j\downarrow}^\dagger,
\\
\eta^-(\bm R_j)
&:=
e^{-i\bm Q\cdot\bm R_j} c_{j\downarrow} c_{j\uparrow}.
\end{align}
They satisfy the following commutation relations:
\begin{align}
[\rho(\bm R_j), \eta^\pm]
&=
\pm 2\eta^\pm(\bm R_j),
\\
[\eta^\pm(\bm R_j), \eta^\pm]
&=
0,
\\
[\eta^\pm(\bm R_j), \eta^\mp]
&=
\pm(\rho(\bm R_j)-1).
\end{align}
Applying the above relations iteratively, we can reduce the $N$-particle density-density correlation function to
\begin{align}
&
\langle \psi_N| \rho(\bm R_j, t)\rho(0,0)|\psi_N\rangle
\notag
\\
&=
4e^2C_{M,N}
\langle 0| \eta^-(\bm R_j, t) \eta^+(0,0) |0\rangle
+4e^2\frac{\frac{N}{2}(\frac{N}{2}-1)}{M(M-1)}.
\end{align}
From this result, we can evaluate the charge response function as
\begin{align}
K^{00}(\bm q, t)
&=
-4i\theta(t)c^2e^2t_h^2 C_{M,N}
\notag
\\
&\quad\times
[\langle \bm Q-\bm q, \bm r=0| e^{-iH(\bm Q-\bm q)t} |\bm Q-\bm q, \bm r=0\rangle
\notag
\\
&\quad
-\langle \bm Q+\bm q, \bm r=0| e^{iH(\bm Q+\bm q)t} |\bm Q+\bm q, \bm r=0\rangle],
\end{align}
where
\begin{align}
H(\bm Q \pm \bm q)
&=
-t_h((1-e^{\pm iq})\Delta_z^+ +(1-e^{\mp iq})\Delta_z^-)
\notag
\\
&\quad
+U(\delta_{\bm r,0}-1).
\end{align}
Thus, the problem reduces to solving the two-particle dynamics in the one-dimensional Hubbard model,
which can be diagonalized numerically or analytically with the Bethe ansatz method.

\begin{figure}[t]
\includegraphics[width=4.2cm]{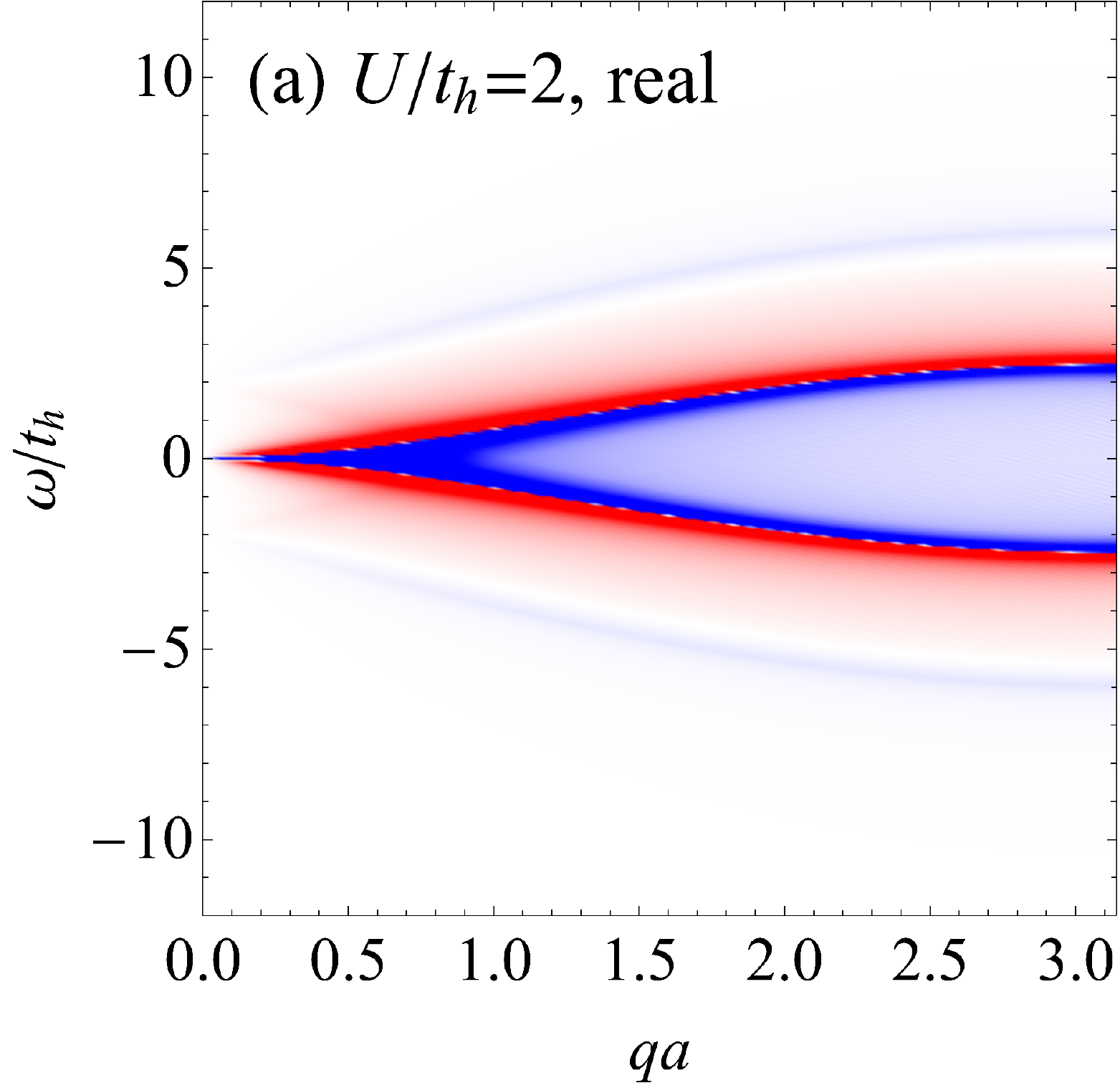}
\vspace{.2cm}
\includegraphics[width=4.2cm]{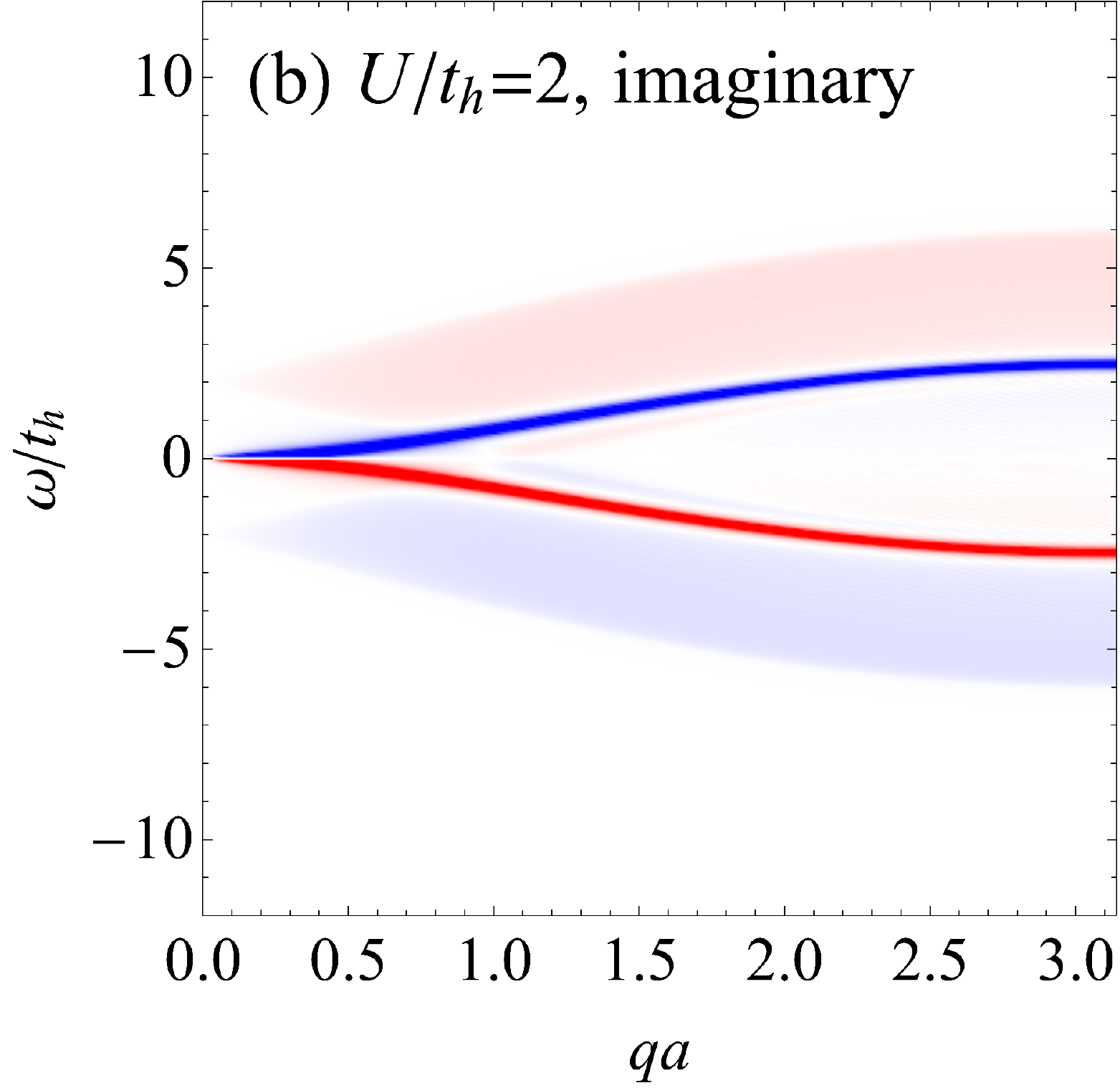}
\includegraphics[width=4.2cm]{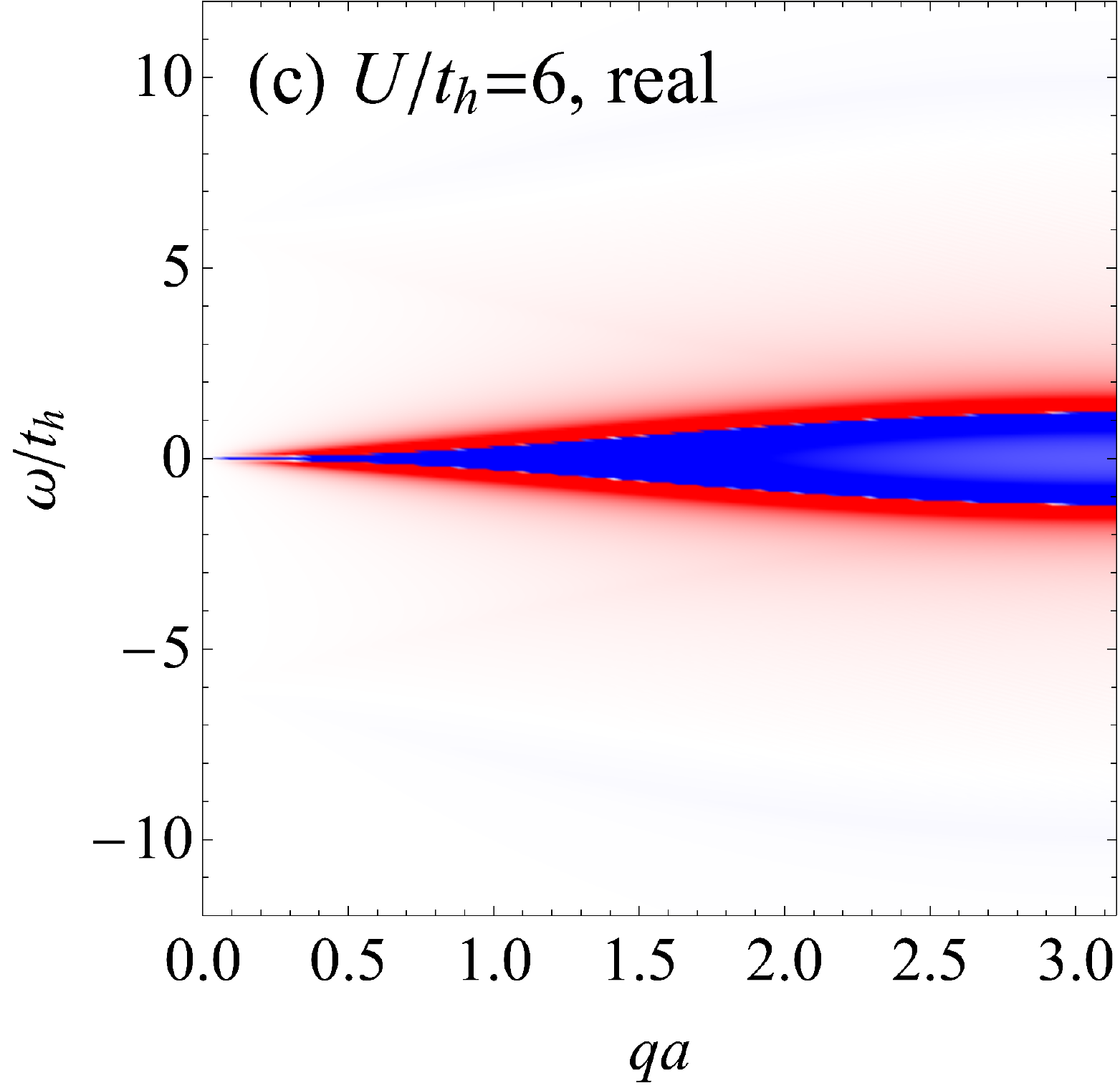}
\vspace{.2cm}
\includegraphics[width=4.2cm]{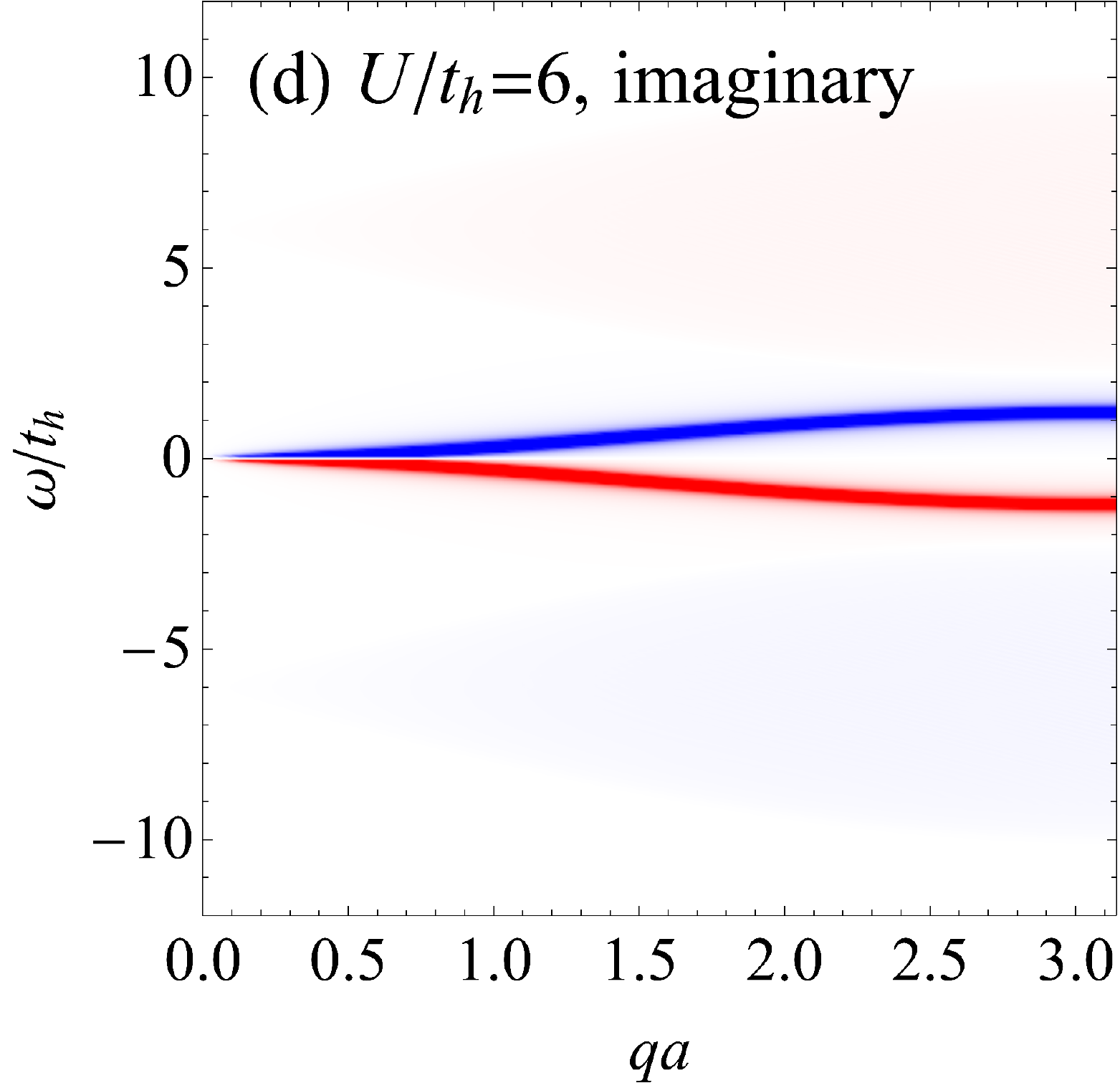}
\includegraphics[width=4.2cm]{color-bar.pdf}
\caption{Charge response function $K^{00}(\bm q,\omega)$ 
for the $\eta$-pairing states in the Hubbard model in units of $4c^2e^2t_hC_{M,N}$.
(a), (b): Real (a) and imaginary (b) parts of $K^{00}(\bm q,\omega)$ for $U/t_h=2$.
(c), (d): Real (c) and imaginary (d) parts of $K^{00}(\bm q,\omega)$ for $U/t_h=6$.}
\label{fig: K00}
\end{figure}

In Fig.~\ref{fig: K00}, we plot the charge response function for $U/t_h=2$ and $6$.
We will see in the next subsection that $K^{00}(\bm q,\omega)$ is related to $K^\parallel(\bm q,\omega)$ due to the symmetry constraint.
Compared with $K^\parallel(\bm q,\omega)$, the contribution of the low-energy sidebands is enhanced in $K^{00}(\bm q,\omega)$,
which can also be understood from the symmetry constraint [Eq.~(\ref{eq: K^00 K^parallel})].
In the low-momentum limit, the charge response function vanishes,
\begin{align}
\lim_{\bm q\to 0} K^{00}(\bm q,\omega)
&=
0,
\end{align}
since $H(\bm Q\pm\bm q)=0$ for $q\to 0$ and $\bm r=0$.

\subsection{B. Symmetry constraint}

Here we see how the symmetry puts a constraint on the electromagnetic response functions \cite{SchriefferBook}.
Our starting point is the continuity equation,
\begin{align}
\frac{d}{dt}\rho(\bm R_j, t)
&=
-\sum_{\mu=x,y,z} (J^\mu(\bm R_j)-J^\mu(\bm R_j-\bm e_\mu)),
\end{align}
which is a direct consequence of the U(1) symmetry of the Hubbard model.
After Fourier transformation, the relation becomes
\begin{align}
-i\omega\rho(\bm q,\omega)
&=
-\sum_{\mu=x,y,z} (e^{\frac{i}{2}\bm q\cdot\bm e_\mu}-e^{-\frac{i}{2}\bm q\cdot\bm e_\mu})J^\mu(\bm q,\omega).
\label{eq: charge conservation}
\end{align}
Taking the expectation value with respect to $|\psi_N\rangle$ and
substituting the electromagnetic response function (\ref{eq: K^munu}) in Eq.~(\ref{eq: charge conservation}), we obtain
\begin{align}
&
-i\frac{\omega}{c}\sum_{\nu=0,x,y,z} K^{0\nu}(\bm q,\omega)A_\nu(\bm q,\omega)
\notag
\\
&=
-\sum_{\mu=x,y,z} (e^{\frac{i}{2}\bm q\cdot\bm e_\mu}-e^{-\frac{i}{2}\bm q\cdot\bm e_\mu}) \sum_{\nu=0,x,y,z} K^{\mu\nu}(\bm q,\omega)
A_\nu(\bm q,\omega).
\end{align}
Since the above relation must hold for arbitrary $A_\nu(\bm q,\omega)$, we conclude that
\begin{align}
i\frac{\omega}{c} K^{0\nu}(\bm q,\omega)
-\sum_{\mu=x,y,z} (e^{\frac{i}{2}\bm q\cdot\bm e_\mu}-e^{-\frac{i}{2}\bm q\cdot\bm e_\mu}) K^{\mu\nu}(\bm q,\omega)
&=
0.
\end{align}
We assume $\bm q\parallel \bm e_z$ without loss of generality. Then the relation becomes
\begin{align}
K^{0\nu}(\bm q,\omega)
&=
\frac{2c\sin\frac{q}{2}}{\omega} K^{z\nu}(\bm q,\omega).
\label{eq: symmetry constraint}
\end{align}
Due to Onsager's reciprocity relation, we also have $K^{\mu\nu}(\bm q,\omega)=K^{\nu\mu}(\bm q,\omega)$.
Therefore, $K^{00}(\bm q,\omega)$ and $K^\parallel(\bm q,\omega)$ are related to each other through
\begin{align}
K^{00}(\bm q,\omega)
&=
\frac{4c^2\sin^2\frac{q}{2}}{\omega^2}K^\parallel(\bm q,\omega).
\label{eq: K^00 K^parallel}
\end{align}

To summarize, all the components of the electromagnetic response function $K^{\mu\nu}(\bm q,\omega)$ can be
expressed in terms of $K^\perp(\bm q,\omega)$ and $K^\parallel(\bm q,\omega)$ as
\begin{align}
&
K^{\mu\nu}(\bm q,\omega)
=
\notag
\\
&
\begin{pmatrix}
\frac{4c^2\sin^2\frac{q}{2}}{\omega^2}K^\parallel(\bm q,\omega) & 0 & 0 & \frac{2c\sin\frac{q}{2}}{\omega}K^\parallel(\bm q,\omega)
\\
0 & K^\perp(\bm q,\omega) & 0 & 0
\\
0 & 0 & K^\perp(\bm q,\omega) & 0
\\
\frac{2c\sin\frac{q}{2}}{\omega}K^\parallel(\bm q,\omega) & 0 & 0 & K^\parallel(\bm q,\omega)
\end{pmatrix}.
\label{eq: K matrix}
\end{align}
One can check that the symmetry constraint (\ref{eq: symmetry constraint}) is consistent with gauge invariance
in the Hubbard model.

\section{III. Dynamical electromagnetic fields}

In this section, we give a detailed description of dynamical electromagnetic fields coupled to the $\eta$-pairing states.
We start with the Maxwell equations,
\begin{align}
&
\nabla\cdot\bm E
=
\frac{1}{\varepsilon_0}\rho,
\label{Maxwell1}
\\
&
\nabla\times\bm E+\frac{\partial}{\partial t}\bm B
=
0,
\label{Maxwell2}
\\
&
\nabla\cdot\bm B
=
0,
\label{Maxwell3}
\\
&
\nabla\times\bm B-\frac{1}{c^2}\frac{\partial}{\partial t}\bm E
=
\mu_0\bm j,
\label{Maxwell4}
\end{align}
where $\bm E$ and $\bm B$ are electric and magnetic fields, 
and $\rho$ and $\bm j$ are the charge density and current, respectively.
As usual, we introduce the scalar potential $\phi$ and the vector potential $\bm A$ through
\begin{align}
\bm E
&=
-\nabla\phi-\frac{\partial}{\partial t}\bm A,
\\
\bm B
&=
\nabla\times \bm A.
\end{align}
In the following, we adopt the Lorenz gauge:
\begin{align}
\frac{1}{c^2}\frac{\partial}{\partial t}\phi+\nabla\cdot\bm A
&=
0.
\label{eq: Lorenz gauge}
\end{align}
Then, the equations for $\phi$ and $\bm A$ become
\begin{align}
\frac{1}{c^2}\frac{\partial^2}{\partial t^2}\phi-\nabla^2\phi
&=
\frac{1}{\varepsilon_0}\rho,
\\
\frac{1}{c^2}\frac{\partial^2}{\partial t^2}\bm A-\nabla^2\bm A
&=
\mu_0\bm j.
\end{align}

To solve these equations, we assume plane-wave solutions,
\begin{align}
\phi(\bm r, t)
&=
\phi_0 e^{-i\omega t+i\bm q\cdot\bm r},
\\
\bm A(\bm r,t)
&=
\bm A_0 e^{-i\omega t+i\bm q\cdot\bm r},
\end{align}
with frequency $\omega$ and momentum $\bm q$. 
Without loss of generality, we assume $\bm q \parallel \bm e_z$.
Combining the linear-response relation $j^\mu(\bm q,\omega)=-K^{\mu\nu}(\bm q,\omega)A_\nu(\bm q,\omega)$
and the gauge condition (\ref{eq: Lorenz gauge}),
we obtain
\begin{align}
-\frac{\omega^2}{c^2}\phi+q^2\phi
&=
\frac{1}{\varepsilon_0}K^{00}(q,\omega)\frac{\phi}{c}
-\frac{1}{\varepsilon_0}K^{0z}(q,\omega)A_z,
\label{eq: longitudinal mode1}
\\
-\frac{\omega^2}{c^2}A^x+q^2A^x
&=
-\mu_0 K^{xx}(q,\omega)A_x,
\\
-\frac{\omega^2}{c^2}A^y+q^2A^y
&=
-\mu_0 K^{yy}(q,\omega)A_y,
\\
-\frac{\omega^2}{c^2}A^z+q^2A^z
&=
-\mu_0 K^{zz}(q,\omega)A_z
+\mu_0 K^{z0}(q,\omega)\frac{\phi}{c},
\label{eq: longitudinal mode2}
\\
\frac{\omega}{c^2}\phi-qA^z
&=
0.
\label{eq: longitudinal mode3}
\end{align}
Note that we use the metric $\eta^{\mu\nu}={\rm diag}(-,+,+,+)$ to write down the above equations.
One can see that the transverse ($A^x, A^y$) and longitudinal ($A^z$) components are decoupled.

In the low-energy and long-wavelength limit, the above field equation can be derived from an effective Lagrangian density
\begin{align}
\mathcal L_{\rm eff}
&=
-\frac{1}{4\mu_0}F_{\mu\nu}F^{\mu\nu}-V_{\rm eff}(A^\mu)
\end{align}
with an effective potential
\begin{align}
V_{\rm eff}(A^\mu)
&=
\frac{1}{2\mu_0}m^2c^2 A_\mu A^\mu,
\end{align}
where $F_{\mu\nu}=\partial_\mu A_\nu-\partial_\nu A_\mu$ is the field strength, and $m$ is the effective mass of the electromagnetic field
corresponding to
\begin{align}
m^2
&=
\frac{\mu_0}{c^2}
\lim_{q\to 0} \lim_{\omega\to 0} K^\perp(q,\omega)
=
\frac{\mu_0}{\pi c^2}
D_s.
\end{align}
Thus, the squared mass of the electromagnetic field is proportional to the Meissner weight.
In ordinary situations ($D_s>0$), the electromagnetic field acquires a positive squared mass
due to the Anderson-Higgs mechanism. In the opposite case ($D_s<0$),
the electromagnetic field becomes ``tachyonic'' with a negative squared mass.

\subsection{A. Transverse mode}

In order for the transverse modes to exist ($A^x, A^y\neq 0$), the dispersion must satisfy
\begin{align}
q^2-\frac{\omega^2}{c^2}
&=
-\mu_0 K^\perp(q,\omega),
\label{eq: transverse dispersion}
\end{align}
which corresponds to Eq.~(\ref{eq: Maxwell}) in the main text. While the condition (\ref{eq: transverse dispersion}) gives
a complicated nonlinear relation between $\omega$ and $q$, 
the situation becomes simplified at low momentum.

In the case of $U>0$, Eq.~(\ref{eq: transverse dispersion}) reduces to
\begin{align}
-\frac{\omega^2}{c^2}
&=
-8\mu_0 e^2t_h^2 C_{M,N}\left(\frac{1}{\omega+U}-\frac{1}{\omega-U}\right)
\label{eq: transverse mode q->0}
\end{align}
in the limit of $q\to 0$ [see Eq.~(\ref{eq: K q->0})]. The solution for $\omega^2$ is given by
\begin{align}
\omega^2
&=
\frac{1}{2}\left(
U^2 \pm \sqrt{U^4-64\mu_0 c^2 e^2 t_h^2 C_{M,N}U}
\right).
\label{eq: omega^2}
\end{align}
In order for $\omega$ to take a real value, the interaction strength $U$ must satisfy
\begin{align}
U^3
\ge
64\mu_0 c^2 e^2 t_h^2 C_{M,N},
\label{eq: U bound}
\end{align}
which is exactly the condition derived in the main text.
In the thermodynamic limit ($M,N\to\infty$ with $N/M$ being fixed),
$C_{M,N}$ approaches $\rho(1-\rho)$, where $\rho:=(N/2)/M$ is the doublon density.
In this limit, the condition (\ref{eq: U bound}) becomes
\begin{align}
\frac{U}{t_h}
\ge
4\left(\frac{\mu_0 c^2 e^2 \rho(1-\rho)}{t_h}\right)^{\frac{1}{3}}.
\end{align}

\begin{figure}[t]
\includegraphics[width=7cm]{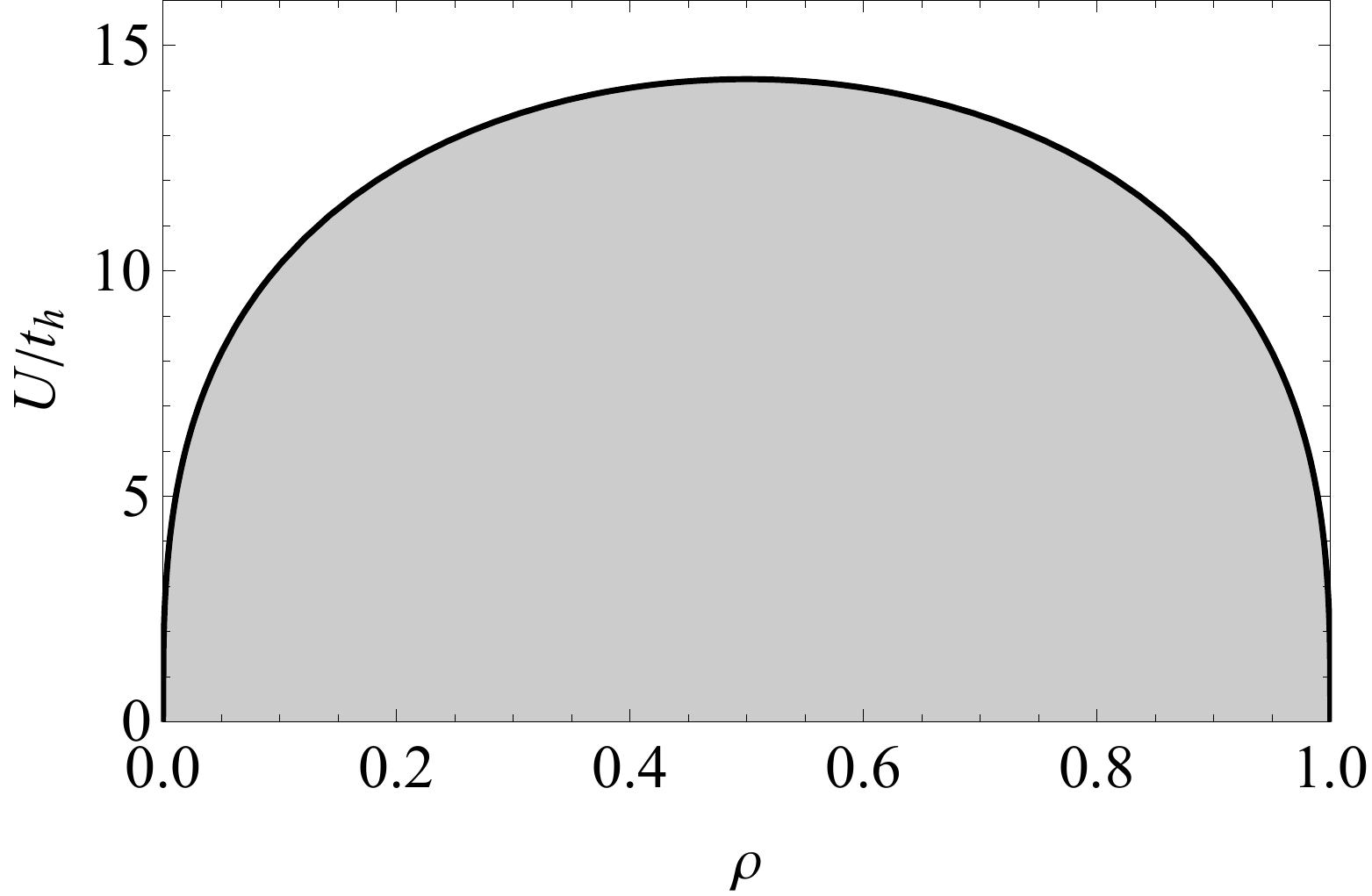}
\caption{Parameter space (shaded region) where the transverse electromagnetic field coupled to the $\eta$-pairing state in the Hubbard model
becomes unstable in the long-wavelength limit ($\bm q\to 0$).}
\label{fig: U range}
\end{figure}

In Fig.~\ref{fig: U range}, we plot the parameter space where the frequency $\omega$ has an imaginary part.
One can see that a wide range of the parameter region shows a dynamical instability of the electromagnetic field
coupled to the $\eta$-pairing state in the long-wavelength limit.
The real part of the frequency at $q\to 0$ is given by
\begin{align}
{\rm Re}\, \omega
&=
\frac{1}{2}\sqrt{
\sqrt{64\mu_0c^2e^2t_h^2C_{M,N}U}+U^2
},
\end{align}
which is proportional to $U^{1/4}$ at small $U$.
The imaginary part of the frequency at $q\to 0$ is given by
\begin{align}
{\rm Im}\, \omega
&=
\frac{1}{2}\sqrt{
\sqrt{64\mu_0c^2e^2t_h^2C_{M,N}U}-U^2
}
\end{align}
for $U^3<64\mu_0c^2e^2t_h^2C_{M,N}$.

More generally, if we take into account arbitrary $\bm q$ modes, we can prove that a dynamical instability exists
for all $U$ and $\rho$. First, we observe that $K^\perp(\bm q,\omega)$ in Eq.~(\ref{eq: K_perp suppl})
takes a real value if and only if $|\omega+U|>4t_h\sin\frac{q}{2}$ and $|\omega-U|>4t_h\sin\frac{q}{2}$.
Let us first consider the case of $0<\omega<U-4t_h\sin\frac{q}{2}$. 
Using the result for $K^\perp(\bm q,\omega)$ in Eq.~(\ref{eq: K_perp suppl}), the dispersion relation (\ref{eq: transverse dispersion})
can be written as
\begin{align}
q^2-\frac{\omega^2}{c^2}
&=
-8\mu_0 e^2t_h^2 C_{M,N}
\left\{
\frac{1}{\sqrt{(\omega+U)^2-16t_h^2\sin^2\frac{q}{2}}}
\right.
\notag
\\
&\quad
\left.
+\frac{1}{\sqrt{(\omega-U)^2-16t_h^2\sin^2\frac{q}{2}}}
\right\},
\end{align}
which is negative definite. Therefore, we have $\omega>cq$.
For the other case of $\omega>U+4t_h\sin\frac{q}{2}$, the condition (\ref{eq: transverse dispersion}) reads
\begin{align}
q^2-\frac{\omega^2}{c^2}
&=
-8\mu_0 e^2t_h^2 C_{M,N}
\left\{
\frac{1}{\sqrt{(\omega+U)^2-16t_h^2\sin^2\frac{q}{2}}}
\right.
\notag
\\
&\quad
\left.
-\frac{1}{\sqrt{(\omega-U)^2-16t_h^2\sin^2\frac{q}{2}}}
\right\},
\end{align}
which is positive definite since $(\omega+U)^2-16t_h^2\sin^2\frac{q}{2}>(\omega-U)^2-16t_h^2\sin^2\frac{q}{2}$.
Therefore, we have $\omega<cq$.

\begin{figure}[t]
\includegraphics[width=7cm]{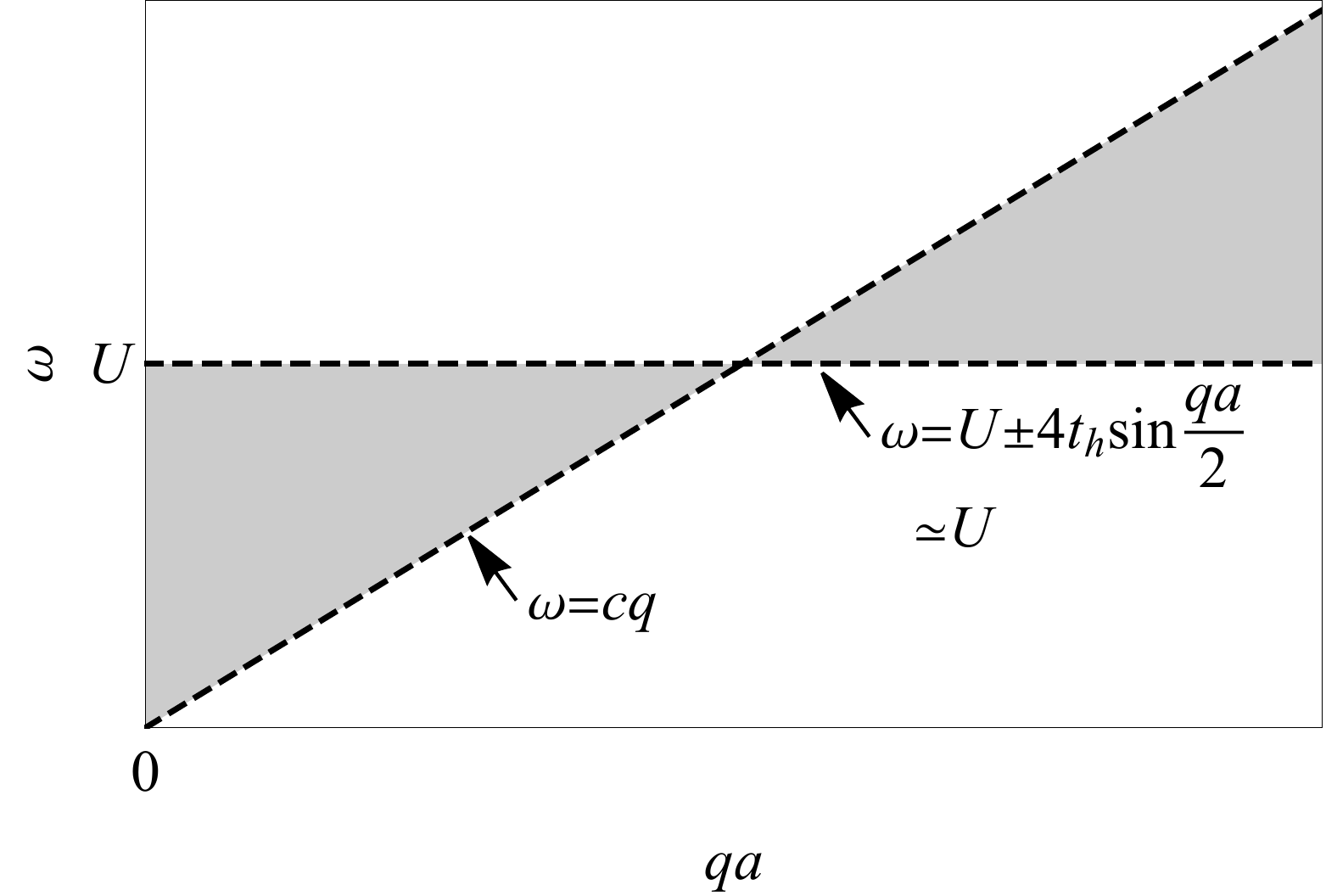}
\caption{Schematic illustration of the range of $(q,\omega)$ in which the real solution for Eq.~(\ref{eq: transverse dispersion}) 
with $U>0$ is allowed as shown by the shaded region for $qa\ll 1$. The dashed lines indicate that the boundary is not included.}
\label{fig: omega q range}
\end{figure}

In Fig.~\ref{fig: omega q range}, we plot the range of $(q, \omega)$ in which the real solution for Eq.~(\ref{eq: transverse dispersion})
is allowed. Since the allowed region is separated into two disjoint islands, 
it is clear from a topological point of view that the real band dispersion $\omega=\omega(q)$ for all $q$ is not possible.
This means that there must always be a region in $(q, \omega)$ where the solution for Eq.~(\ref{eq: transverse dispersion}) 
becomes complex. Thus, there is a dynamical instability for arbitrary $U(>0)$ and $\rho$.

In the attractive case ($U<0$), the mode equation in the long-wavelength limit is similarly given by Eq.~(\ref{eq: transverse mode q->0}).
The solution for $\omega^2$ is the same as Eq.~(\ref{eq: omega^2}). When $U<0$, $\omega^2$ is always real. However, 
there exists a solution with $\omega^2<0$ when one chooses the minus sign in Eq.~(\ref{eq: omega^2}).
Hence the frequency becomes imaginary for $\bm q\to 0$. The inverse of the imaginary part of $\omega$ is given by
\begin{align}
\frac{1}{{\rm Im}\,\omega}
&=
\frac{1}{\sqrt{
\frac{1}{2}\left(
\sqrt{U^4-64\mu_0 c^2 e^2 t_h^2 C_{M,N}U}-U^2
\right)}}.
\label{eq: tau}
\end{align}
In the attractive case, the electromagnetic field is dynamically unstable against the $\bm q=0$ mode.
This corresponds to the fact that the electromagnetic field has a negative squared mass.
Physically, the electromagnetic field with a long wavelength penetrates deeply inside the $\eta$ pairing state,
transferring the kinetic energy of doublons to the electromagnetic field.
The time scale of the growth of this instability is determined by Eq.~(\ref{eq: tau}).

Combining the arguments for the two cases ($U>0$ and $U<0$), we have established that 
the electromagnetic field coupled to the $\eta$-pairing state is always dynamically unstable for all $U$ and $\rho$.

\subsection{B. Longitudinal mode}

The mode equation for the longitudinal components can be derived from Eqs.~(\ref{eq: longitudinal mode1}),
(\ref{eq: longitudinal mode2}), and (\ref{eq: longitudinal mode3}). To simplify the situation, we focus on the low-momentum region ($qa\ll 1$).
In this region, the dispersion is determined by
\begin{align}
q^2-\frac{\omega^2}{c^2}
&=
-\mu_0 K^\parallel(q,\omega)
\left(1-\frac{c^2q^2}{\omega^2}\right),
\end{align}
where we have used the relation (\ref{eq: K matrix}).
In the limit of $q\to 0$, the mode equation becomes
\begin{align}
-\frac{\omega^2}{c^2}
&=
-\mu_0 \lim_{q\to 0} K^\parallel(q,\omega).
\label{eq: longitudinal dispersion}
\end{align}
If we recall the relation (\ref{eq: K q->0}), the dispersion (\ref{eq: longitudinal dispersion}) is the same
as that of the transverse mode (\ref{eq: transverse dispersion}). Therefore, the longitudinal mode
has the same dynamical instability as the transverse one at low momentum.

\section{IV. Magnetic properties of tachyonic superconductors}
\label{sec: magnetic properties}

In this section, we describe static magnetic properties of tachyonic superconductors realized as the $\eta$-pairing states
in the Hubbard model with $U<0$.
As shown in the main text and in the preceding section, the tachyonic superconductors are dynamically unstable.
Here we focus on the response of the $\eta$-pairing states against static magnetic fields within the linear-response regime,
and do not consider their decay dynamics.

The static magnetic field $\bm B$ obeys the following Maxwell equations:
\begin{align}
\nabla \cdot \bm B
&=
0,
\label{eq: static Maxwell1}
\\
\nabla \times \bm B
&=
\mu_0 \bm j.
\label{eq: static Maxwell2}
\end{align}
We introduce a static vector potential $\bm A$ as $\bm B=\nabla\times\bm A$.
If we take the Coulomb gauge ($\nabla\cdot\bm A=0$), the equation for $\bm A$ becomes
\begin{align}
\nabla^2\bm A
&=
-\mu_0 \bm j.
\end{align}
To solve the equation, we assume a plane-wave form,
\begin{align}
\bm A(\bm r)
&=
\bm A_0 e^{i\bm q\cdot\bm r},
\label{eq: A plane wave}
\end{align}
with amplitude $\bm A_0$ and wave number $\bm q$.
Without loss of generality, we choose $\bm q =q\bm e_z$.
We apply the linear-response theory to the $\eta$-pairing states to obtain the equations for $\bm A_0$:
\begin{align}
\bm q\cdot \bm A_0
&=
0,
\label{eq: A0 1}
\\
-q^2 A_0^\mu
&=
-\mu_0 j^\mu(\bm q)
\notag
\\
&=
\mu_0\sum_\nu K^{\mu\nu}(\bm q, \omega=0) A_{0\nu}
\quad
(\mu,\nu=x,y,z).
\label{eq: A0 2}
\end{align}
From Eq.~(\ref{eq: A0 1}), we find $A_0^z=0$.
In order for the solution $A_0^\mu\neq 0$ to exist, $q$ must satisfy a nonlinear equation,
\begin{align}
q^2
&=
-\mu_0 K^\perp(\bm q,\omega=0)
\notag
\\
&=
-16\mu_0 e^2t_h^2 C_{M,N}
\notag
\\
&\quad\times
\begin{cases}
\frac{{\rm sgn}(U)}{\sqrt{U^2-16t_h^2\sin^2\frac{q}{2}}} & \mbox{for}\; |U|>4t_h\sin\frac{q}{2};
\\
0 & \mbox{for}\; |U|<4t_h\sin\frac{q}{2},
\end{cases}
\label{eq: q}
\end{align}
where we have used the result (\ref{eq: K_perp suppl}).

\begin{figure}[t]
\includegraphics[width=4.2cm]{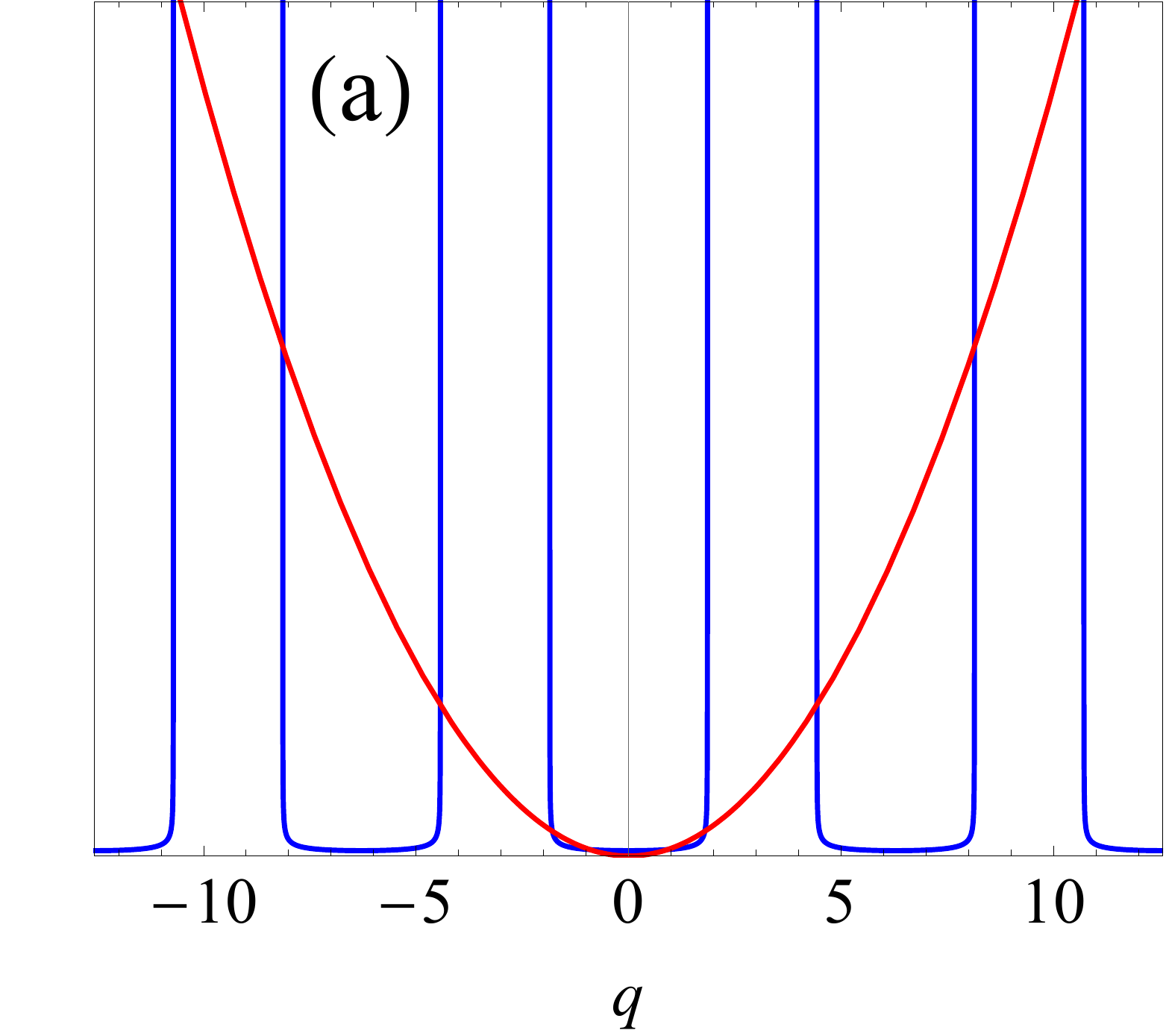}
\includegraphics[width=4.2cm]{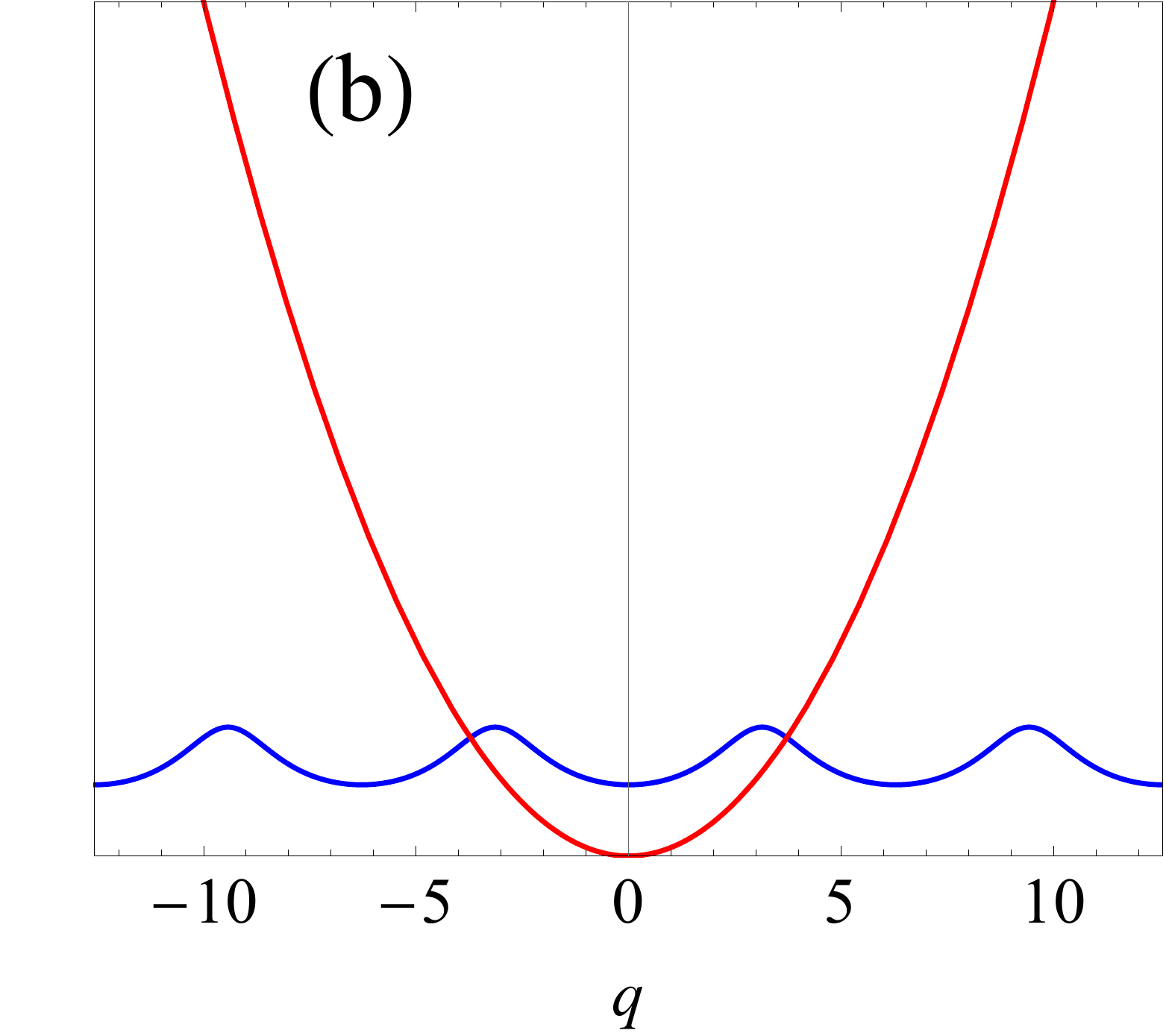}
\caption{Graphical illustration of Eq.~(\ref{eq: q}) for (a) $-4t_h<U<0$ and (b) $U<-4t_h$.
The left-hand side of Eq.~(\ref{eq: q}) is shown by the red curve, while the right-hand side is shown by the blue curve.}
\label{fig: q}
\end{figure}

One can immediately see that a real solution for Eq.~(\ref{eq: q}) does not exist for $U>0$.
This is nothing but the Meissner effect; that is, a magnetic field cannot propagate freely into superconductors.
In fact, the magnetic field decays exponentially in space with the penetration depth
\begin{align}
\lambda
&=
\sqrt{\frac{U}{16\mu_0e^2t_h^2 C_{M,N}}}.
\end{align}
On the other hand, when $U<0$ a real solution is possible as is clear from the graphical illustration of Eq.~(\ref{eq: q}) in Fig.~\ref{fig: q}.
Thus, a magnetic field {\it can} penetrate into tachyonic superconductors without decay.

\begin{table}[t]
\begin{center}
\begin{tabular}{cc}
\hline
& \# of real solutions for Eq.~(\ref{eq: q})
\\
\hline
\hline
$U>0$ & 0
\\
$-4t_h<U<0$ & $\infty$
\\
$U<-4t_h$ & 2
\\
\hline
\end{tabular}
\caption{List of the number of real solutions for Eq.~(\ref{eq: q}) with $16\mu_0e^2t_ha/\hbar^2\ll 1$.}
\label{table: q solution}
\end{center}
\end{table}

From Fig.~\ref{fig: q}, we can see that the number of real solutions for Eq.~(\ref{eq: q}) changes at the boundary of $U=-4t_h$.
In Table.~\ref{table: q solution}, we list the number of real solutions for Eq.~(\ref{eq: q}). This result suggests that
the number of modes of magnetic fields that can propagate inside tachyonic superconductors for $-4t_h<U<0$
is different from that for $U<-4t_h$. Following the main text, we call the former a type-I tachyonic superconductor, and the latter 
a type-II tachyonic superconductor.

In the case of the type-I tachyonic superconductor, we further classify the solutions into two types according to the number of
real solutions in the range of $-\pi<q<\pi$. When $U$ belongs to the range $(-4t_h<)$ $U_\ast<U<0$ with a certain boundary $U_\ast$,
real solutions for Eq.~(\ref{eq: q}) do not exist in $-\pi<q<\pi$. 
Since $16\mu_0e^2t_ha/\hbar^2\ll 1$ for ordinary materials ($t_h\sim$ 1[eV] and $a\sim$ 1[\AA]), the threshold $U_\ast$
is approximately given as
\begin{align}
\frac{U_\ast}{t_h}
&\approx
-2\sqrt{3}(4\mu_0 e^2t_hC_{M,N})^{\frac{1}{3}}.
\end{align}
For $-4t_h<U<U_\ast$, there exist four real solutions in $-\pi<q<\pi$, which are denoted by $\pm q_1$ and $\pm q_2$
with $0<q_1<q_2$. Using $16\mu_0e^2t_ha/\hbar^2\ll 1$ again, we can approximately evaluate $q_1$ and $q_2$ as
\begin{align}
q_1
&\approx
\sqrt{\frac{16\mu_0e^2t_h^2C_{M,N}}{-U}},
\label{eq: q_1}
\\
q_2
&\approx
2\sin^{-1}\left(\frac{-U}{4t_h}\right).
\label{eq: q_2}
\end{align}
Note that $q_1$ corresponds to the inverse of the analytically continued London's penetration depth ($q_1=|\lambda|^{-1}$).
If we extend the range of $q$ to $-\infty<q<\infty$, there are infinitely many real solutions.
They are approximately given as $\pm q_1$ and $\pm q_2+2n\pi$ ($n\in\mathbb Z$).

\begin{figure}[t]
\includegraphics[width=7cm]{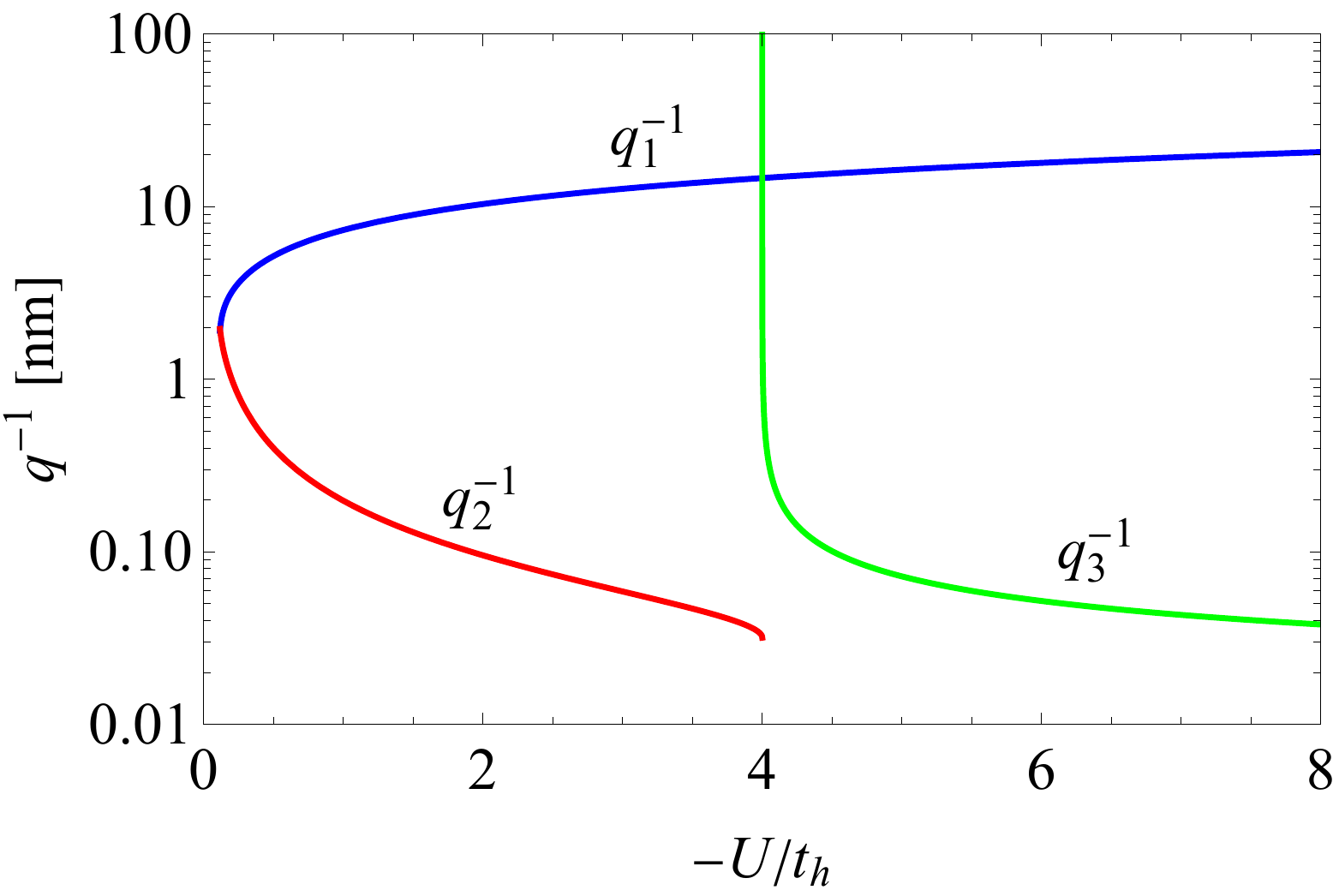}
\caption{Log plot of $q_1^{-1}$, $q_2^{-1}$, and $q_3^{-1}$ for $t_h=$ 1[eV], $a=$ 1[\AA], and $\rho=0.5$
as a function of $-U/t_h$.}
\label{fig: q inverse}
\end{figure}

In the case of the type-II tachyonic superconductor, there are only two real solutions $\pm q_1$ approximately given by (\ref{eq: q_1}).
Instead, there emerge infinitely many complex solutions approximately given by
$\pm iq_3+2n\pi$ ($n\in\mathbb Z$) with
\begin{align}
q_3
&\approx
2\cosh^{-1}\left(\frac{-U}{4t_h}\right).
\label{eq: q_3}
\end{align}
Physically, these solutions correspond to a magnetic field localized near the surface of a tachyonic superconductor.
The localization length $q_3^{-1}$ diverges at $U=U_c'=-4t_h$ as $q_3^{-1}\sim |U-U_c'|^{-1/2}$.

In Fig.~\ref{fig: q inverse}, we plot $q_1^{-1}$, $q_2^{-1}$, and $q_3^{-1}$ for typical parameters.
The $q_1$ mode produces a long-period magnetic structure with the period of the order of $|\lambda|\gg a$.
On the other hand, the $q_2$ mode provides a short-period magnetic structure with the period length of the order of the lattice constant $a$.
In type-I tachyonic superconductors, both the long- and short-period structures are allowed to exist,
whereas in type-II tachyonic superconductors the short-period magnetic structure is screened, and it can penetrate
only near the surface. 

\begin{figure}[t]
\includegraphics[width=4.2cm]{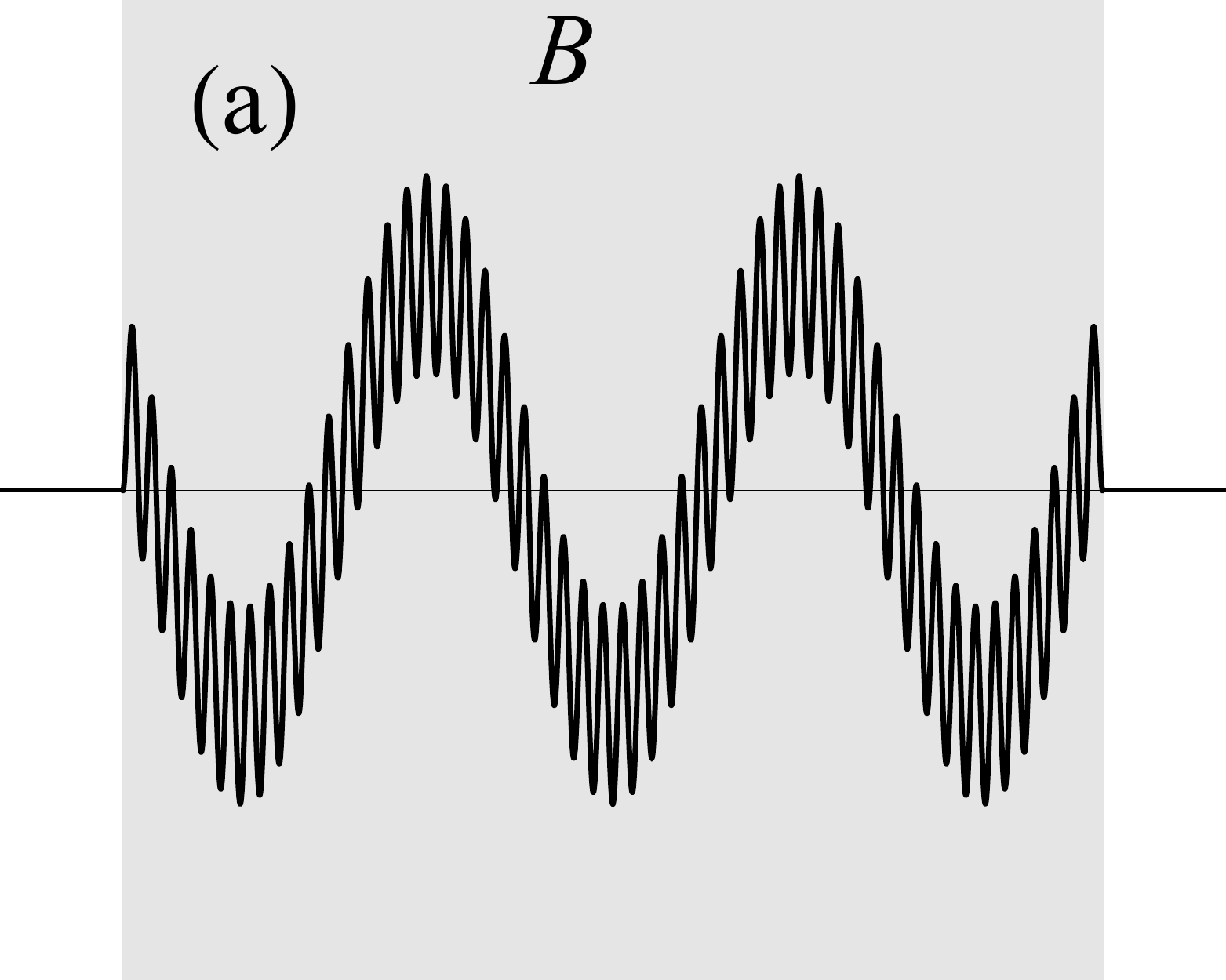}
\includegraphics[width=4.2cm]{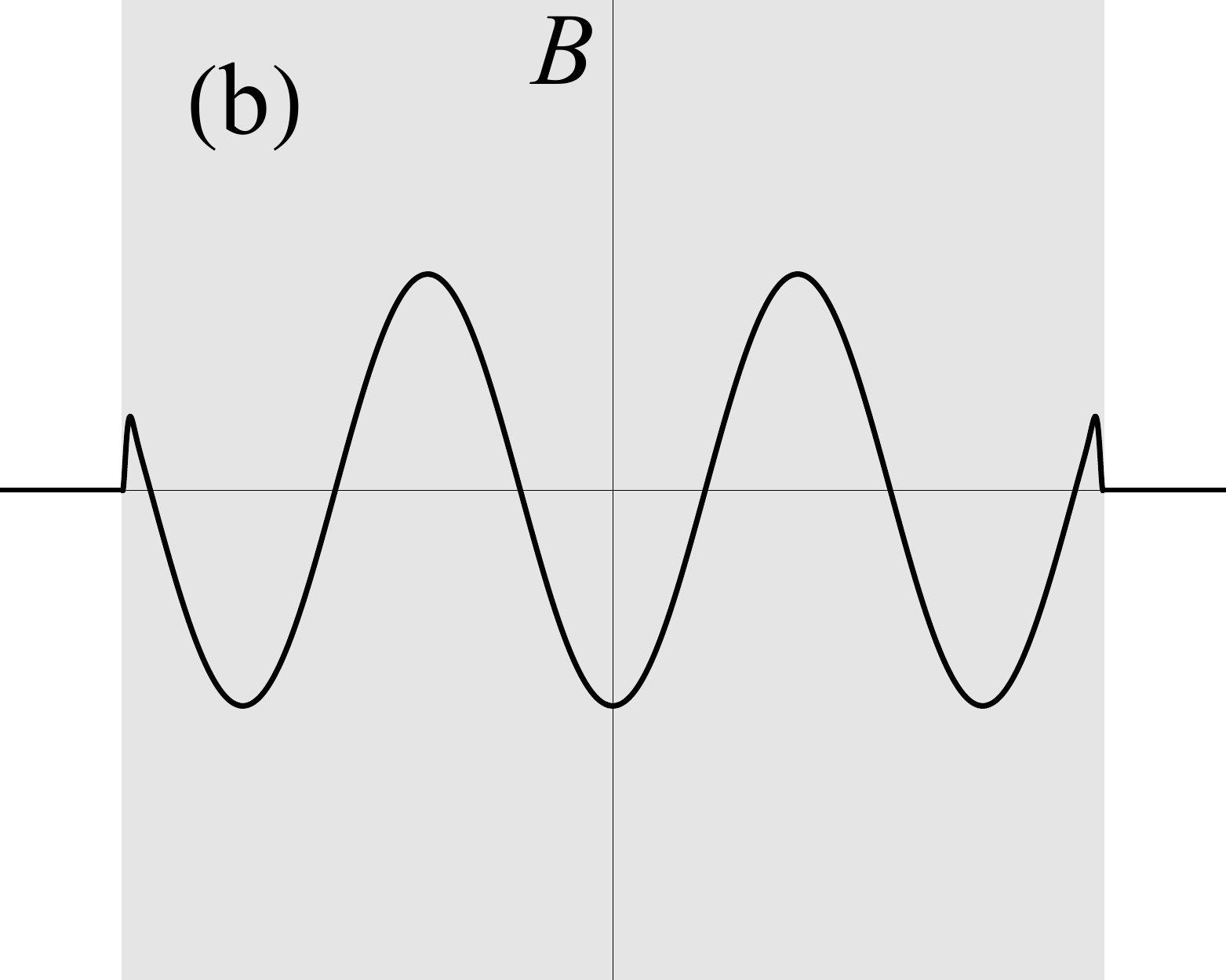}
\caption{Examples of static magnetic fields $B=B(x)$ that can exist inside
(a) type-I and (b) type-II tachyonic superconductors shown by shaded regions.}
\label{fig: magnetic structure}
\end{figure}

In Fig.~\ref{fig: magnetic structure}, we show examples of magnetic fields that can be
realized in type-I [Fig.~\ref{fig: magnetic structure}(a)] and type-II (b) tachyonic superconductors.
Even when there is no magnetic field outside of tachyonic superconductors,
nonzero magnetic fields can be trapped statically inside tachyonic superconductors.
To support those magnetic fields, constant electric currents are flowing in the bulk of tachyonic superconductors.
In the type-II tachyonic superconductor, there is also a surface current to satisfy the boundary condition.

\begin{figure}[t]
\includegraphics[width=4.cm]{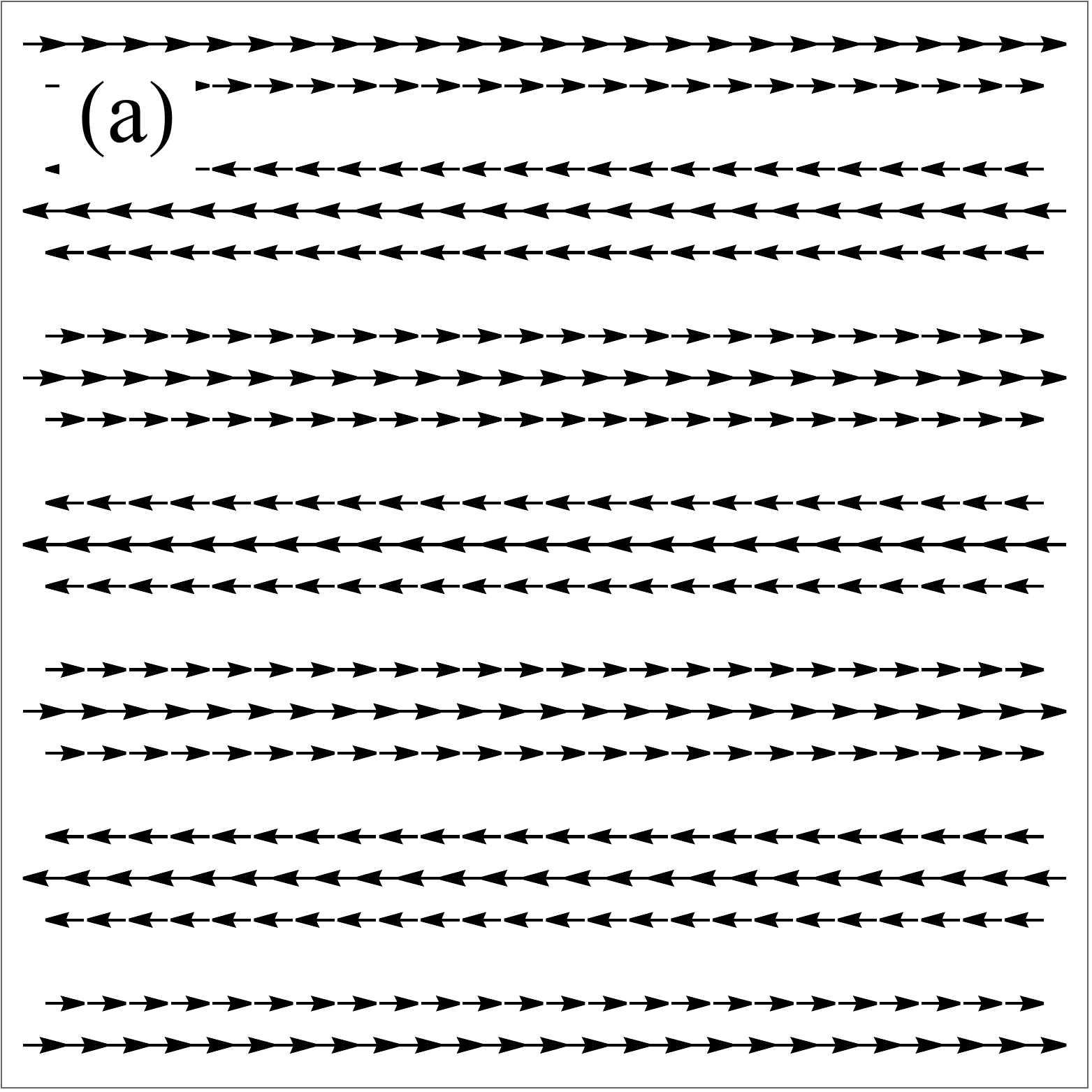}
\hspace{.1cm}
\includegraphics[width=4.cm]{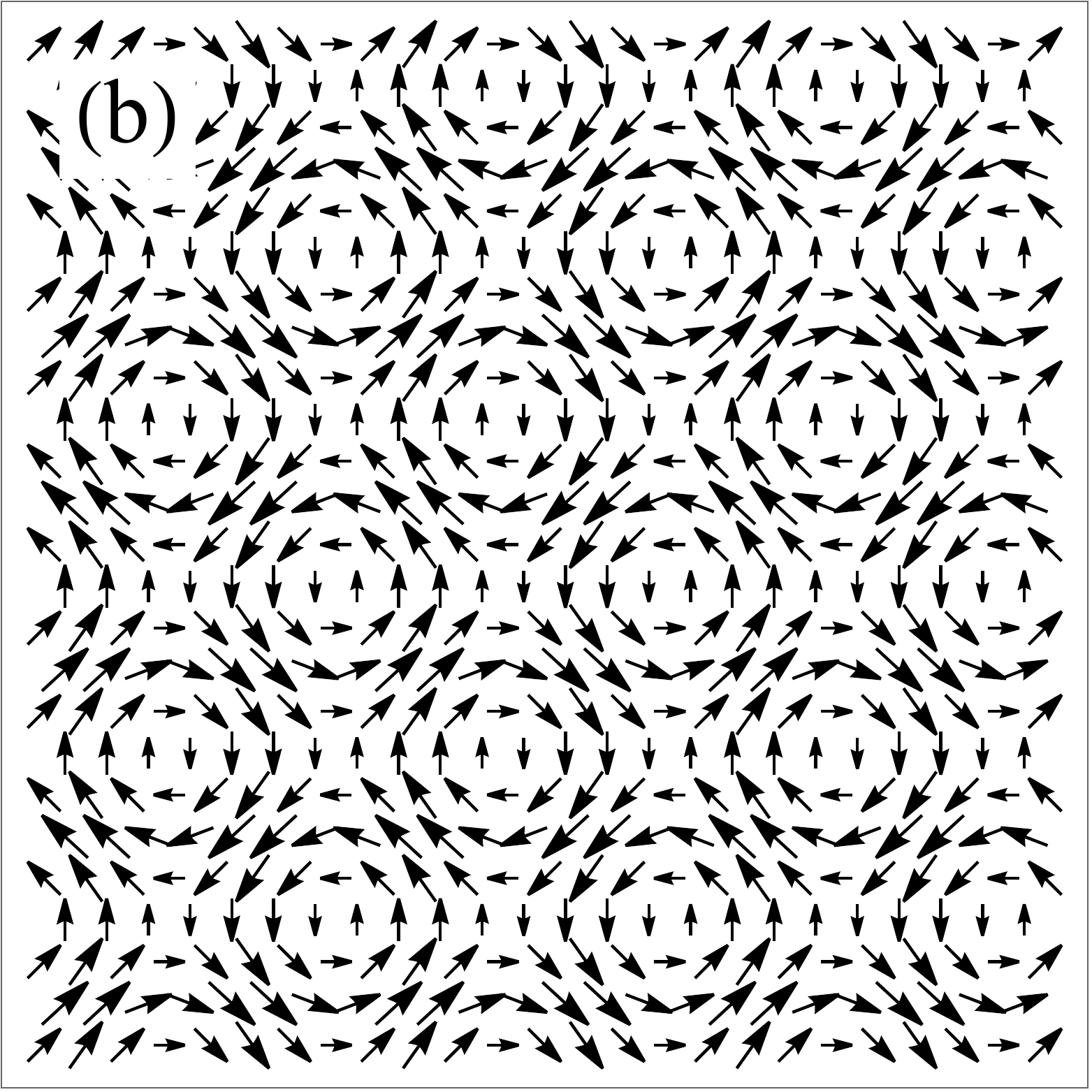}
\caption{Examples of (a) stripe-like and (b) vortex-antivortex-like magnetic structures in tachyonic superconductors
shown in a two-dimensional plane.}
\label{fig: magnetic structure 2d}
\label{fig: vortex}
\end{figure}

Various $q$ modes can be linearly superposed in several different directions.
In Fig.~\ref{fig: vortex}, we show two examples of magnetic structures that can be realized
in tachyonic superconductors. If one only takes a single $q_1$ mode, it gives a stripe-like structure
as shown in Fig.~\ref{fig: vortex}(a). Here we neglect short-period structures ($q_2$ modes).
If one superposes two $q_1$ modes in $x$ and $y$ directions, one obtains a square lattice with alternating
vortex and antivortex structures as shown in Fig.~\ref{fig: vortex}. This is to be contrasted with Abrikosov's triangular lattice
of vortices in type-II superconductors.
One can also superimpose three $q_1$ modes
in three different directions, creating a three-dimensional magnetic structure (not shown).
In this way, various configurations of magnetic fields can be trapped in tachyonic superconductors.
We note, however, that these structures are not dynamically stable as shown in the main text and
in the preceding section.

\section{V. Spontaneous light emission}

In this section, we evaluate the rate of spontaneous light emission for $\eta$-pairing states in the Hubbard model with $U>0$.
In the repulsive case, doublons can decay spontaneously into pairs of single particles by emitting light with frequency $\omega=U$.
After emitting light, the $\eta$-pairing state $|\psi_N\rangle$ is transformed into
\begin{align}
J^\mu(\bm q=0)|\psi_N\rangle
&=
\frac{e}{\sqrt{\mathcal N_N}}N(\eta^+)^{\frac{N}{2}-1}
\sum_{\bm k} v^\mu(\bm k)c_{\bm k\uparrow}^\dagger c_{\bm Q-\bm k\downarrow}^\dagger |0\rangle,
\label{eq: J|psi_N>}
\end{align}
where $v^\mu(\bm k)=\frac{\partial\varepsilon_{\bm k}}{\partial k^\mu}=2t_h\sin k^\mu$ is the group velocity 
($\varepsilon_{\bm k}=-2t_h\sum_{\mu=x,y,z} \cos k^\mu$ is the band dispersion).

Let us define a one-doublon-broken state \cite{Yang1989}
\begin{align}
|\zeta_{N,\bm a}\rangle
&=
\frac{1}{\sqrt{\mathcal N_{N,\bm a}}}(\eta^+)^{\frac{N}{2}-1} \eta_{\bm a}^+ |0\rangle,
\end{align}
where $\mathcal N_{N,\bm a}$ is the normalization constant 
(such that $\langle \zeta_{N,\bm a}|\zeta_{N,\bm a'}\rangle=\delta_{\bm a,\bm a'}$), 
$\bm a$ represents a lattice-site coordinate, and
\begin{align}
\eta_{\bm a}^+
&=
\sum_{\bm k} e^{-i\bm k\cdot\bm a} c_{\bm k\uparrow}^\dagger c_{\bm Q-\bm k\downarrow}^\dagger.
\end{align}
The state $|\zeta_{N,\bm a}\rangle$ consists of $\frac{N}{2}-1$ doublons with momentum $\bm Q$ and
two unpaired particles with the lattice spacing $\bm a$. One can show that $|\zeta_{N,\bm a}\rangle$ with $\bm a\neq 0$ is
an exact eigenstate of the Hubbard model [Eq.~(\ref{eq: Hubbard}) in the main text] with the eigenenergy $-U$.
At $\bm a=0$, we have $|\zeta_{N,\bm a}\rangle=|\psi_N\rangle$.
Using $|\zeta_{N,\bm a}\rangle$, we can write the one-photon emitted state (\ref{eq: J|psi_N>}) as
\begin{align}
J^\mu(\bm q=0)|\psi_N\rangle
&=
e\sqrt{\frac{\mathcal N_{N,\bm a}}{\mathcal N_N}}\frac{N}{M}
\sum_{\bm k} v^\mu(\bm k) \sum_{\bm a\neq 0} e^{i\bm k\cdot\bm a} |\zeta_{N,\bm a}\rangle.
\end{align}
Therefore, all the states that are accessible by one-photon emission are covered by the eigenstates 
$|\zeta_{N,\bm a}\rangle$ ($\bm a\neq 0$).

The rate of spontaneous emission is given by Einstein's A coefficient \cite{LoudonBook}:
\begin{align}
\Gamma
&=
\frac{\omega^3}{3\pi\varepsilon_0 c^3\hbar} \sum_\mu
|\langle e|P^\mu|\sigma\rangle|^2,
\end{align}
where $P^\mu$ is the polarization operator, $|\sigma\rangle$ is an initial state, and $|e\rangle$
is a one-phonon emitted state. In the present case, we take $|\sigma\rangle=|\psi_N\rangle$ and $|e\rangle=|\zeta_{N,\bm a}\rangle$.
Since $\frac{d}{dt} P^\mu=J^\mu$, the rate $\Gamma$ is rewritten as
\begin{align}
\Gamma
&=
\frac{\omega}{3\pi\varepsilon_0 c^3\hbar}
\sum_\mu \sum_{\bm a}
|\langle \zeta_{N,\bm a}|J^\mu|\psi_N\rangle|^2
\notag
\\
&=
\frac{e^2\omega}{3\pi\varepsilon_0 c^3\hbar}
\sum_\mu \sum_{\bm a}
\frac{\mathcal N_{N,\bm a}}{\mathcal N_N} \frac{N^2}{M^2}
\left|
\sum_{\bm k} v^\mu(\bm k) e^{i\bm k\cdot\bm a}
\right|^2.
\end{align}
Using Eq.~(\ref{eq: normalization}) and
\begin{align}
\mathcal N_{N,\bm a}
&=
\frac{M(M-2)!(\frac{N}{2}-1)!}{(M-\frac{N}{2}-1)!}
\quad
(\bm a\neq 0),
\end{align}
we obtain
\begin{align}
\Gamma
&=
\frac{8e^2\omega t_h^2a^2dC_{M,N}}{3\pi\varepsilon_0 c^3\hbar^3}
M,
\label{eq: Gamma}
\end{align}
where $d$ is the dimension of the system.
One can see that the rate $\Gamma$ is proportional to the system size, which is natural because
the doublon decay can take place at any lattice site with equal probability.

For ordinary three-dimensional materials, we substitute $t_h=$ 1[eV], $\omega=U=$ 1[eV], $a=$ 1[\AA], $\rho=0.5$, 
and $d=3$ in Eq.~(\ref{eq: Gamma}), obtaining
\begin{align}
\frac{\Gamma}{M}
&=
2.3\times 10^7\; [{\rm s}^{-1}].
\end{align}



\end{document}